\documentclass[twocolumn,superscriptaddress,floatfix,preprintnumbers,prd ,nofootinbib,hyperref]{revtex4-2}
\pdfoutput=1
\usepackage[colorlinks=true,breaklinks=true]{hyperref}
\usepackage[normalem]{ulem}
\usepackage[utf8]{inputenc}
\hypersetup{allcolors=[rgb]{0.0 0.0 0.6},linkcolor=[rgb]{0.75 0.05 0.05}}
\usepackage{amsmath,amssymb}
\usepackage{mathtools}
\usepackage{cancel}
\usepackage{tensor}
\usepackage{amsmath, amssymb}
\usepackage{braket}
\usepackage{dcolumn}
\usepackage{multirow}
\usepackage{booktabs}
\usepackage[dvipsnames]{xcolor}
\usepackage{float}
\usepackage{bbold}
\usepackage{letltxmacro}
\LetLtxMacro{\oldcite}{\cite}
\renewcommand{\cite}[1]{\mbox{\oldcite{#1}}}
\usepackage{comment}
\long\def\exclude#1{}

\DeclareMathOperator{\GeV}{GeV}

\newcommand{\beq}{\begin{equation}}
\newcommand{\eeq}{\end{equation}}

\def\ga{\,\,\raise0.14em\hbox{$>$}\kern-0.76em\lower0.28em\hbox
{$\sim$}\,\,}

\newcommand{\ie}{\emph{i.e.},~}
\newcommand{\eg}{\emph{e.g.},~}

\long\def\exclude#1{}

\newcommand{\OO}[1]{\mathcal O \left( #1 \right)}

\allowdisplaybreaks

\bibpunct{[}{]}{,}{n}{}{,}

\hypersetup{
    colorlinks=true,
    citecolor=[rgb]{.1, .7, .6},
    linkcolor=[rgb]{.2, .55, .95},
    filecolor=magenta,
    urlcolor=[rgb]{.1, .7, .6},
}

\definecolor{cornellGreen}{HTML}{6EB43F}
\definecolor{cornellRed}{HTML}{B31B1B}
\definecolor{gcsblue}{HTML}{4169E1}

\begin{document}
\preprint{DESY-26-156}
\title{Toward the minimal set of states in axion superradiance}
\author{Enrico Cannizzaro}
\email{enrico.cannizzaro@tecnico.ulisboa.pt}
\affiliation{CENTRA, Departamento de F\'{\i}sica, Instituto Superior T\'ecnico -- IST, Universidade de Lisboa -- UL, Avenida Rovisco Pais 1, 1049 Lisboa, Portugal}
\affiliation{Max Planck Institute for Gravitational Physics (Albert Einstein Institute), Am Mühlenberg 1, 14476 Potsdam, Germany}
\author{Thomas F.~M.~Spieksma}
\email{spieksma@ias.edu}
\affiliation{Rudolf Peierls Centre for Theoretical Physics, University of Oxford, Parks Road, Oxford OX1 3PU, United Kingdom}
\affiliation{School of Natural Sciences, Institute for Advanced Study, 1 Einstein Drive, Princeton, NJ 08540, USA}
\author{Samuel J.~Witte}
\email{samuel.witte@physics.ox.ac.uk}
\affiliation{Rudolf Peierls Centre for Theoretical Physics, University of Oxford, Parks Road, Oxford OX1 3PU, United Kingdom}
\affiliation{Deutsches Elektronen-Synchrotron DESY, Notkestraße 85, 22607 Hamburg, Germany}
\affiliation{II. Institute of Theoretical Physics, Universität Hamburg, 22761, Hamburg, Germany}

\begin{abstract} 
Observational searches for black hole superradiance require the ability to predict the growth and evolution of quasi-bound states on astrophysical timescales. For axions (and more broadly speaking, for most classes of light spin-0 particles), self-interactions efficiently redistribute energy and angular momentum from the leading-order superradiant states (\ie the fastest growing) into higher-order, more weakly bound states, and they do so on short timescales. This has made it difficult to compute even the most mundane of observables across large regions of parameter space. In this work we extend the study of self-interacting systems, following the evolution of nearly 300 coupled quasi-bound states. We solve for the evolution of a $10 \, M_\odot$ black hole over a Salpeter timescale (\ie the timescale on which accretion can feasibly begin to alter the properties of the black hole), and of $10^4$ and $ 10^8\,M_\odot$ black holes over a Gyr timescale, identifying for each axion mass the minimal set of quasi-bound states required to describe the evolution of the black hole and the shape of the superradiant cloud. In most cases, the evolution converges as increasingly weakly bound states are added; we identify a set of no less than 203 distinct quasi-bound states that play a fundamental role in establishing this convergence across the parameter space of interest. In small regions of parameter space, however, we are unable to identify converged solutions, and we show that wide resonant features and nearly degenerate scattering rates favor a redistribution of mass to bound states with principal quantum numbers $n \gtrsim 45$.  We further show that the cloud can extend to radii orders of magnitude larger than usually assumed, that gravitational-wave level transitions can prevent the growth of superradiant states, and that both the gravitational-wave spectrum and the interaction with the environment are far richer than previously appreciated.
\end{abstract}
\maketitle
\section{Introduction}
\label{sec:intro}
\begin{figure*}
    \includegraphics[width=\textwidth]{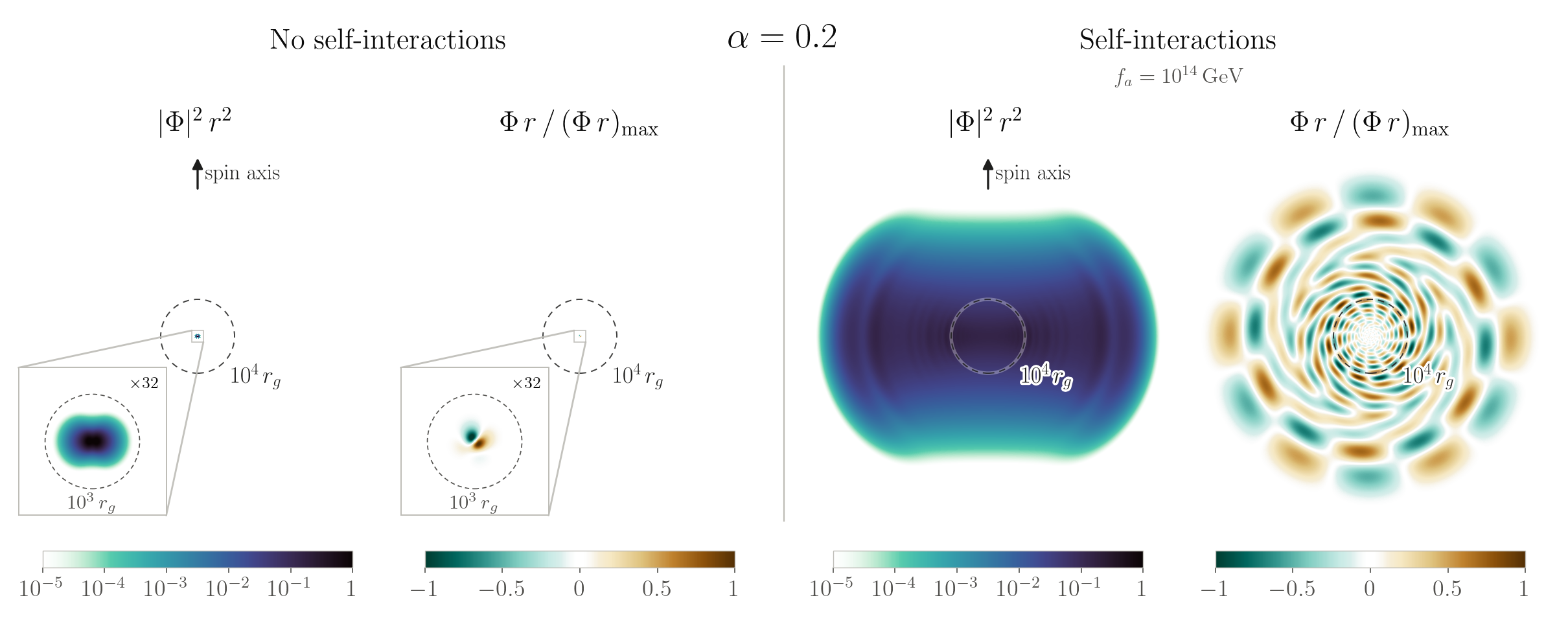}
    \caption{Radially-weighted energy density (side view) and axion field value $\Phi$ (projected in the equatorial plane) for a superradiant cloud formed around an initial black hole with mass $M = 10 \, M_\odot$ and $\tilde{a} = 0.95$ at a time $t=3 \times 10^4$ years. The gravitational coupling is $\alpha = 0.2$. Left panel shows the result for a free scalar field, while the right panel shows the structure of the cloud for an axion with self-interactions and $f_a = 10^{14}$ GeV. For reference, we have included a radial scale illustrating $r=10^3 \, r_g$ and $10^4 \, r_g$.}
\label{fig:overview_fig}
\end{figure*}

Axions are among the best-motivated extensions of the Standard Model, as they can explain the origin of charge-parity (CP) conservation in QCD \cite{Peccei:1977hh, Weinberg:1977ma, Wilczek:1977pj}, provide a compelling dark matter candidate \cite{Preskill:1982cy, Abbott:1982af,Dine:1982ah}, and emerge (in abundance) from some of our most well-motivated UV extensions of the Standard Model, such as String Theory  \cite{Svrcek:2006yi, Arvanitaki:2009fg}. A generic challenge in probing such particles is that their couplings to the Standard Model fields are typically extremely weak. Extreme astrophysical environments (\eg the near-field regime of black holes and neutron stars), however, offer a powerful and promising arena in the search for these elusive particles. This is because there exist various mechanisms operating in these environments that allow for the efficient production and confinement of axions, leading to large axion field values (see, \eg \cite{ Arvanitaki:2010sy, Baryakhtar:2022hbu, Noordhuis:2023wid}) which can offset the feeble nature of the interactions and leave strong, detectable, imprints in these systems.
In this context, black hole superradiance stands out as one of the most powerful probes~\cite{ZelDovich1971,ZelDovich1972,Starobinsky:1973aij,Brito:2015oca}. A massive bosonic field around a rotating Kerr black hole admits quasi-bound states, labeled by quantum numbers $\ket{n\ell m}$ in analogy with the hydrogen atom, whose occupation numbers grow exponentially for low-frequency modes that satisfy the superradiant condition
\begin{equation}
    \omega < m \Omega_{\rm H} \,,
\label{eq:SR_cond}
\end{equation}
where $m$ is the azimuthal number and $\Omega_{\rm H}$ the black hole horizon angular velocity. For an appropriate range of boson masses, a black hole of a given mass is unstable to the bosonic field, and the instability timescale can be remarkably short, allowing the rotational energy of the black hole to be efficiently extracted on sub-astrophysical timescales. The observation of rapidly spinning black holes -- through either electromagnetic or gravitational-wave (GW) observations -- can therefore be used to place stringent constraints on the existence of ultralight bosons (see, \eg \cite{Witte:2024drg,Caputo:2025oap,Aswathi:2025nxa} for the leading superradiant constraints on axions). The extracted energy populates a macroscopic ``boson cloud'', which can contain up to $\mathcal{O}(10\%)$ of the black hole mass \cite{Brito:2014wla,East:2017ovw,Herdeiro:2021znw}, and which can produce a variety of observational signatures, including (nearly) monochromatic GW emission \cite{Tsukada:2018mbp,Palomba:2019vxe,Ng:2020jqd,LIGOScientific:2021rnv,Yuan:2022bem,Ellis:2026gwt} and an imprint on the GWs emitted from binary systems \cite{Arvanitaki:2010sy,Arvanitaki:2014wva,Arvanitaki:2016qwi,Ferreira:2017pth,Hannuksela:2018izj,Zhang:2019eid,Palomba:2019vxe,Baumann:2018vus,Tsukada:2020lgt,Baumann:2019ztm,Baumann:2021fkf,Baumann:2022pkl,Tong:2022bbl,Siemonsen:2022yyf,Fan:2023jjj,Tomaselli:2023ysb,Duque:2023seg,Boskovic:2024fga,Tomaselli:2024bdd,Tomaselli:2024dbw,May:2024npn,Boskovic:2025ixx,Roy:2025qaa,Boskovic:2026eth}.
Much of the early work studying black hole superradiance focused on understanding the behavior and evolution of free bosonic fields (\ie the limit in which \emph{all} interactions are assumed to be negligible). While this may seem at first glance to be a reasonable approximation to broad classes of light bosonic fields, the existence of even tiny couplings with the ambient medium or between the bosonic fields themselves can have profound implications for the evolution of the system (in effect, this is simply due to the fact that occupation numbers $N$ of a fully saturated cloud can exceed $10^{70}$, meaning the enormous densities can offset the smallness of the interactions). Generally speaking, interactions tend to dissipate energy from the high density axion field configuration, potentially quenching the superradiant growth, and in some cases generating new observational signatures of these systems \cite{Arvanitaki:2009fg, Arvanitaki:2010sy, Arvanitaki:2014wva, Rosa:2017ury, Ikeda:2018nhb, Baryakhtar:2020gao, Blas:2020kaa, Mathur:2020aqv, Caputo:2021efm, East:2022ppo, Cannizzaro:2022xyw, Siemonsen:2022ivj, Spieksma:2023vwl,Cheng:2026dsq}.
In the case of axions, the existence of a discrete shift symmetry restricts the effective form of the potential, leading to the presence of self-interactions.\footnote{Similar arguments also apply to dilatons, or other light spin-0 fields with an approximate symmetry that serves to protect the mass.} At leading order, these interactions mediate 2-to-2 scattering processes, and provide the dominant mechanism for dissipating energy and angular momentum from the system -- their efficiency stems from the fact that they are strongly Bose enhanced, with an effective scattering rate scaling as $N^3$. Early investigations suggested that these interactions would become important as the axion began to feel the non-linearity of the potential, occurring for axion field values $\Phi$ near the decay constant $\Phi \sim f_a$, leading to an explosive non-linear event known as a ``bosenova'' \cite{Arvanitaki:2010sy,Yoshino:2012kn, Yoshino:2015nsa}; it is now understood, however, that bosenovae almost certainly do not occur (at least for conventional axion potentials and for the limiting regimes which have been studied to date). Instead, self-interactions tend to drive the rapid growth of higher order, more weakly bound states, quickly pushing the system towards quasi-equilibrium configurations where the occupation numbers of all states sit far below the bosenova threshold \cite{Gruzinov:2016hcq, Baryakhtar:2020gao, Witte:2024drg} -- this is illustrated in Fig.~\ref{fig:overview_fig}, where the left panel shows the structure of a cloud without self-interactions, and the right panel shows the cloud at the same point in the evolutionary history with self-interactions. This has a number of important implications, including the relaxation of superradiant constraints derived from spin down \cite{Witte:2024drg}, and a suppression in the sensitivity of all superradiant searches that rely on large axion field values. 

Baryakhtar \textit{et al.}~\cite{Baryakhtar:2020gao} first studied the interplay between superradiant growth and self-interactions in detail, using semi-analytic rates derived in the non-relativistic regime to understand the equilibrium configurations and growth rates of individual states. They found that for small gravitational couplings\footnote{We take $\hbar = c = 1$.} ($\alpha \equiv GM\mu \lesssim 0.22$), the cloud is well-described at early times by just the two fastest-growing modes; higher-order modes could grow at later times, but were argued not to disrupt the equilibrium between the leading states. This picture is fragile, though: the competing rates differ by order-one factors, not orders of magnitude -- so relativistic corrections, which change individual rates by $\OO{10-100}$ even at relatively modest values of $\alpha$, can easily reshuffle which channels dominate. 
Witte \& Mummery~\cite{Witte:2024drg} tested this by computing the full relativistic rates for all bound states with principal quantum number $n \leq 5$. The corrections became relevant at gravitational couplings as low as $\alpha \gtrsim 0.15$; here, it was shown that higher-order states become populated on short timescales, and the cloud passes through a sequence of quasi-equilibria involving not two, but many modes. The spin-down, however, turns out to be more robust than the occupation numbers, since the higher-$n$ states barely feed back on the dominant $m=1,2$ modes. It was therefore argued that $n \leq 5$ suffices to follow the $m=1,2$  spin-down of a typical stellar-mass black hole, which allowed the constraints on axions to be extended into the relativistic regime up to $m=2$ and $\alpha \approx 0.6$.
A number of important questions still remain. First, the state space must be extended to establish convergence of the $m=2$ spin-down at larger $\alpha$, and to include higher-$m$ superradiant levels. Second, the complete population history is itself essential for charting the broader phenomenology of self-interacting superradiant clouds, such as GW emission and signatures in binary systems. 
In this work, we address both questions by systematically extending the analysis of \cite{Witte:2024drg} to progressively higher-$n$ states and studying the convergence of the evolution across the $(\alpha, f_a)$ parameter space. Specifically, we start by defining a complete system of coupled states up to a fixed value of $n=n_{\rm max}$, including all allowed scattering permutations among these states. The restriction to such a subset is motivated by the fact that energy and angular momentum are transferred via a contact interaction: the scattering rate for a process $\ket{1} \times \ket{2} \rightarrow \ket{3} \times \ket{4} $ scales as $\gamma \propto |\int \mathrm{d}^3x \,  \psi_1 \, \psi_2 \, \psi_3^* \, \psi_4^*|^2$ ($\ket{4}$ is typically a bound oscillation that dissipates into the black hole, or a relativistic free axion), so that in the limit $n_3 \rightarrow \infty$ (with $n_3$ the principal quantum number of state $3$), $\gamma \propto n_3^{-3}$. One therefore expects energy and angular momentum transfer to become increasingly inefficient in this limit -- although, as we show in Sec.~\ref{subsec:ocillations}, this is not always the case.
Yet it is not sufficient to retain only states below a threshold value of $n_{\rm max}$. The reason is that the large-$n$ states will be efficiently driven by the low-$n$ states, but they themselves may not be able to efficiently dissipate their energy. As such, in some cases they can grow in an uncontrolled manner. In order to avoid this unphysical feature (which merely arises from the need to truncate at some threshold), we also include a series of ``sink'' states, such that $\ket{n\ell m} \times \ket{n\ell m} \rightarrow \ket{2\ell+1 \: \, 2\ell \, \, 2m} \times X$ and its back-reaction process ($\ket{2\ell+1 \: \, 2\ell \, \, 2m} \times \ket{2\ell+1 \: \, 2\ell \, \, 2m} \rightarrow \ket{n\ell m}  \times X$) are always present in the system. 
Our goal is to search for converged solutions, meaning that the inclusion of additional higher-order states does not alter the evolution. In this vein, we differentiate two distinct notions of convergence: one defined by the evolution of the properties of the black hole (meaning the black hole mass and spin remain stable), and one defined by the properties of the superradiant cloud (meaning, do the occupation numbers remain invariant, and no new states with non-negligible occupation numbers appear, as new states and scattering processes are included\footnote{In practice, this requires implementing a threshold on the occupation numbers of states. We specifically take this to be $\epsilon_{\rm th} = 10^{-20}$, where $\epsilon_{n\ell m} \equiv N_{n\ell m} / (G M^2)$ with $N_{n\ell m}$ the occupation number. A state could in principle matter by mediating the growth of another state while itself staying below threshold and so not appearing in the figure; this needs the sink state to drain almost as fast as it is fed, which is exceptional. Across the parameter space we verified that closing the evolution on the tabulated states alone reproduces the converged result.}). We call these two concepts ``black-hole convergence'' and ``cloud convergence'', respectively. While establishing convergence was our primary goal, in its pursuit we found a number of new, previously unappreciated phenomena in black hole superradiance.

We consider three benchmark black hole masses: $M=10\,M_\odot$, evolved for a Salpeter time $\tau_{\rm Sal}=5\times10^7\,{\rm yr}$, and $M=10^4\,M_\odot$ and $10^8\,M_\odot$, each evolved for $1\,{\rm Gyr}$. The Salpeter time is the natural reference timescale for spin-down constraints before sustained accretion substantially changes the black hole \cite{1964ApJ...140..796S}; the longer integrations for intermediate-mass and supermassive black holes allow the slower growth of higher-order levels to be followed.

\begin{figure*}
    \centering
    \includegraphics[width=\linewidth]{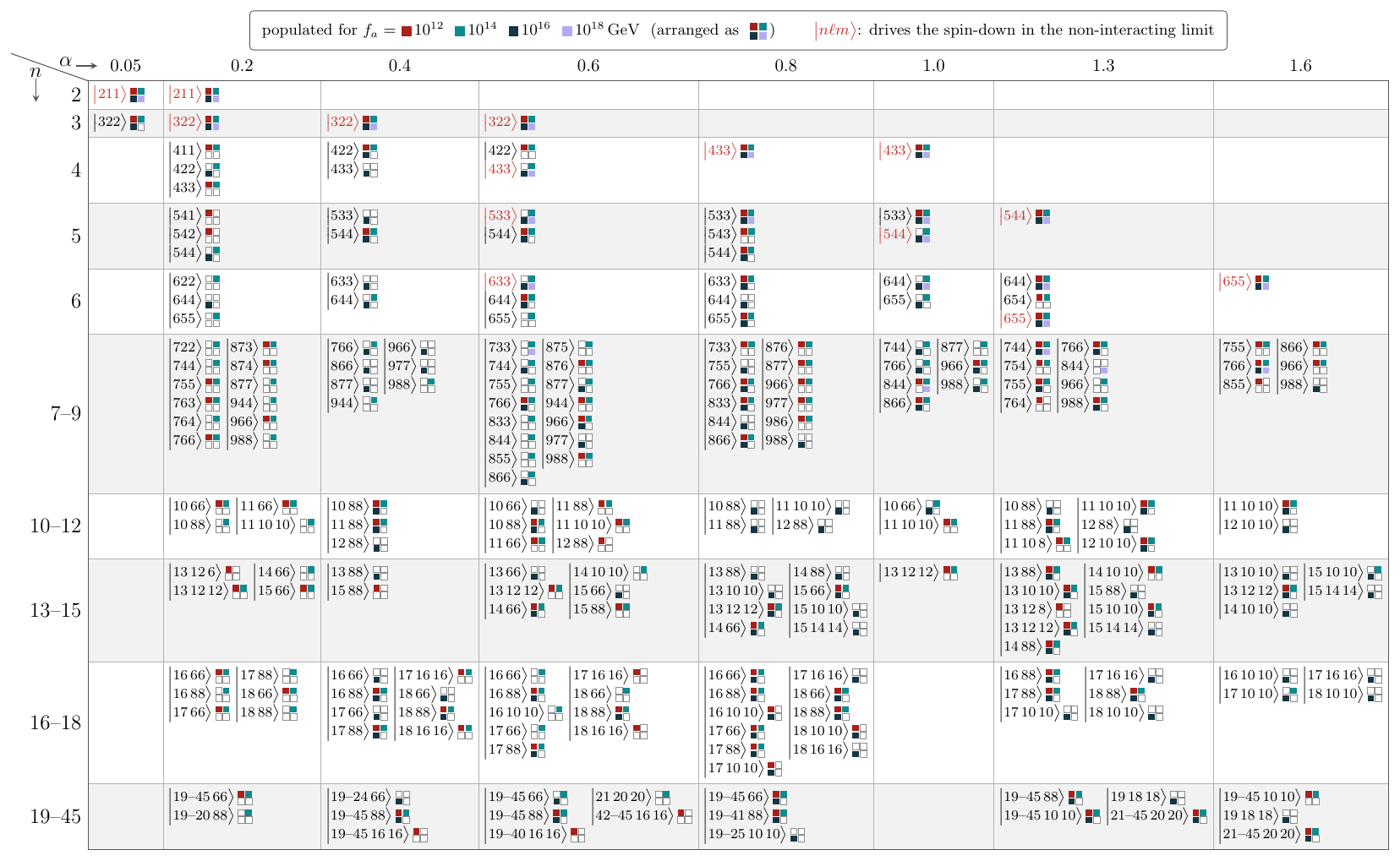}
    \caption{Quasi-bound states reaching $\epsilon_{n\ell m} \equiv N_{n\ell m} / (G M^2) \geq \epsilon_{\rm th} = 10^{-20}$ for $M = 10\,M_\odot$, $\tilde a_i = 0.95$ and $\tau = 5\times10^{7}\,$yr, as a function of $\alpha$ (columns) and $n$ (rows). Here, $N_{n\ell m}$ is the occupation number of state $\ket{n\ell m}$. Filled squares mark the values of $f_a$ for which a state is populated. Red kets drive the spin-down in the non-interacting limit. Ranges such as $|19\text{--}45\,16\,16\rangle$ denote towers of consecutive $n$ found in large $n_{\max}$ runs. All runs are black-hole-converged [Eq.~\eqref{eq:bh_conv}], but some regions of parameter space are not cloud-converged -- these are indicated in Fig.~\ref{fig:cloudconv}. The corresponding tables for $M = 10^4$ and $10^8\,M_\odot$ are shown in App.~\ref{app:evolutions}, and the complete set of states for each mass is collected in Table~\ref{tab:minimal_sets}.}
    \label{fig:tab_tot}
\end{figure*}

\begin{figure}
    \centering
    \includegraphics[width=0.99\linewidth]{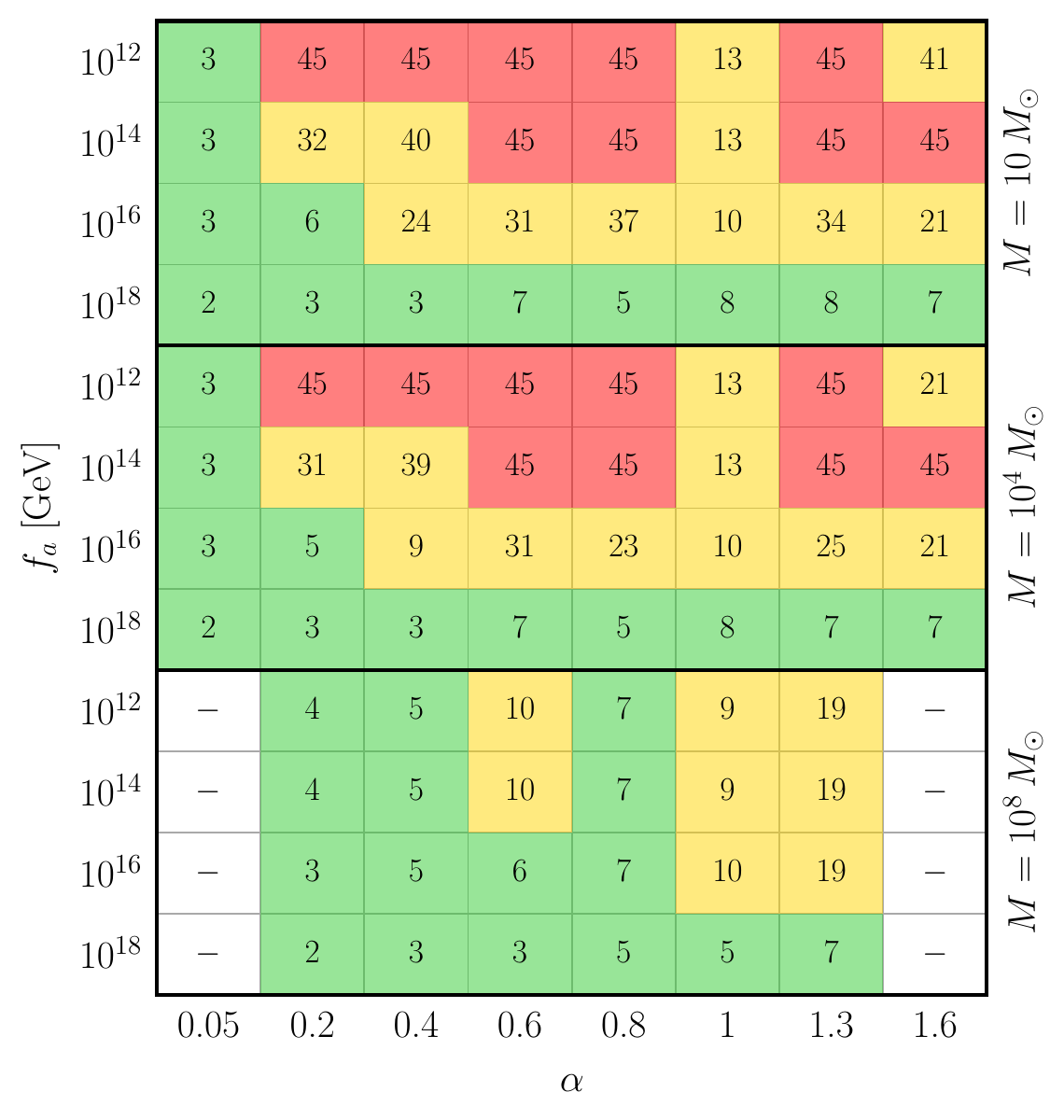}
    \caption{The number inside each grid point corresponds to the maximum principal quantum number realized at each point in parameter space -- when superradiance is not active, we instead place a ``--''. The top, middle, and bottom panels correspond to a black hole mass of $10 \, M_\odot$, $10^4 \, M_\odot$, and $10^8 \, M_\odot$, respectively. Black hole convergence is obtained at all points. Cloud convergence is obtained in the green and yellow regions, but not the red regions. Above $n = 8$ we evolve only selected families of states (and, for $n > 18$, only specific towers), so the yellow points are converged with respect to the states included; states or scattering processes outside this set could still perturb them. This caveat is milder for $n \le 18$, where the basis is much larger, than for $n > 18$.}
    \label{fig:cloudconv}
\end{figure}

\begin{figure*}
    \centering    
    \includegraphics[width=\linewidth]{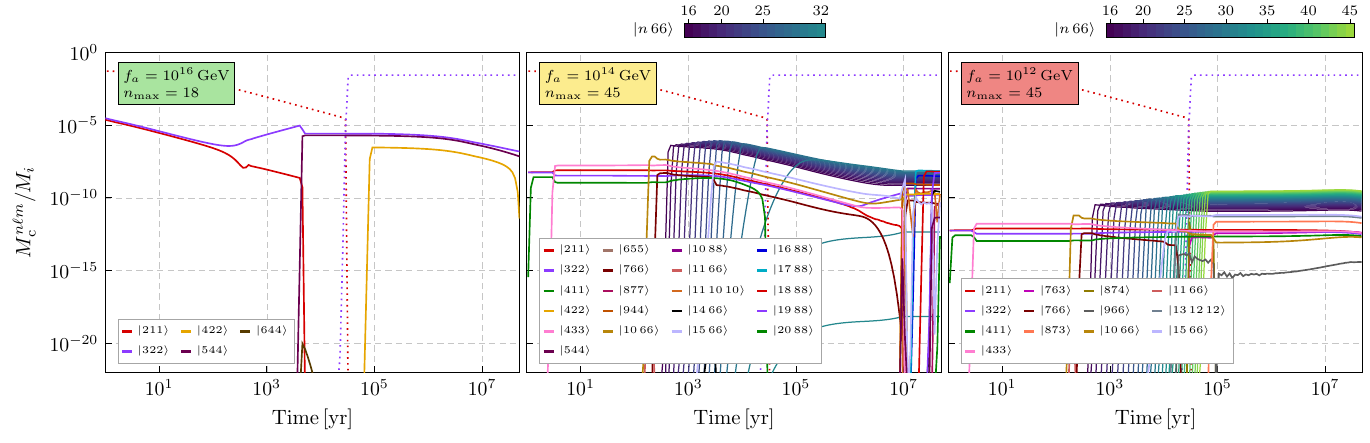}
    \caption{Evolution of the cloud for $M_i = 10\,M_\odot$ and $\alpha = 0.2$ for $f_a = 10^{16}\,$GeV, $10^{14}\,$GeV and $10^{12}\,$GeV, illustrating the three degrees of convergence of Fig.~\ref{fig:cloudconv}. Each panel displays the fractional mass stored in a given quasi-bound state, $M_{\rm c}^{n\ell m} = \alpha \epsilon_{n\ell m} M_i$ with $M_i$ the initial black hole mass. We show all levels which achieve $\epsilon_{n\ell m} > 10^{-20}$ at some point during the evolution; dotted lines show the non-interacting evolution. The high-$n$ levels $\ket{n\,66}$ ($n \geq 16$) are colored by $n$. For $f_a = 10^{16}\,$GeV no level beyond $n = 6$ is populated. For $f_a = 10^{14}\,$GeV the low-$n$ levels have converged, but a $\ket{n\,66}$ tower develops up to $n = 32$. For $f_a = 10^{12}\,$GeV the tower populates every level up to the truncation, $n = 45$, so that even higher states are expected to be occupied.}
\label{fig:states_convergence}
\end{figure*}

Our main results are the following.
\begin{enumerate}
\item \emph{The minimal set.} We identify the set of quasi-bound states required to evolve an isolated $10 \, M_\odot$ black hole on a Salpeter timescale, and a $10^4$ and $10^8 \, M_\odot$ black hole on a Gyr timescale, for values of $\alpha \lesssim 1.6$ (we use a dimensionless spin of $\tilde{a} = 0.95$, which prevents superradiant growth on these timescales at larger $\alpha$). This result is summarized in Table~\ref{tab:minimal_sets} and Figure~\ref{fig:tab_tot}. For a $10 \, M_\odot$ black hole evolving on a Salpeter timescale (and all values of $\alpha$ considered), we find the evolution is described by at least 203 distinct quasi-bound states. Since we do not find cloud-convergence in all parts of parameter space, the minimal set of states is guaranteed to be larger than this.

\item \emph{Black-hole convergence.} The mass and spin evolution of the black hole is set by the lowest-lying, most unstable levels; their growth is stabilized by the nearest states, and the high-$n$ states typically do not alter their evolution significantly. Black-hole convergence is therefore reached well before cloud convergence. In quantifying the convergence of the mass and spin of the black hole, it is not sufficient to simply compare these quantities at fixed times -- a tiny temporal shift in the spin-down threshold might yield a large instantaneous effect, but one which could be absorbed with a small perturbative shift in $f_a$ (since $f_a$ perturbs the occupation numbers of the states driving the spin-down, see \eg Fig.~\ref{fig:spin-down-fa}). Instead, we compare the evolutions within a shifted time window, allowing the spin-down to move by up to $30\%$ in time while requiring the spin and mass to agree to within $1\%$. Assuming the states driving the spin-down are in equilibrium, this time shift roughly corresponds to rescaling $f_a$ by a factor of $1.14$ (see Sec.~\ref{sec:convergence} for the precise definition). With this choice, the evolution of the black hole is converged across our entire parameter space at $n_{\max}=8$.

\item \emph{Cloud convergence.} Cloud convergence is obtained across a broad range of parameter space, but not everywhere. Figure~\ref{fig:cloudconv} summarizes the largest principal quantum number populated at each point in parameter space, and Fig.~\ref{fig:states_convergence} illustrates the three situations we encounter. Points shaded green are cloud-converged in our complete analyses, which contain all states and scattering permutations with $n \leq n_{\rm max}$ -- this convergence is considered robust. Points shaded yellow are cloud-converged only in our extended analyses ($n_{\rm max} = 15$, $18$ and $45$), which retain only the subset of higher-$n$ states (and scattering permutations) that we identify as most relevant; small shifts from the identified equilibrium configurations are therefore possible. Points highlighted in red are not cloud-converged: they populate states out to the truncation of our analysis, $n = 45$, and we expect even higher states to be occupied. Even where cloud convergence is obtained, states with $n \gg 10$ frequently appear with sizable occupation numbers.
\item \emph{Extended clouds with intricate frequency structure.} Interestingly, these high-$n$ states often carry most of the cloud mass. Since the characteristic radius of a level scales as $r_n \simeq n^2 a_0$, with $a_0 = (\alpha\mu)^{-1} = GM/\alpha^2$ the Bohr radius of the cloud [Eq.~\eqref{eq:bohr}], the cloud can extend to radii tens, hundreds, or even thousands of times larger than is often assumed (Fig.~\ref{fig:density_extended}). This modifies the phenomenology in a number of important ways, including: $(i)$ the GW signal from isolated objects -- annihilation is weaker, level transitions add many more spectral lines, and the lowest-frequency line sits many orders of magnitude below what was previously appreciated; $(ii)$ the sensitivity of indirect searches for energy exchange between a grown cloud and its environment (\eg through dynamical friction); and $(iii)$ the interplay with binary systems, where resonant excitations occur across a much broader range of frequencies and at far more distant radii.
\item \emph{Resonant scattering.} The failure of cloud convergence can be traced to the scattering rates themselves. Contrary to the naive $n^{-3}$ expectation, the rate for populating a tower of states $\ket{n\, 2\ell\, 2m}$ from a pair of $\ket{n'\ell m}$ states can be resonantly enhanced, out to very large $n$ (see Fig.~\ref{fig:1188_direction}).  We also identify a class of ``super-resonant'' channels (see Fig.~\ref{fig:Horizon} and Table~\ref{tab:superres}) in which the fourth leg cannot be the dominant $\ell =0$ monopole yet a higher multipole lands almost exactly on a quasi-bound state. These are enhanced by orders of magnitude.
\item \emph{Suppressing superradiant growth with GW level transitions.} We identify a regime in which level transitions, \eg processes of the form $\ket{n\ell m} \rightarrow \ket{n^\prime \, \ell^\prime \, m^\prime} + {\rm GW}$, can prevent the growth of superradiant states. In particular, we provide an example (see Fig.~\ref{fig:GW_impact}) in which the growth of the $\ket{433}$ can be inhibited by a pre-existing large occupation of $\ket{322}$ -- this occurs preferentially at large $f_a$ and small $\alpha$. Gravitational-wave emission is often assumed to play a negligible role in, \eg the black hole spin-down; we show explicitly that this is not always the case. 
\end{enumerate}
These results lead to an unavoidable conclusion: self-interactions are crucial to axion superradiance and cannot be ignored. They are necessarily present for axions,\footnote{One could attempt to fine-tune the numerical pre-factor of the $a^4$ term to zero via delicate cancellations, however higher-order corrections arising at $a^6$ are only mildly suppressed, and a similar phenomenology would arise from these terms.} and Bose enhancement ensures that scattering is both the dominant process and relevant on short timescales. The strength of the interaction is effectively set by the factor $(M_{\rm Pl}/f_a)^4$, with $M_{\rm Pl}$ the Planck mass. At small decay constants ($f_a\sim10^{12}$--$10^{14}\,\GeV$) the cloud settles almost immediately into a quasi-equilibrium shared among many states at very low occupation, and the spin-down is delayed by orders of magnitude, or suppressed altogether. At larger decay constants, the total black hole spin down is comparable to the non-interacting case, but the cloud it leaves behind looks different from the standard non-interacting picture of isolated growth and decay: the dominant level is depleted by orders of magnitude, and states out to $n \gtrsim 15$ are populated (Fig.~\ref{fig:tab_tot}). Even at $f_a = 10^{18}\,\GeV$, additional states can appear and GW transitions can alter which states grow at all (Fig.~\ref{fig:GW_impact}). A large body of work on the observational signatures of scalar superradiance (including work done by the authors here) has been built on the simplified picture of non-interacting fields --  the predictions and observational consequences of these works will need to be revisited.
Our code and data are publicly available on GitHub and Zenodo folder~\cite{AxionSR:github,AxionSR:zenodo}. This includes the complete set of superradiant and self-interaction rates, computed to high principal quantum number with several independent methods. We stress that this was to a large degree a computational problem rather than a conceptual one -- resolving imaginary parts across thirty orders of magnitude, computing $\OO{10^5}$ scattering rates across a wide parameter space, and integrating a complex system of equations through sharp transient features and over long timescales. We hope that the release of this code and dataset removes the need for anyone to repeat it. 
The remainder of this paper is organized as follows. Section~\ref{sec:formalism} reviews the basics of axion superradiance, self-interactions and the resulting evolution equations. Section~\ref{sec:convergence} then shows the convergence study of the cloud evolution and of the spin-down. Section~\ref{sec:obsv} discusses some of the observational implications of our results. We conclude in Sec.~\ref{sec:conclusions}. Additional material is collected in the appendices: the computation of the spectrum (App.~\ref{app:superradiance}), a review of the non-relativistic limit (App.~\ref{app:NonRel}), the calculation of the GW rates (App.~\ref{app:gw_matrix}), and additional figures showing the evolution of the system (App.~\ref{app:evolutions}).

\begin{table}
\centering
\begin{tabular}{@{}lll cc@{}}
\toprule
\multicolumn{3}{c}{$n$} & & \\
\cmidrule{1-3}
$10\,M_\odot$ & $10^4\,M_\odot$ & $10^8\,M_\odot$ & $\ell$ & $m$ \\
\midrule
$2,\,4$ & $2,\,4$ & $2,\,4$ & $1$ & $1$ \\
$5$ & -- & -- & $4$ & $1$ \\
$3,\,4,\,6,\,7$ & $3,\,4$ & $3,\,4$ & $2$ & $2$ \\
$5$ & $5$ & -- & $4$ & $2$ \\
$4\text{--}8$ & $4\text{--}8$ & $4,\,5$ & $3$ & $3$ \\
$5$ & -- & -- & $4$ & $3$ \\
$7$ & $7$ & -- & $6$ & $3$ \\
$8$ & $8$ & -- & $7$ & $3$ \\
$5\text{--}9$ & $5\text{--}9$ & $5\text{--}7$ & $4$ & $4$ \\
$6,\,7$ & -- & -- & $5$ & $4$ \\
$7$ & $7$ & -- & $6$ & $4$ \\
$8$ & $8$ & -- & $7$ & $4$ \\
$6\text{--}8$ & $6\text{--}8$ & $6,\,7$ & $5$ & $5$ \\
$8$ & $8$ & -- & $7$ & $5$ \\
$7\text{--}11,\,13\text{--}45$ & $7\text{--}11,\,14\text{--}45$ & $7\text{--}10$ & $6$ & $6$ \\
$8$ & $8$ & -- & $7$ & $6$ \\
$9$ & -- & -- & $8$ & $6$ \\
$13$ & -- & -- & $12$ & $6$ \\
$8,\,9$ & $8,\,9$ & -- & $7$ & $7$ \\
$9\text{--}45$ & $9\text{--}45$ & $9\text{--}19$ & $8$ & $8$ \\
$11$ & -- & -- & $10$ & $8$ \\
$13$ & -- & -- & $12$ & $8$ \\
$11\text{--}45$ & $11\text{--}45$ & $12\text{--}14$ & $10$ & $10$ \\
$13$ & $13$ & -- & $12$ & $12$ \\
$15$ & -- & -- & $14$ & $14$ \\
$17\text{--}45$ & $17\text{--}45$ & -- & $16$ & $16$ \\
$19$ & -- & -- & $18$ & $18$ \\
$21\text{--}45$ & $22\text{--}45$ & -- & $20$ & $20$ \\
\midrule
203 & 189 & 29 & \multicolumn{2}{c@{}}{total} \\
\bottomrule
\end{tabular}
\caption{Minimal set of quasi-bound states for each black-hole mass: for every $(\ell, m)$ family, the values of $n$ of all states that reach $\epsilon_{n\ell m} \equiv N_{n\ell m}/(G M^2) > \epsilon_{\rm th} = 10^{-20}$ at some time, for any $\alpha \in \{0.05, 0.2, 0.4, 0.6, 0.8, 1.0, 1.3, 1.6\}$ and $f_a \in \{10^{12}, 10^{14}, 10^{16}, 10^{18}\}\,$GeV ($\tilde a_i = 0.95$; $\tau = 5\times10^{7}$, $10^{9}$ and $10^{9}\,$yr). Of the quoted ``total states'', 145, 143 and 4 appear only in the $n_{\max} = 45$ runs: at $M = 10$ and $10^4\,M_\odot$, 138 of them lie in the $|n\,6\,6\rangle$, $|n\,8\,8\rangle$, $|n\,10\,10\rangle$, $|n\,16\,16\rangle$ and $|n\,20\,20\rangle$ towers, and at $10^8\,M_\odot$ all four are $|n\,8\,8\rangle$ states with $n = 16$--$19$ (Sec.~\ref{subsec:ocillations}). Without them the sets contain 58, 46 and 25 states.}
\label{tab:minimal_sets}
\end{table}

%
\section{Black hole superradiance and self-interactions}
\label{sec:formalism}
We now summarize the formalism used throughout the paper, following closely \cite{Witte:2024drg}. We first review the quasi-bound spectrum of a massive scalar on Kerr, then describe the quartic processes that couple different levels, and finally assemble the corresponding rate equations.
We consider a scalar field $\Phi$ of mass $\mu$ on a Kerr spacetime with black hole mass $M$ and angular momentum $J$, so that the dimensionless spin is $\tilde a\equiv J/(GM^2)$. The relevant dimensionless coupling is the ratio of the gravitational radius $r_g$ to the scalar Compton wavelength $\lambda_c$,
\begin{equation}
    \alpha\equiv\frac{r_g}{\lambda_c} = GM\mu\,,
\label{eq:alpha}
\end{equation}
which plays the role of a gravitational fine-structure constant. In the non-relativistic limit, the size of a bound state is set by the associated gravitational \emph{Bohr radius},
\begin{equation}
    a_0 \equiv (\alpha \mu)^{-1} = \frac{GM}{\alpha^2}\,.
\label{eq:bohr}
\end{equation}
\subsection{Scalar quasi-bound states}
\label{sec:spec}
Neglecting self-interactions and the backreaction of the cloud on the geometry, the scalar field obeys the homogeneous Klein-Gordon equation
\begin{equation}
\left(\Box - \mu^2 \right)\Phi = 0\,.
\label{eq:KG}
\end{equation}
Exploiting the stationarity and axial symmetry of the background, the field admits the standard separable ansatz:
\begin{equation}
    \Phi(\mathbf{r}, t) = R(r) \, S(\theta) \, e^{-i\omega t} \, e^{i m \phi}\,,
\label{eq:ansatz}
\end{equation}
where $m$ is the azimuthal quantum number. This reduces the Klein-Gordon equation \eqref{eq:KG} to coupled radial and angular equations,
\begin{equation}
\begin{aligned}
     \frac{\mathrm{d}}{\mathrm{d}r}  \left( \Delta \frac{\mathrm{d} R}{\mathrm{d} r} \right) + \bigg[ \frac{\omega^2 (r^2 + a^2)^2 - 4 G M a m \omega r + m^2 a^2 }{\Delta}  \\ - \left( \omega^2 a^2 + \mu^2 r^2 + \Lambda_{\ell m} \right) \bigg] R(r) &= 0 \,,\\
    \frac{1}{\sin \theta} \frac{\mathrm{d}}{\mathrm{d} \theta} \left(\sin \theta \frac{\mathrm{d} S}{\mathrm{d} \theta} \right) +   \bigg[ a^2 (\omega^2 - \mu^2) \cos^2\theta - \\ \frac{m^2}{\sin^2 \theta} + \Lambda_{\ell m} \bigg] S(\theta)  &= 0\,,
\end{aligned}
\label{eq:radial_angular}
\end{equation}
where $\Delta=r^2-2GMr+a^2$ and $\Lambda_{\ell m}$ is the angular separation constant. The angular solutions are oblate spheroidal harmonics \cite{Berti:2005gp}. Imposing physically appropriate boundary conditions -- purely ingoing waves at the horizon and exponential decay at spatial infinity -- turns the radial equation into an eigenvalue problem. Its solutions are the \emph{quasi-bound states}, labeled by $\ket{n\ell m}$ with $n$, $\ell$, and $m$ the principal, orbital, and azimuthal quantum numbers. The eigenfrequency $\omega_{n\ell m}=\omega^{\rm R}_{n\ell m}+i\omega^{\rm I}_{n\ell m}$ is complex, and a state grows or decays according to the sign of $\omega^{\rm I}_{n\ell m}$.
In the absence of interactions, each growing mode extracts energy and angular momentum until the black hole reaches its superradiant threshold, $\omega^{\rm R}_{n\ell m}=m\Omega_{\rm H}$~\eqref{eq:SR_cond}. Standard superradiance calculations need only the handful of low-$n$ modes that dominate this process. Here the situation is different: self-interactions redistribute energy across many levels, so we need accurate frequencies \emph{and} wavefunctions for a large set of states, including weakly unstable, high-$n$ modes with very small imaginary parts. This is numerically demanding. The scattering rates below are built from overlap integrals of four wavefunctions, so an error in the far-field tail of a high-$n$ state feeds directly into the dynamics, and the imaginary parts we must resolve span more than thirty orders of magnitude (Fig.~\ref{fig:SRrates}).
Analytical expressions for the spectrum are available in the asymptotic regimes $\alpha \ll 1$ \cite{Detweiler:1980uk, pani:2012bp} and $\alpha \gg 1$ \cite{Zouros:1979iw}. The phenomenologically relevant regime is instead $\alpha \lesssim \mathcal{O}(1)$, where the instability is strongest and numerical methods are required. We use two complementary ones. Leaver's continued-fraction method \cite{Leaver:1985ax, Dolan:2007mj} is efficient across much of the spectrum, but becomes unstable when $\omega^{\rm I}$ is small. There we switch to a pseudospectral method based on Chebyshev interpolation, which is more expensive but resolves the full spectrum accurately \cite{Baumann:2019eav}, and construct the corresponding wavefunctions in terms of Heun functions. Details are given in App.~\ref{app:superradiance}; the resulting data and code are publicly available \cite{AxionSR:github}.
\subsection{Self-interactions and level mixing}
The picture above treats the levels as independent; as highlighted in the introduction, self-interactions couple the quasi-bound states. We adopt the standard cosine potential for the axion, invariant under the shift $\Phi \rightarrow \Phi + 2\pi f_a$ (as mentioned above, other potentials respecting the shift symmetry have self-interactions of comparable strength), whose leading-order self-interaction appears in the small-field limit $\Phi \ll f_a$:
\begin{equation}
\begin{aligned}
    V(\Phi)&=\mu^2 f_a^2\left[1-\cos\left(\frac{\Phi}{f_a}\right)\right]
    \\&=\frac{1}{2}\mu^2\Phi^2-\frac{1}{4!}\frac{\mu^2}{f_a^2}\Phi^4+\cdots
\end{aligned}
\end{equation}
 We keep the leading quartic term\footnote{Cubic self-interactions can be present in more general scalar potentials, but are not expected to matter here \cite{Baryakhtar:2020gao}.} and parameterize its strength by
\begin{equation}
    \lambda \equiv \frac{\mu^2}{f_a^2}\,.
\label{eq:lambda}
\end{equation}
For the QCD axion $\mu$ and $f_a$ are related, but we treat them as independent to keep the discussion general.
The quartic term sources an inhomogeneous Klein-Gordon equation,
\begin{equation}
    \left( \Box - \mu^2 \right) \Phi = -\frac{\lambda}{6}\Phi^3 \,,
\label{eq:inhomogeneousKG}
\end{equation}
which we solve perturbatively in $\lambda$. At zeroth order the solution is a superposition of quasi-bound states,
\begin{equation}\label{eq:phi0_decomp}
    \Phi^{(0)}=\sum_{n\ell m}\sqrt{\frac{N_{n\ell m}}{2\mu}}\psi_{n\ell m}(r,\theta,\phi)e^{-i\omega_{n\ell m}t}+{\rm c.c.}\,,
\end{equation}
with $N_{n\ell m}$ the occupation number of $\ket{n\ell m}$. At first order, the cubic source $(\Phi^{(0)})^3$ then contains several harmonic components, describing three distinct effects: (i) frequency shifts of the bound states, (ii) relativistic $3\to1$ emission, and (iii) $2\to2$ scattering.
We focus on the $2\to2$ processes, which are the dominant effect for the systems of interest \cite{Witte:2024drg} -- the $3\to1$ processes are parametrically suppressed, and negligible throughout the parameter space we study \cite{Baryakhtar:2020gao, Witte:2024drg},\footnote{This suppression stems from the disparate spatial scales of the relativistic unbound axion and the quasi-bound state wavefunctions, which lead to strong cancellations in their overlap.} while the frequency shifts do not appreciably alter the population evolution (although they may be important for binary systems \cite{Lucas_prep}).
A quartic scattering process has the generic form
\begin{equation}
    \ket{a}\times\ket{b}\rightarrow \ket{c}\times X\,,
\end{equation}
where $\ket{a}$, $\ket{b}$, $\ket{c}$ are quasi-bound levels and $X$ is the fourth leg sourced by the interaction. Its frequency is fixed by energy conservation, $\omega_X = \omega_a + \omega_b - \omega_c$. If $\omega_X < \mu$, the fourth leg is a forced bound oscillation absorbed by the black hole ($X={\rm BH}$); if $\omega_X > \mu$, it is an axion radiated to infinity ($X=\infty$). Either way, the cloud loses energy.
Two representative examples make this concrete. In $\ket{211}\times\ket{211}\rightarrow\ket{322}\times {\rm BH}$, the fourth leg is a forced bound oscillation, predominantly $\ell=0$, that is absorbed at the horizon (horizon absorption is generically dominated by the monopole component of the forced mode, since the absorption of higher multipoles at the horizon is suppressed by the centrifugal barrier).\footnote{Exceptions arise when the fourth leg is nearly on shell, in which case $\ell, m \neq 0$ contributions can become appreciable (see Sec.~\ref{sec:superres}).}  In $\ket{322}\times\ket{322}\rightarrow\ket{211}\times \infty$, occupation is fed back to $\ket{211}$ while the fourth leg is radiated to infinity as an unbound mode (with $\ell=m=3$).
\subsection{Relativistic scattering rates}
In the hydrogenic limit $\alpha \ll 1$, the problem reduces to a non-relativistic Schr\"odinger problem with hydrogen-like wavefunctions. This approximation is useful for intuition and is reviewed in App.~\ref{app:NonRel}. However, since the hierarchy among the different rates is often mild, relativistic corrections can play an important role in determining how energy and angular momentum are transferred through the system. These corrections have two origins: (i) the relativistic Kerr wavefunctions differ from hydrogenic wavefunctions, modifying the overlap integrals that determine the rates; (ii) the relativistic problem no longer admits the simple hydrogen-like spectral decomposition and the set of quasi-bound and quasi-normal modes do not form a complete basis. The full solution then also receives contributions from prompt and late-time components associated with the dissipative nature of the spacetime \cite{Cannizzaro:2023jle, Lestingi:2026peq, Guo:2026qfb}. The perturbed evolution cannot therefore be formulated as a mode expansion, and a Green's-function calculation, which incorporates both effects, is needed once $\alpha$ is not very small.
At first order in the quartic coupling, each harmonic component of the source obeys an inhomogeneous version of the radial equation~\eqref{eq:radial_angular}, with the angular projection of the source on the right-hand side,
\begin{equation}\label{eq:sourcefull}
    \mathcal{T}_{\ell m\omega}(r)= -\frac{\lambda}{6}\int \mathrm{d}\phi\,\mathrm{d}(\cos\theta)\,\Sigma\,(\Phi^{(0)}_\omega)^3 S^*_{\ell m\omega}(\theta)e^{-im\phi}\,,
\end{equation}
where $\Sigma\equiv r^2+a^2\cos^2\theta$ and $S^*_{\ell m\omega}(\theta)$ are the spheroidal harmonics. The angular integral enforces selection rules: only mode combinations with a non-vanishing angular overlap contribute.
The solution to the sourced Klein-Gordon equation is built from two homogeneous solutions, $R^{\rm in}_{\ell m\omega}$ and $R^{\rm up}_{\ell m\omega}$, which are, respectively, ingoing at the horizon and outgoing at infinity. The radial solution satisfying the physical boundary conditions can then be written as
\begin{equation}
\begin{aligned}
R_{\ell m}(r) &= \frac{1}{W_{\ell m \omega}} \bigg[ R^{\rm up}_{\ell m \omega}(r) \int_{r_+}^{r} \mathrm{d}r' \, R^{\rm in}_{\ell m \omega}(r') \, \mathcal{T}_{\ell m \omega}(r') \\ &+ R^{\rm in}_{\ell m \omega}(r) \int_{r}^{\infty} \mathrm{d}r' \, R^{\rm up}_{\ell m \omega}(r') \, \mathcal{T}_{\ell m \omega}(r') \bigg] \,,
\end{aligned}
\label{eq:radial_sol}
\end{equation}
where $W_{\ell m \omega}$ is the (radially conserved) Wronskian,
\begin{equation}
W_{\ell m \omega} = \Delta \left(R^{\rm in}_{\ell m \omega} \frac{\mathrm{d} R^{\rm up}_{\ell m \omega}}{\mathrm{d}r} - R^{\rm up}_{\ell m \omega} \frac{\mathrm{d} R^{\rm in}_{\ell m \omega}}{\mathrm{d}r} \right)\,.
\label{eq:wronskian}
\end{equation}
The energy carried by the sourced solution is obtained from the scalar stress-energy tensor,
\begin{equation}
    \frac{\mathrm{d}E}{\mathrm{d}t}=\int_{\partial \Sigma} T_0^\mu n_\mu \mathrm{d}\Omega\,,
\end{equation}
where the surface $\partial \Sigma$ is either the horizon or a two-sphere at large radii, $n_\mu$ is the normal vector to the hypersurface and $T^{\mu\nu}$ is the stress-energy tensor of a scalar field. 
From the flux, one can determine the scattering rate using energy balance. For a process
\begin{equation}
    \ket{n\ell m}\times\ket{n'\ell' m'}\rightarrow \ket{n''\ell''m''}\times {\rm BH}\,,
\end{equation}
we write schematically
\begin{equation}
    \dot N_{n\ell m}\simeq -\frac{1}{\mu}\frac{\mathrm{d}E}{\mathrm{d}t}\equiv-\Gamma^{n''\ell''m''\times {\rm BH}}_{n\ell m\times n'\ell' m'} N_{n\ell m}N_{n'\ell' m'}N_{n''\ell''m''}\,,
\label{eq:number_evol}
\end{equation}
where we introduced the scattering rate $\Gamma$, following the notation of previous works \cite{Baryakhtar:2020gao, Witte:2024drg}. 
In computing the differential energy loss we implicitly convert energy to occupation number using the rest-mass energy $\mu$. A more exact conversion would use the real part of the relevant bound-state frequency, $\omega^{\rm R}_{n\ell m}$; the difference is a binding-energy correction of order $(\mu-\omega_{\rm R})/\mu$ and negligible in the evolutions.
It is convenient to introduce and work with dimensionless occupation numbers
\begin{equation}
    \epsilon_{n\ell m} \equiv \frac{N_{n\ell m}}{G M^2}\, .
\label{eq:epsilon_re}
\end{equation}
We also factor out the universal coupling dependence of the quartic rates. Since the flux is quadratic in the source, every quartic-induced scattering rate scales as $\lambda^2$. Using Eqs.~\eqref{eq:alpha}, \eqref{eq:lambda} and \eqref{eq:epsilon_re}, this factor can be written as
\begin{equation}
    \lambda^2 G^2 M^4 =\alpha^4\left(\frac{M_{\rm Pl}}{f_a}\right)^4\,, \qquad M_{\rm Pl}\equiv G^{-1/2}\,.
\label{eq:coupling}
\end{equation}
Converting the differential rate given in Eq.~\eqref{eq:number_evol}, and denoting the scattering coefficient as $\gamma$, one obtains an equation of the form
\begin{equation}
    \dot{\epsilon}_{n\ell m} \supset -\gamma_{{n\ell m}\times {n'\ell'm'}}^{{n''\ell''m''}\times {\rm BH}}\,\epsilon_{n\ell m} \epsilon_{n'\ell'm'} \epsilon_{n''\ell''m''} \, ,
\end{equation}
with analogous terms that describe emission to infinity. We denote superradiant growth and decay rates by $\gamma^{\rm SR}_{n\ell m}$ and GW depletion rates by $\gamma^{\rm GW}$.
Finally, when computing the rates, we normalize the bound-state wavefunctions with the standard non-relativistic convention, $\int \mathrm{d}V\,|\psi_{n\ell m}|^2=1$, and define $N_{n\ell m}$ through Eq.~\eqref{eq:phi0_decomp}. Strictly speaking, a relativistic treatment should instead fix the normalization from the cloud mass computed from the stress-energy tensor.\footnote{For a complex scalar field, the cloud mass can be related to the occupation number through the Noether charge $Q$ of the global $U(1)$ symmetry, $M_{\rm c}=\omega Q$ \cite{Brito:2023pyl}.} We carefully checked that this affects the rates at most at the percent level, and thus can be safely neglected. We therefore adopt the simpler non-relativistic normalization throughout for computational convenience.
\subsection{Rate equations}
\label{sec:rateeqs}
Collecting the ingredients above, the system is governed by a set of coupled, non-linear ordinary differential equations. In full generality, the occupation of a level $\ket{a}$ evolves as
\begin{equation}
\begin{aligned}
\dot{\epsilon}_{a} = \gamma^{\rm SR}_{a}\,\epsilon_{a}
&+ \sum_{b,c,X}\big[ \mathcal{S}^{a}_{bc}\,\gamma_{abcX} \big]\,\epsilon_{a}\epsilon_{b}\epsilon_{c} \\ &
 -\sum_b\big[(1+\delta_{ab})\gamma^{\rm GW}_{a\times b} + \gamma^{\rm GW}_{a\rightarrow b} - \gamma^{\rm GW}_{b\rightarrow a}\big]\epsilon_a\epsilon_b\, ,
\label{eq:general_rate}
\end{aligned}
\end{equation}
where the first term describes superradiant growth (or, for levels that do not satisfy the superradiance condition~\eqref{eq:SR_cond}, absorption by the horizon), the second term runs over every quartic channel that connects $\ket{a}$ to a pair of retained levels $\ket{b},\ket{c}$ with the fourth leg $X=\{{\rm BH},\infty\}$ ($\gamma_{abcX}$ therefore includes both $\gamma^{c\times X}_{a\times b}$ and $\gamma^{a\times X}_{b\times c}$), and the third term collects GW losses from annihilation $\gamma^{\rm GW}_{a\times b}$ and level transitions $\gamma^{\rm GW}_{a\rightarrow b}$ (the calculation of these rates is included in App.~\ref{app:gw_matrix}); note that Eq.~\eqref{eq:general_rate} includes level transitions describing both excitations and de-excitations, but in vacuum only de-excitations have a non-zero rate.  The integer $\mathcal{S}^{a}_{bc}$ counts, with sign, how many quanta of $\ket{a}$ a given channel creates or destroys, and is therefore $\pm1$ or $\pm2$ depending on whether $\ket{a}$ appears once or twice among the legs. Written out over the full state space, Eq.~\eqref{eq:general_rate} is a large and highly connected system; which channels can be dropped without changing the answer is precisely the question of Sec.~\ref{sec:convergence}.
The structure is easiest to read off from the minimal two-level system built from $\ket{211}$ and $\ket{322}$. Keeping (i) superradiance, (ii) the leading quartic scattering channels, and (iii) the leading GW losses,\footnote{Gravitational-wave emission is the other intrinsic dissipation channel available to the cloud. It is typically subdominant to self-interactions throughout the parameter space we consider, but we retain it as we find that in some cases, small perturbations to the energy dissipation rate (\eg from GW emission) can change the composition of the cloud. We adopt the leading order contribution to the flat space energy loss rates -- see App.~\ref{app:gw_matrix}. } one obtains
\begin{align}
\dot{\epsilon}_{211} ={}& \gamma^{\rm SR}_{211}\epsilon_{211} -2\gamma^{322\times{\rm BH}}_{211\times211}\epsilon_{211}^2\epsilon_{322} +\gamma^{211\times\infty}_{322\times322}\epsilon_{322}^2\epsilon_{211} \nonumber\\
&-2\gamma^{\rm GW}_{211\times211}\epsilon_{211}^2+\gamma^{\rm GW}_{322\to211}\epsilon_{211}\epsilon_{322}\,,\\[4pt]
\dot{\epsilon}_{322} ={}& \gamma^{\rm SR}_{322}\epsilon_{322} +\gamma^{322\times{\rm BH}}_{211\times211}\epsilon_{211}^2\epsilon_{322} -2\gamma^{211\times\infty}_{322\times322}\epsilon_{322}^2\epsilon_{211} \nonumber\\
&-2\gamma^{\rm GW}_{322\times322}\epsilon_{322}^2 -\gamma^{\rm GW}_{322\to211}\epsilon_{211}\epsilon_{322} \, .
\end{align}
The corresponding evolution of the black hole spin and mass is 
\begin{align}
\dot{\tilde{a}} &=-\gamma^{\rm SR}_{211}\epsilon_{211}-2\gamma^{\rm SR}_{322}\epsilon_{322} \, - 2 \tilde{a} \frac{\dot{M}}{M} \label{eq:timed_a},\\[4pt]\frac{\dot{M}}{\mu G M^2} &=-\gamma^{\rm SR}_{211}\epsilon_{211}-\gamma^{\rm SR}_{322}\epsilon_{322}+\gamma^{322\times{\rm BH}}_{211\times211}\epsilon_{211}^2\epsilon_{322}\, . \label{eq:timed_M}
\end{align}
The final term of Eq.~\eqref{eq:timed_a} is usually neglected, as $\dot{M}/M$ is typically small. Eq.~\eqref{eq:timed_M} also assumes that the energy of the quasi-bound states is $\omega_{n\ell m} \simeq \mu$; this approximation is excellent at small $\alpha$, but introduces a small error at larger $\alpha$. 
This two-level system is only a pedagogical example and, as we will see, a poor description of the real dynamics outside a narrow corner of parameter space. In the full analysis we include all allowed leading-order quartic $2\to2$ processes up to the chosen truncation, together with the relevant superradiant and GW rates.
\section{Stable evolution}
\label{sec:convergence}
The formalism above leads to a large set of coupled, non-linear rate equations. As described in Sec.~\ref{sec:formalism}, three of the four legs are populated bound states, while the fourth is a forced mode that either carries energy into the black hole or radiates it to infinity. The number of allowed channels therefore grows rapidly as more bound states are retained.
The purpose of this section is to determine how many states are actually needed to characterize the evolution. We enlarge the state space systematically and ask when the evolution stops changing on the astrophysical timescale of interest, using the criteria defined in Sec.~\ref{sec:intro}: adding the next set of more weakly bound states, together with all of the self-interaction channels they bring with them, must introduce no new mode above $\epsilon_{\rm th}$ (\emph{cloud convergence}) and must leave the black hole spin and mass unchanged, up to a small shift in time (black-hole convergence; see Eq.~\eqref{eq:bh_conv} below).
We organize the calculation by a maximum principal quantum number $n_{\rm max}$. For a given $n_{\rm max}$, we include all superradiant bound states with $n\leq n_{\rm max}$ and compute both their superradiant rates and the leading-order scattering rates among them. The restriction to superradiant states (with $m > 0$) is well motivated for isolated black holes in vacuum, since superradiance only populates $m >0$, and selection rules then inhibit the ability to spread the energy and angular momentum to $m \leq 0$ states (states with $m=0$ are also unstable and decay on short timescales). This conclusion will not hold when environmental factors are considered, but it provides an estimate of the leading order behavior of these systems. Restricting attention to the set of $m > 0$ states, the calculation of interest covers $\mathcal{O}(10^3)$ quasi-bound states and $\mathcal{O}(10^5)$ scattering rates, which we pre-compute on dense grids in parameter space and interpolate for the time evolutions.
An important subtlety arises when truncating the system at a given $n_{\rm max}$. In order to illustrate the problem, consider for example what happens with $n_{\rm max} = 3$. The $\ket{211}$ state grows first, and carries $\ket{322}$ with it, establishing an equilibrium. Eventually $\ket{211}$ spins down the black hole and begins to decay, while $\ket{322}$, which is still superradiant, grows. Without higher-order states, $\ket{322}$ grows unimpeded (see, \eg Fig.~7 in \cite{Witte:2024drg}), and behaves as in the non-interacting limit. This is unphysical. One can certainly write down a process, such as $\ket{322} \times \ket{322} \rightarrow \ket{544} \times {\rm BH}$, which efficiently populates $\ket{544}$ and drives the system to a new equilibrium. While other states may also grow, there is always a minimal state with $n > n_{\rm max}$ that must be included in order to prevent the run-away growth of states with $n$ near, or equal to, $n_{\rm max}$. To avoid such spurious growth (which would complicate the interpretation of convergence with increasing $n_{\rm max}$), we always include a minimal set of higher-lying ``sink'' channels that keep the lower-$n$ evolution self-consistent. In practice, the most important examples are double-annihilation channels, in which two particles in the same state scatter into a higher state while the remaining energy is absorbed by the black hole. For example, even in a calculation nominally truncated at $n_{\rm max}=4$, the evolution of $\ket{322}$ and $\ket{433}$ depends on channels such as
\begin{equation}
\begin{aligned}
    \ket{322}\times\ket{322} &\to \ket{544}\times {\rm BH}\,,\\
    \ket{433}\times\ket{433} &\to \ket{766}\times {\rm BH}\,.\nonumber
\end{aligned}
\end{equation}
The sink channels alone are not enough, though -- one must also include the corresponding reverse processes (\eg $\ket{544}\times\ket{544} \to \ket{322}\times \infty$), which regulate the evolution and establish an equilibrium. Throughout the rest of this work, an evolution labeled by $n_{\rm max}$ thus includes all allowed leading-order quartic processes among the superradiant states with $n\leq n_{\rm max}$, plus these sink channels. We then raise $n_{\rm max}$ and recompute until the occupation-number histories -- and therefore the spin-down history -- are stable.
In practice, black hole convergence over the full parameter space is reached at $n_{\max} = 8$, while obtaining cloud convergence in many cases requires retaining states at $n \sim 40$ -- in some cases, cloud convergence even requires states above this threshold (see Fig.~\ref{fig:cloudconv}). There exist, in fact, special points in parameter space where extremely high $n$ states are populated -- in these cases, the suppression of the scattering rate that is expected to occur as $n$ increases (which arises because the spatial separation of the wavefunctions increases, see the integrand in Eq.~\eqref{eq:sourcefull}) is offset by either a shift in the phase of the wave functions, or by a resonant feature that only arises for more weakly bound states. We comment more on these special cases below.
Finally, the numerical integration itself is challenging. The system often evolves through long quasi-equilibrium phases, during which superradiant growth and self-interaction losses approximately balance, followed by much shorter transient phases triggered by the delayed growth of another mode. These transitions can redistribute a significant fraction of the cloud energy among several states. The relevant timescale therefore changes dynamically during the evolution. We handle this by continuously adjusting the tolerances and time steps according to the shortest relevant dynamical timescale in the system. 
\subsection{Included and neglected processes}
\label{sec:superres}
\begin{figure*}
    \centering    
    \includegraphics[width=\linewidth]{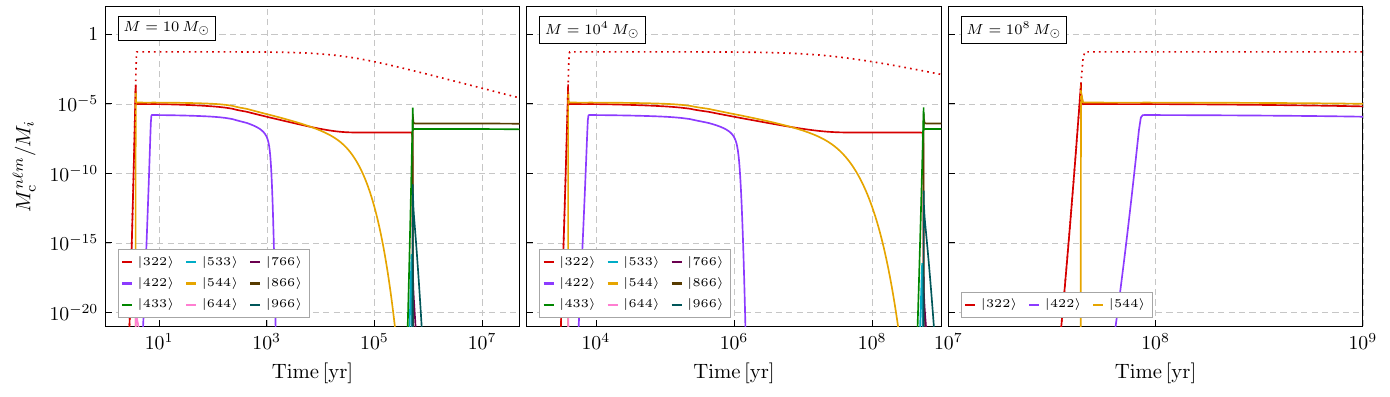}
    \caption{Fractional cloud mass $M_{\rm c}^{n\ell m}/M_i$ per level for $f_a = 10^{16}\,$GeV, $\alpha = 0.4$, $\tilde a_i = 0.95$ and $n_{\max} = 15$, for $M_i = 10$, $10^4$ and $10^8\,M_\odot$ (left to right), evolved for $5\times10^{7}\,$yr, 1\,Gyr and 1\,Gyr. Dotted lines show the non-interacting limit. Apart from a small growth delay due to the mass-dependent seed, the evolutions follow the same history when time is rescaled by the black hole mass. }
\label{fig:overview_states}
\end{figure*}

\begin{figure}
    \centering    
    \includegraphics[width=0.98\linewidth]{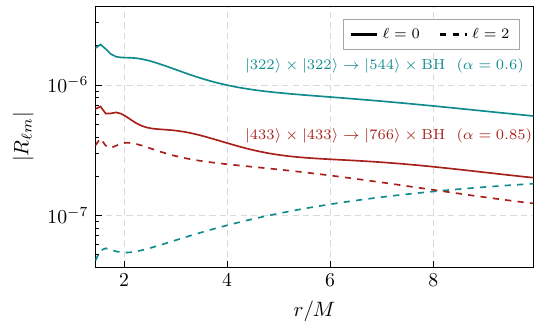}
    \caption{Absolute value of the $\ell = 0$ and $\ell = 2$ inhomogeneous wavefunctions for two processes dissipating at the horizon. The quadrupole is strongly suppressed at $\alpha = 0.6$, but comparable to the monopole at $\alpha = 0.85$.}
    \label{fig:Horizon}
\end{figure}

Before discussing which processes we include, it is useful to explain how the evolution depends on the black hole mass. At fixed $\alpha$, $f_a$ and $\tilde a$, the rate coefficients entering Eq.~\eqref{eq:general_rate} all scale as $(GM)^{-1}$. The self-interaction prefactor $\alpha^4(M_{\rm Pl}/f_a)^4$ is also independent of $M$ [Eq.~\eqref{eq:coupling}]. Written in terms of the dimensionless occupations $\epsilon_{n\ell m}$ and time in units of $GM$, the equations therefore have no explicit dependence on the overall mass scale. The only difference comes from the initial seed which introduces a small dependence via $\epsilon_{n\ell m}(0)=1/(GM^2)$. Heavier black holes thus require more e-foldings to reach the same dimensionless occupation, with a growth delay that depends logarithmically on the seed.
Figure~\ref{fig:overview_states} shows this for $f_a=10^{16}\,$GeV, $\alpha=0.4$ and $\tilde a_i=0.95$. When compared at the rescaled time $t_{\rm ref}=t\,(10\,M_\odot/M)$, the three systems follow the same sequence of populated levels over the intervals covered by each calculation. The peak occupation of the dominant level agrees to within $\sim5\%$ for $10^4\,M_\odot$ and $\sim13\%$ for $10^8\,M_\odot$ across the parameter space; larger differences occur only for levels that remain strongly subdominant. The spin agrees to within $0.2\%$ (not shown here). Changing the mass thus simply changes how much of the evolution fits within the integration time. Over $1\,$Gyr, a $10^4\,M_\odot$ ($10^8\,M_\odot$) black hole covers roughly the first $10^6\,$yr ($10^2\,$yr). This explains why fewer states are populated for the heavier black holes (189 and 29 states for $10^4$ and $10^8\,M_\odot$, against 203 for $10\,M_\odot$ (compare Figs.~\ref{fig:table_M1e4} and~\ref{fig:table_M1e8} with Fig.~\ref{fig:tab_tot}), and why the supermassive case generally requires a lower $n_{\max}$ to reach convergence. Since the stellar-mass calculations cover the rescaled time intervals of both heavier cases, we focus for the rest of this paper on $M=10\,M_\odot$. The corresponding evolution figures for $M=10^4$ and $10^8\,M_\odot$ are available online~\cite{AxionSR:zenodo}.
In our analysis, we {\emph{do not}} include the following effects:
\begin{enumerate}
    \item[(i)] $2 \to 2$ processes involving at least one non-superradiant bound state -- for isolated black holes, the occupation numbers of such states are expected to always remain small, and thus their scattering rates are heavily suppressed;
    \item[(ii)] $3\to1$ processes, which are suppressed by the mismatch in spatial scale between the relativistic emitted axion and the bound wavefunctions;
    \item[(iii)] the frequency shifts induced by the quartic term, which do not appreciably alter the populations;

    \item[(iv)] accretion (and all environmental effects), which in the case of $10^4$ and $10^8 \, M_\odot$ could alter the mass and spin evolution (and thus also indirectly alter which states are superradiant, or how efficient various scattering processes are). This would also break the aforementioned time scaling. 
\end{enumerate}
Everything else -- every $2\to2$ channel among the retained superradiant levels, together with the sink channels described above -- is kept. There is, however, one class of process that requires more care.
When the fourth leg is a forced bound oscillation that is absorbed at the horizon, dissipation is typically most efficient for an $\ell=m=0$ mode, since higher multipoles must tunnel through the centrifugal barrier to reach the horizon. The horizon absorption probability of a fourth leg with angular momentum $\ell$ is suppressed relative to the monopole by $\sim \alpha^{4\ell}$, so at small $\alpha$ only the monopole contributes appreciably. This is why previous work retained the $\ell=0$ leg alone \cite{Baryakhtar:2020gao, Witte:2024drg}. As $\alpha\to\mathcal{O}(1)$, however, this suppression weakens, and higher multipoles can no longer be dismissed automatically. This is illustrated in Fig.~\ref{fig:Horizon}, which shows the absolute value of the fourth-leg wavefunction as a function of radius for two processes: at $\alpha=0.6$ the quadrupole remains strongly suppressed relative to the monopole, whereas at $\alpha=0.85$ the two become comparable.
In some cases the $\ell=2$ contribution can even dominate over the monopole, by several orders of magnitude, due to resonant enhancement. The rate is determined by the full sourced solution \eqref{eq:radial_sol}, whose denominator contains the Wronskian \eqref{eq:wronskian} evaluated at the forced frequency $\omega_X$, and the Wronskian vanishes whenever $\omega_X$ coincides with a quasi-bound eigenfrequency. For some channels, the fourth leg lands \emph{nearly on shell}, $\omega_X \simeq \omega^{\rm R}_{n\ell m}$, so that $W_{\ell m\omega_X}$ becomes anomalously small and the rate is resonantly enhanced. Such resonances were already noticed in \cite{Baryakhtar:2020gao}, where the fourth leg of $\ket{211} \times \ket{311} \rightarrow \ket{322} \times {\rm BH}$ was found to be nearly degenerate with the $\ket{200}$ state, enhancing the rate by $\alpha^{-4}$ (typically $|\omega_X - \omega_{n00}| = \mathcal{O}(\alpha^2)$, but in this case $|\omega_X - \omega_{n00}| = \mathcal{O}(\alpha^4)$). We find several cases in which the $\ell = 2$ leg is resonant while the $\ell = 0$ leg is not. For this reason, we systematically include both $\ell=0$ and $\ell=2$ contributions in all calculations.
The rare and unexpected resonant enhancement of the $\ell=2$ scattering processes raises two additional points. First, higher multipoles could in principle also benefit from resonant enhancements. We have checked representative $\ell=4$ channels, however, and found that even at large $\alpha$ the centrifugal suppression remains too strong for resonances to make them competitive; we therefore neglect $\ell\geq4$ contributions, leaving their systematic inclusion to future work. Second, some processes may have their monopole contribution forbidden by selection rules, while a bound mode with $\ell, m \neq 0$ is allowed and resonantly enhanced. Such channels can exhibit particularly large rates despite the absence of a monopole contribution, giving rise to what we call \emph{super-resonances}\footnote{The term ``super'' is used to differentiate these from the resonant processes highlighted in~\cite{Baryakhtar:2020gao}, with the former being far more effective in amplifying the rate. We thank Lucas Gathercole for first noticing this effect, which will be described in more detail in upcoming work \cite{Lucas_prep}.} -- we discuss these super-resonances in the following subsection.
\subsection{Super-resonances}
Restricting our attention to superradiant bound states, we identify approximately $60$ super-resonant channels, whose contributions can significantly affect the cloud evolution -- these rates are listed in Table~\ref{tab:superres}. A representative example is
\begin{equation}
\ket{533} \times \ket{533} \rightarrow \ket{544} \times \mathrm{BH}\nonumber\,,
\end{equation}
for which the forced mode lies extremely close to the $\ket{522}$ quasi-bound state. These channels have a further striking feature. Unlike the $\ell=2$ contributions discussed above, which have $m=0$, super-resonant channels can carry $m_X\neq0$. When the resonant mode carries $m>0$ and the black hole spins fast enough, the mode itself satisfies the superradiant condition \eqref{eq:SR_cond}. The horizon flux then \emph{reverses}: the process extracts rotational energy from the black hole rather than dissipating cloud energy into it. Super-resonances therefore mediate not only enhanced horizon absorption, but also enhanced superradiant scattering. Because these forced legs carry $m_X\neq0$, they also exchange angular momentum with the black hole, and this contribution enters the evolution of $\tilde{a}$ alongside the superradiant terms.
\begin{table*}
\centering
\small
\renewcommand{\arraystretch}{1.2}
\setlength{\tabcolsep}{9pt}
\begin{tabular}{lll}
\toprule
\multicolumn{3}{c}{\textbf{Super-resonant channels}} \\
\midrule
  $\ket{422}\times\ket{422}\to\ket{411}\times\mathrm{BH}$ & $\ket{744}\times\ket{755}\to\ket{733}\times\mathrm{BH}$ & $\ket{844}\times\ket{877}\to\ket{855}\times\mathrm{BH}$ \\
  $\ket{422}\times\ket{422}\to\ket{433}\times\mathrm{BH}$ & $\ket{744}\times\ket{755}\to\ket{766}\times\mathrm{BH}$ & $\ket{844}\times\ket{877}\to\ket{866}\times\mathrm{BH}$ \\
  $\ket{522}\times\ket{522}\to\ket{511}\times\mathrm{BH}$ & $\ket{755}\times\ket{755}\to\ket{744}\times\mathrm{BH}$ & $\ket{855}\times\ket{855}\to\ket{833}\times\mathrm{BH}$ \\
  $\ket{522}\times\ket{522}\to\ket{533}\times\mathrm{BH}$ & $\ket{755}\times\ket{755}\to\ket{766}\times\mathrm{BH}$ & $\ket{855}\times\ket{855}\to\ket{844}\times\mathrm{BH}$ \\
  $\ket{533}\times\ket{533}\to\ket{522}\times\mathrm{BH}$ & $\ket{765}\times\ket{765}\to\ket{766}\times\mathrm{BH}$ & $\ket{855}\times\ket{855}\to\ket{866}\times\mathrm{BH}$ \\
  $\ket{533}\times\ket{533}\to\ket{544}\times\mathrm{BH}$ & $\ket{833}\times\ket{833}\to\ket{844}\times\mathrm{BH}$ & $\ket{855}\times\ket{855}\to\ket{877}\times\mathrm{BH}$ \\
  $\ket{633}\times\ket{633}\to\ket{644}\times\mathrm{BH}$ & $\ket{833}\times\ket{833}\to\ket{855}\times\mathrm{BH}$ & $\ket{855}\times\ket{866}\to\ket{844}\times\mathrm{BH}$ \\
  $\ket{633}\times\ket{633}\to\ket{655}\times\mathrm{BH}$ & $\ket{833}\times\ket{844}\to\ket{855}\times\mathrm{BH}$ & $\ket{855}\times\ket{866}\to\ket{877}\times\mathrm{BH}$ \\
  $\ket{633}\times\ket{644}\to\ket{655}\times\mathrm{BH}$ & $\ket{833}\times\ket{844}\to\ket{866}\times\mathrm{BH}$ & $\ket{866}\times\ket{866}\to\ket{855}\times\mathrm{BH}$ \\
  $\ket{644}\times\ket{644}\to\ket{633}\times\mathrm{BH}$ & $\ket{833}\times\ket{866}\to\ket{844}\times\mathrm{BH}$ & $\ket{866}\times\ket{866}\to\ket{877}\times\mathrm{BH}$ \\
  $\ket{644}\times\ket{644}\to\ket{655}\times\mathrm{BH}$ & $\ket{833}\times\ket{866}\to\ket{855}\times\mathrm{BH}$ & $\ket{966}\times\ket{966}\to\ket{988}\times\mathrm{BH}$ \\
  $\ket{733}\times\ket{733}\to\ket{744}\times\mathrm{BH}$ & $\ket{833}\times\ket{866}\to\ket{877}\times\mathrm{BH}$ & $\ket{10\,66}\times\ket{10\,66}\to\ket{10\,88}\times\mathrm{BH}$ \\
  $\ket{733}\times\ket{733}\to\ket{755}\times\mathrm{BH}$ & $\ket{833}\times\ket{877}\to\ket{844}\times\mathrm{BH}$ & $\ket{11\,66}\times\ket{11\,66}\to\ket{11\,88}\times\mathrm{BH}$ \\
  $\ket{733}\times\ket{744}\to\ket{755}\times\mathrm{BH}$ & $\ket{833}\times\ket{877}\to\ket{866}\times\mathrm{BH}$ & $\ket{11\,66}\times\ket{11\,66}\to\ket{11\,10\,10}\times\mathrm{BH}$ \\
  $\ket{733}\times\ket{744}\to\ket{766}\times\mathrm{BH}$ & $\ket{844}\times\ket{844}\to\ket{833}\times\mathrm{BH}$ & $\ket{11\,66}\times\ket{11\,88}\to\ket{11\,10\,10}\times\mathrm{BH}$ \\
  $\ket{733}\times\ket{766}\to\ket{744}\times\mathrm{BH}$ & $\ket{844}\times\ket{844}\to\ket{855}\times\mathrm{BH}$ & $\ket{11\,88}\times\ket{11\,88}\to\ket{11\,66}\times\mathrm{BH}$ \\
  $\ket{733}\times\ket{766}\to\ket{755}\times\mathrm{BH}$ & $\ket{844}\times\ket{844}\to\ket{866}\times\mathrm{BH}$ & $\ket{11\,88}\times\ket{11\,88}\to\ket{11\,10\,10}\times\mathrm{BH}$ \\
  $\ket{744}\times\ket{744}\to\ket{733}\times\mathrm{BH}$ & $\ket{844}\times\ket{844}\to\ket{877}\times\mathrm{BH}$ & $\ket{12\,66}\times\ket{12\,66}\to\ket{12\,10\,10}\times\mathrm{BH}$ \\
  $\ket{744}\times\ket{744}\to\ket{755}\times\mathrm{BH}$ & $\ket{844}\times\ket{866}\to\ket{833}\times\mathrm{BH}$ & $\ket{13\,66}\times\ket{13\,66}\to\ket{13\,10\,10}\times\mathrm{BH}$ \\
  $\ket{744}\times\ket{744}\to\ket{766}\times\mathrm{BH}$ & $\ket{844}\times\ket{866}\to\ket{877}\times\mathrm{BH}$ & $\ket{14\,66}\times\ket{14\,66}\to\ket{14\,10\,10}\times\mathrm{BH}$ \\
\bottomrule
\end{tabular}
\caption{\label{tab:superres} Super-resonant channels $\ket{a}\times\ket{b}\to\ket{c}\times\mathrm{BH}$ discussed in Sec.~\ref{sec:superres}, ordered by $n$; in every channel all three states share the same $n$.}
\end{table*}
One might worry that these resonances are fragile, in the sense that perturbations to the energy eigenvalues from, \eg self-gravity or self-interactions could de-tune them. We have verified that this does not appear to be the case for the systems studied here, but we cannot rule out that environmental perturbations are relevant; conversely, such perturbations could also create resonances where there otherwise would not be one. We leave a more detailed investigation of these effects to future work. 

\subsection{Evolution of the states}
\begin{figure}
    \centering    
    \includegraphics[width=0.98\linewidth]{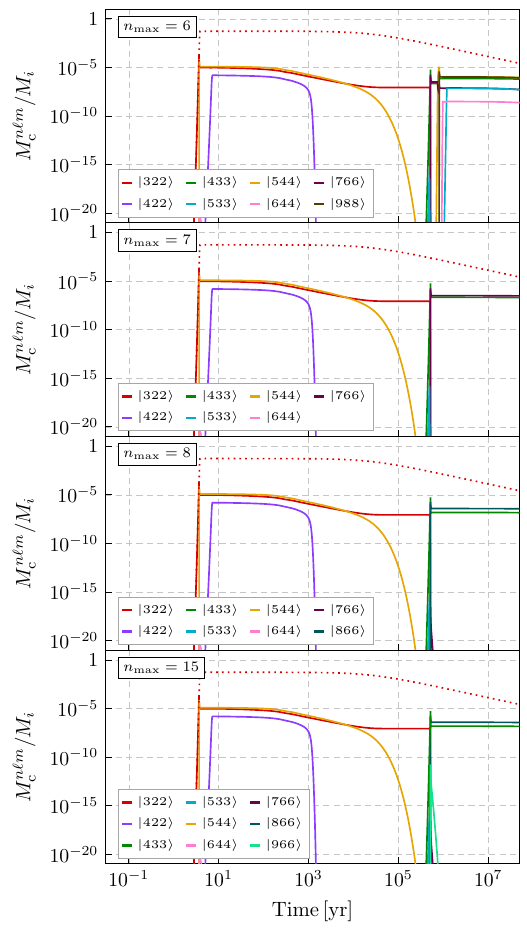}
    \caption{Fractional cloud mass $M_{\rm c}^{n\ell m}/M$ per level for $M_i = 10\,M_\odot$, $\tilde a_i = 0.95$, $\alpha = 0.4$ and $f_a = 10^{16}\,$GeV, for $n_{\max} = 6$, 7, 8 and 15 (top to bottom). Dotted lines show the non-interacting limit.}
    \label{fig:states_fa1e16}
\end{figure}
\begin{figure}
    \centering    \includegraphics[width=0.98\linewidth]{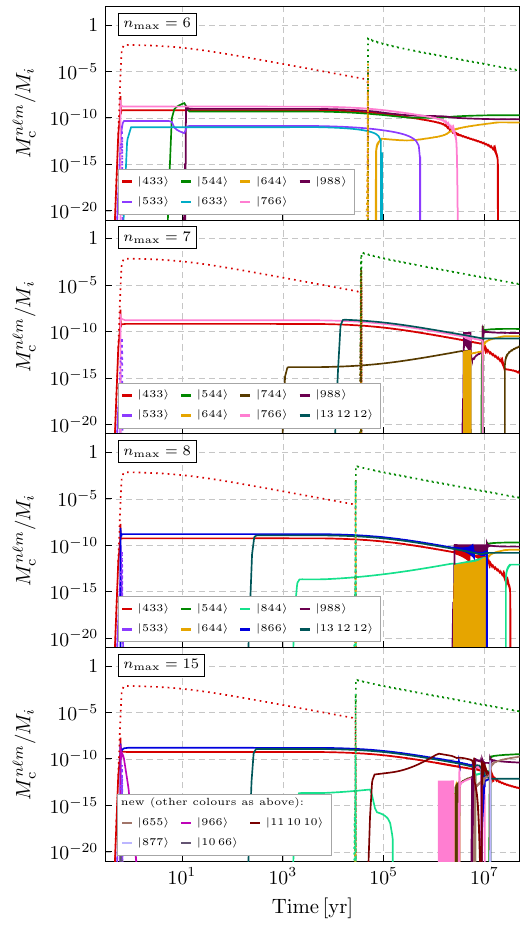}
    \caption{Same as Fig.~\ref{fig:states_fa1e16}, for $f_a = 10^{14}\,$GeV and $\alpha = 1.0$. In the bottom panel only states not present at lower $n_{\max}$ are labeled; colors are as in the panels above.}
    \label{fig:states_fa1e14}
\end{figure}
\begin{figure}
    \centering    
    \includegraphics[width=0.98\linewidth]{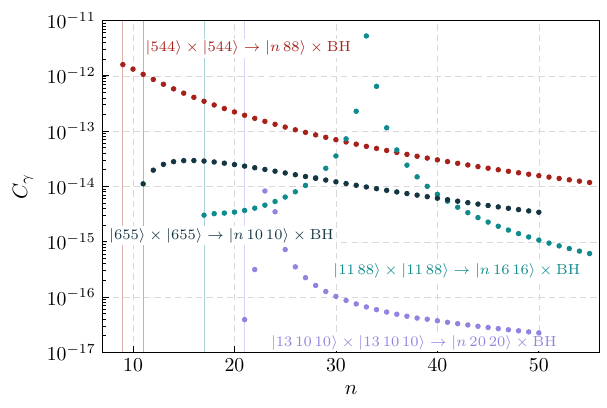}
    \caption{Coefficient $C_\gamma$ of the non-relativistic scattering rate, $\gamma \propto C_\gamma (M_{\rm Pl}/f_a)^4 \alpha^{p} \mu$ with $p$ scattering-channel-dependent, as a function of the final-state $n$ for four double-annihilation channels. Vertical lines mark the lowest allowed $n$. }
    \label{fig:1188_direction}
\end{figure}

We consider three black hole masses, $M= 10$, $10^4$ and $10^8\,M_\odot$, each starting from $\tilde a_i = 0.95$, and follow the evolution for $\tau = 5\times 10^7$, $10^9$ and $10^9\,\mathrm{yr}$, respectively, over the range $0.05 \lesssim \alpha \lesssim 1.6$, chosen to roughly coincide with where superradiant growth is fast enough to spin down the black hole on these timescales. At each $\alpha$ we scan the decay constant over $f_a = 10^{12}-10^{18}\,\mathrm{GeV}$, and raise $n_{\rm max}$ until the evolution converges.  
We begin by computing the evolution with all states, and all possible permutations of scattering processes, among states with $n \leq 8$. This was sufficient to establish black hole convergence, but not cloud convergence. Across much of the parameter space, towers of states in $n$ -- \eg  $\ket{n \, 66}$, $\ket{n \, 88}$, $\ket{n \, 10 \, 10}$, and $\ket{n \, 16 \, 16}$ -- were populated up to $ n = n_{\rm max}$. On the other hand, only a small number of states with $\ell \neq m$ appeared. Based on the $n_{\rm max} = 8$ results, and using the fact that superradiance is too inefficient in this range to populate $n \geq 9$ states, we extended the state space to include all states with $\ell = m$ at $n = 9$ (with the associated ``sink'' states), and include all possible scattering permutations of the new state space. Analyzing these states, it becomes clear that the occupied states extending to high values of $n$ (and preventing cloud convergence) are ``towers'' of states with  $\ell, m \in \{6,8,10, \ldots\}$. As discussed below, these states are populated via scattering processes which remain degenerate, or are even resonantly enhanced at high $n$ (see Fig.~\ref{fig:1188_direction}). Therefore, in order to push to higher order, we construct these towers up to $n_{\rm max} = 15$ and $n_{\rm max} = 18$, including, as before, all scattering processes and sink states. In some cases, this is sufficient to establish cloud convergence -- in many cases, however, it is not. We therefore extend specific towers to $n \leq 45$, including in these special cases only the minimal required scattering processes, \eg $\ket{n33} \times \ket{n33} \rightarrow \ket{n^\prime 66} \times {\rm BH}$ and $\ket{n^\prime 66} \times \ket{n^\prime 66} \rightarrow  \ket{n33} \times \infty$ but no additional cross-scattering processes between the newly added states. Our complete set of runs therefore consists of\footnote{Sink states are always implicitly included.}
\begin{itemize}
    \item $n_{\rm max}= 3-8$: complete set of states  and scattering processes;
    \item $n_{\rm max}= 9$: added $\ket{9 \ell m}$ with $\ell=m$ only, and all scattering processes;
    \item $n_{\rm max}= 15,18$: added $\ket{n 66}$, $\ket{n 88}$, $\ket{n \, 10 \, 10}$ states, and all scattering processes;
    \item $n_{\rm max}= 45$: $\ket{n 66}$, $\ket{n 88}$, $\ket{n \, 10 \, 10}$, $\ket{n \, 16 \, 16}$, $\ket{n \, 20 \, 20}$ states with minimal set of scattering processes (annihilation to, and inverse processes scattering from, these states with the lower $(\ell,m)$ states that are responsible for sourcing them). 
\end{itemize}
While our higher $n_{\rm max}$ runs do not strictly include all states or scattering permutations, we see no indication in the existing runs that states outside this set would be populated with a significant occupation number.
We call a run with truncation $n_{\max}$ black-hole converged with respect to the next smaller truncation in this list, $n'_{\max}$, if
\begin{equation}
  \min_{|\ln(t'/t)|\le\tau} \frac{\bigl|X_{n_{\max}}(t)-X_{n'_{\max}}(t')\bigr|}{X_{n_{\max}}(t)} \le \delta\,, \qquad X \in \{\tilde a, M\}\,,
  \label{eq:bh_conv}
\end{equation}
for all $t$ covered by both runs. For $\tau = 0$ this reduces to the pointwise criterion, while a finite $\tau$ allows a spin-down step to move by a factor $e^{\tau}$ in time. Assuming the states driving the spin-down are in equilibrium, their occupation scales as $f_a^2$ and the spin-down time as $t_{\rm sd} \propto f_a^{-2}$, so such a shift corresponds to a change $\delta\log_{10} f_a = \tau/(2\ln 10)$ in the decay constant. We take $\delta = 1\%$ and $\tau = 0.26$, \ie a time shift of up to $30\%$ (at most $15\,$Myr for $M = 10\,M_\odot$), equivalent to rescaling $f_a$ by a factor of $1.14$. 
The parameter space is too large to describe point by point, so we focus on two representative configurations that exhibit non-trivial dynamics, characterized by the appearance of higher-order states. Further examples are presented in App.~\ref{app:evolutions} and in an online repository~\cite{AxionSR:zenodo}, while an overview of all relevant states across the parameter space is provided in Fig.~\ref{fig:tab_tot}.
Specifically, we consider the evolution of a stellar-mass black hole for $\alpha=0.4$ and $f_a=10^{16}\,\mathrm{GeV}$, shown in Fig.~\ref{fig:states_fa1e16}, and for $\alpha=1.0$ and $f_a=10^{14}\,\mathrm{GeV}$, shown in Fig.~\ref{fig:states_fa1e14}. For both configurations, we examine the evolution at $n_{\rm max}=6,7,8$, and $15$. By $n_{\rm max}=15$, the evolution has converged, with results indistinguishable from those obtained at $n_{\rm max}=18$.
\paragraph{Moderate coupling} At $\alpha = 0.4$ the $m=1$ modes are stable, and the dominant superradiant mode is $|322\rangle$. It grows first through superradiance and, once it reaches a sufficiently large occupation number, sources the growth of $|544\rangle$ through the process
\begin{equation*}
    |322\rangle \times |322\rangle \rightarrow |544\rangle \times \mathrm{BH}.
\end{equation*}
As the $|544\rangle$ occupation increases, the inverse depletion channel
\begin{equation*}
    |544\rangle \times |544\rangle \rightarrow |322\rangle \times \infty
\end{equation*}
also becomes important. The two modes then enter a quasi-equilibrium phase, in which the black hole spin evolution is controlled mainly by the superradiant growth of the $|322\rangle$ mode.
Shortly afterwards, the $|422\rangle$ mode also grows via superradiance.
This growth does not continue, however, as the channel
\begin{equation*}
    |422\rangle \times |422\rangle \rightarrow |544\rangle \times \mathrm{BH}
\end{equation*}
depletes it, driving an equilibrium phase similar to that of $|322\rangle$--$|544\rangle$. Moreover, by the time $|422\rangle$ becomes important, the black hole has already been partially spun down by the dominant $|322\rangle$ mode, and the remaining spin is less favorable for a second, efficient $m=2$ growth phase. In the absence of other modes, $|422\rangle$ would simply approach its superradiant threshold. In the full system, however, the lower-frequency $|322\rangle$ and $|544\rangle$ modes continue to extract angular momentum from the black hole; the spin consequently drops below the threshold for $|422\rangle$, after which this mode is reabsorbed by the black hole.
At later times the $|433\rangle$ mode grows through superradiance, driving the population of the higher-lying $|n66\rangle$ modes. For $n_{\max}=7$, the $|766\rangle$ mode becomes significantly populated, and since is the fundamental mode, one could be tempted to think that it could be the dominant state of this family. However, the channel
\begin{equation*}
    |433\rangle \times |433\rangle \rightarrow |n66\rangle \times \mathrm{BH},
\end{equation*}
and its associated backscattering processes remain efficient for a range of higher-$n$ states. In practice, the rates for several states with $n=7,8,\ldots,15$ are comparable, so the evolution cannot be reliably captured by retaining only the lowest allowed final state.
Indeed, increasing the truncation to $n_{\max}=8$ reveals that $|866\rangle$ becomes more populated than $|766\rangle$. The evolution with $n_{\max}=15$, however, shows that the higher-lying modes, starting from $|966\rangle$, are unable to sustain their growth and eventually decay. The late-time evolution is therefore characterized by a quasi-equilibrium between the $|433\rangle$ and $|866\rangle$ modes.
Finally, at $n_{\rm max}=6$, the $\ket{544}$ state is re-populated at late times through the GW level transition $\ket{766}\rightarrow\ket{544}+{\rm GW}$. At higher $n_{\rm max}$ this feature disappears, because the growth of $\ket{544}$ is quenched by the scattering process 
\begin{equation*}
\ket{544}\times\ket{766}\rightarrow\ket{433}\times\infty\,.
\end{equation*}
This channel is absent at $n_{\rm max}=6$, where $\ket{766}$ enters only as a sink state and therefore couples to the rest of the system solely through its double-annihilation channel and its reverse.
\paragraph{Stronger couplings} Figure~\ref{fig:states_fa1e14} shows a point in parameter space where the evolution is considerably more complex, $f_a=10^{14}\,\mathrm{GeV}$ and $\alpha=1.0$. At this value of $\alpha$, $m=2$ modes are stable, so the dominant superradiant modes are $m=3$. As shown in the plot, while at $n_{\max}=6$ all of $\ket{433}$, $\ket{533}$ and $\ket{633}$ are able to grow efficiently almost simultaneously, at $n_{\rm max}=7$ the latter two states do not appear because, at this stage, the cross-process 
\begin{equation*}
\ket{533}\times\ket{633}\rightarrow\ket{766}\times\mathrm{BH}
\end{equation*}
suppresses their growth. Furthermore, following the same reasoning as in the weak-coupling case, $\ket{433}$ drags $\ket{866}$ rather than $\ket{766}$ for $n_{\rm max}\geq 8$. However, unlike in the previous regime, the strong-coupling dynamics also leads to the growth of $\ket{13\,12\,12}$, which was previously absent.  At later times, the evolution is further complicated by the superradiant growth of $\ket{n44}$ modes, particularly $\ket{844}$. In the presence of self-interactions, this mode grows earlier because the reduced occupation number of $\ket{433}$ suppresses the associated spin-down, allowing the black hole to retain a higher spin, at which the superradiant growth rate of $\ket{844}$ is enhanced. At $n_{\rm max}=15$, $n=8$ modes populate $\ket{11\,10\,10}$ via 
\begin{equation*}
\ket{844}\times \ket{866}\rightarrow\ket{11\,10\,10}\times \rm BH\,.
\end{equation*}
Finally, higher-$n$ states populate lower-lying modes via backscattering. In particular, $\ket{13\,12\,12}$ populates $\ket{966}$ and $\ket{10\,6\,6}$, while $\ket{11\,10\,10}$ populates $\ket{655}$. Note that superradiance alone would not be efficient enough to populate $\ket{655}$, as can be seen in the $n_{\max}<15$ case. The final configuration is a delicate equilibrium between these states.
Appendix~\ref{app:evolutions} shows the two ends of the range of decay constants, $f_a = 10^{18}$ and $10^{12}\,\mathrm{GeV}$. In the latter the cloud settles into long sequences of quasi-equilibria and no single level ever reaches a large occupation. The cloud is correspondingly light, but many high-$n$ states stay occupied at once, which may still leave an imprint on GW emission or on binary systems.
These examples are representative, not exhaustive. Other regions of parameter space are richer still, with more modes populated at once. The way to read them, however, is always the same: identify which scattering channels feed and drain each mode, and how they compete with the superradiant rates.
\subsubsection{Oscillations and directions of non-convergence}
\label{subsec:ocillations}
The examples above show that, in most cases, the evolution converges as progressively higher-lying states are included. Our analysis also reveals two peculiar and unexpected features, however, that deserve further discussion.
First, consider a generic double-annihilation process $\ket{n\ell m} \times \ket{n\ell m} \rightarrow \ket{n^\prime \, 2\ell \, 2m} \times {\rm BH}$. Its rate is naively expected to decrease as $n^\prime$ increases from its minimal value $2\ell + 1$, simply because the wavefunctions of $\ket{n\ell m}$ and $\ket{n^\prime \, 2\ell \, 2m}$ become increasingly separated in space. A clear example is shown in Fig.~\ref{fig:1188_direction} for $\ket{544} \times \ket{544} \rightarrow \ket{n\,8\,8} \times {\rm BH}$ (red dots). This expected fall-off is what justifies the $n_{\rm max} \to n_{\rm max}+1$ convergence test used in this work. The problem is that for certain processes the rate is flat, or even enhanced, over an extended range of $n^\prime$ (although at sufficiently large $n^\prime$ it must eventually vanish).
\begin{figure}
    \centering    
    \includegraphics[width=0.98\linewidth]{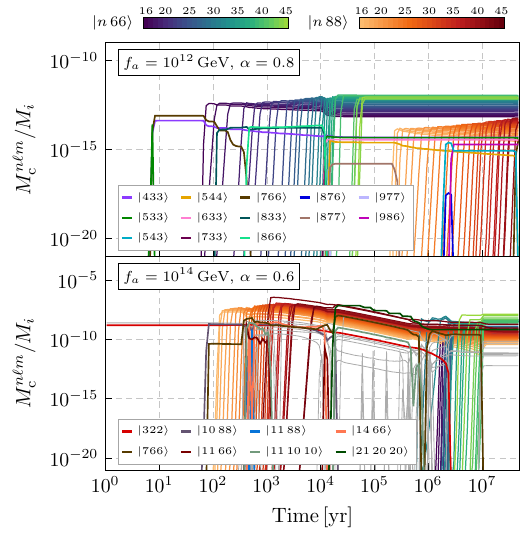}
    \caption{Fractional cloud mass with $n_{\max} = 45$ and $M_i = 10\,M_{\odot}$, for $f_a = 10^{12}\,$GeV, $\alpha = 0.8$ (top) and $f_a = 10^{14}\,$GeV, $\alpha = 0.6$ (bottom). In both cases two high-$n$ towers, $\ket{n\,66}$ and $\ket{n\,88}$ ($n \geq 16$, color scales), are populated up to $n \simeq n_{\max}$, indicating that states beyond the truncation are populated as well (these belong to red entries in Fig.~\ref{fig:cloudconv}). In the bottom panel, only the eight non-tower states with the largest peak occupation are labeled; the remaining ones are shown in grey. For visual purposes, we average out oscillations faster than $\Delta \log_{10}(t/\mathrm{yr}) = 0.05$.}
    \label{fig:high_n_example}
\end{figure}

\begin{figure}
    \centering    
    \includegraphics[width=0.98\linewidth]{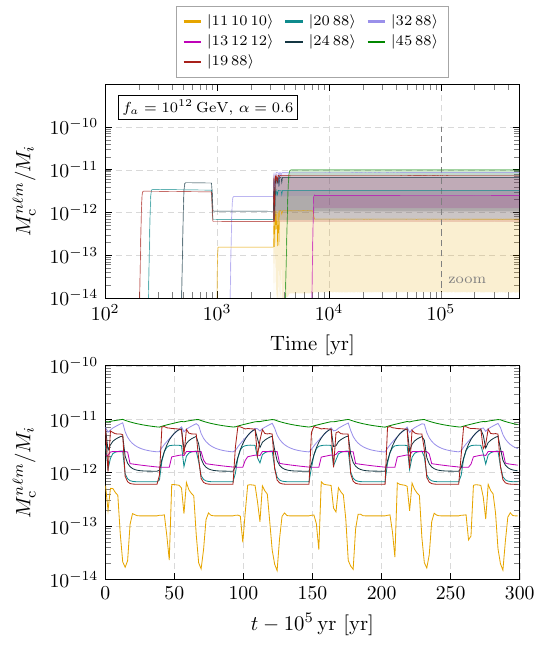}
    \caption{Oscillating occupation numbers discussed in Sec.~\ref{subsec:ocillations}. Result is shown for $M_i = 10\,M_\odot$, $f_a = 10^{12}\,$GeV and $\alpha = 0.6$, with $n_{\max} = 45$. Top: full evolution, with each level shown as its minimum--maximum envelope per logarithmic time bin; the dashed line marks the zoom. Bottom: a $300\,$yr window starting at $t = 10^5\,$yr, showing the individual oscillations. The $\ket{n88}$ tower oscillates collectively with a period of $\simeq 55\,$yr; shown are a selection of its levels together with $\ket{11\,10\,10}$ and $\ket{13\,12\,12}$.}
    \label{fig:oscillations}
\end{figure}

\begin{figure*}
    \centering
    \includegraphics[width=0.98\linewidth]{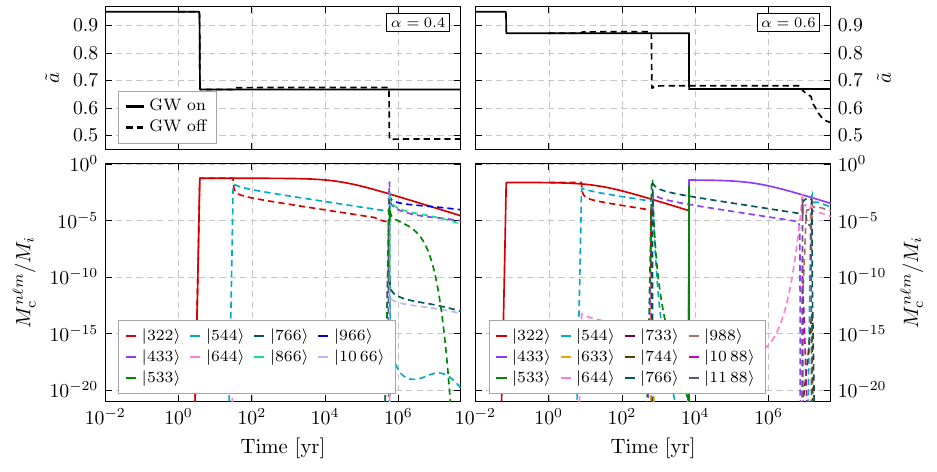}
    \caption{Impact of GW emission for $M_i = 10\,M_\odot$, $\tilde a_i = 0.95$, $f_a = 10^{18}\,$GeV and $n_{\max} = 15$, at $\alpha = 0.4$ (left) and $0.6$ (right): spin (top) and $M_{\rm c}^{n\ell m}/M_i$ (bottom), with GW annihilation and level transitions included (solid) or switched off (dashed).}
    \label{fig:GW_impact}
\end{figure*}

There are a few reasons this can occur. In some cases, the suppression from the larger spatial separation is compensated by a shift in the relative phase of the wavefunctions, which reduces the cancellation between positive and negative contributions to the source term -- an example of this behavior was already identified in Fig.~11 of~\cite{Witte:2024drg}. In other cases, the frequency of the fourth ``BH’' leg, $\omega_X = 2\omega_{n\ell m} - \omega_{n’,2\ell,2m}$, slowly drifts as $n’$ increases and can approach an $\ell = m = 0$ quasi-bound eigenfrequency, at which point the Wronskian in Eq.~\eqref{eq:radial_sol} becomes small and the rate is resonantly enhanced. In addition, since $\ell = 0$ modes are efficiently absorbed, the resonance is broad, with the counter-intuitive result that high-$n$ states are preferentially occupied. Fig.~\ref{fig:1188_direction} provides an example: the process $\ket{11 \, 8 \, 8} \times \ket{11 \, 8 \, 8} \rightarrow \ket{n \, 16 \, 16} \times {\rm BH}$ has a strong resonance at $n=33$, such that the $n=40$ state is populated as efficiently as the $n=17$ state. These situations appear predominantly at small $f_a$, and are dealt with on a case-by-case basis. Whenever such behavior is observed, we extend the tower of states at fixed $(\ell,m)$ to high $n$. For simplicity we then include only the channels $\ket{n \ell \ell} \times \ket{n \ell \ell} \rightarrow \ket{n^\prime \, 2\ell \, 2\ell} \times {\rm BH}$ and $\ket{n^\prime \, 2\ell \, 2\ell} \times \ket{n^\prime \, 2\ell \, 2\ell} \rightarrow \ket{n \ell \ell} \times \infty$, rather than all scattering permutations; we do not expect the cross terms to change the conclusions significantly. Two examples are shown in Fig.~\ref{fig:high_n_example}, for a $10\,M_\odot$ black hole with $f_a = 10^{12}\,$GeV, $\alpha = 0.8$ and $f_a = 10^{14}\,$GeV, $\alpha = 0.6$: in both cases the $\ket{n66}$ and $\ket{n88}$ towers are populated with comparable occupation all the way out to $n \simeq n_{\max} = 45$, where we stop for computational reasons, indicating that even higher states are occupied. These are the red entries of Fig.~\ref{fig:cloudconv}.
The second peculiar feature is the appearance of strong and rapid oscillations in the occupation numbers of the states (Fig.~\ref{fig:oscillations}). This behavior was also observed in the initial calculations of~\cite{Witte:2024drg}, and appears to be physical: it stems from an instability in how energy and angular momentum circulate among the quasi-bound states. These oscillations considerably complicate the evolution as resolving them slows the integration down by a factor of $\sim 10^2$. Their presence is, however, fragile. Small changes to the system can remove them, or introduce new ones where none were present before. For example, at $f_a = 10^{12}\,$GeV and $\alpha = 0.2$, regular oscillations with a period of $\simeq 200\,$yr appear at $n_{\max} = 15$ and $18$ but disappear once the $\ket{n66}$ tower is extended to $n = 45$. Instead, at $\alpha = 0.6$ (Fig.~\ref{fig:oscillations}), they are absent for $n_{\max} \le 18$, and only appear once the $\ket{n88}$ tower is included. In some cases, a small amount of additional dissipation is enough to stabilize the evolution (for example, from any environmental effect); in other cases they appear to be robust. Should these oscillations survive environmental effects, they could provide a ``smoking gun'' signature in the form of a time-modulated GW or relativistic-axion spectrum.
\subsubsection{The impact of GW level transitions}
A surprising feature that emerged from this analysis is that GW level transitions can fully prevent the growth of higher-$m$ superradiant states. While we have not analyzed this effect extensively, we have identified a number of points where it takes place. Fig.~\ref{fig:GW_impact} shows an example at $f_a = 10^{18}\,\mathrm{GeV}$: at $\alpha = 0.4$ (left panels), the pre-existing long-lived $\ket{322}$ state fully prevents the growth of $\ket{433}$, which in the absence of GW emission would spin the black hole down to $\tilde a \simeq 0.5$. This can be understood from the growth rate of $\ket{433}$, which is roughly governed by
\begin{eqnarray}
    \dot{\epsilon}_{433} \sim \left[ \gamma^{{\rm SR}}_{433}  - \gamma_{433 \rightarrow 322}^{\rm GW}  \epsilon_{322} \right] \epsilon_{433} \,.
\end{eqnarray}
In the non-relativistic limit the transition rate is $\gamma_{433 \rightarrow 322}^{\rm GW} \sim 5.5 \times 10^{-7} \alpha^{10} \mu$ (App.~\ref{app:gw_matrix}) and the leading-order superradiant rate is $\gamma^{{\rm SR}}_{433} \sim 3.97 \times 10^{-7} \alpha^{16} \mu$, so $\ket{433}$ can only grow once $\epsilon_{322} < \epsilon_{322}^{\rm crit} \sim 0.72\, \alpha^6$. The effect is therefore not specific to self-interacting clouds: in the non-interacting limit the saturated $\ket{322}$ cloud is even larger, and the spin history of Fig.~\ref{fig:GW_impact} (left) indeed coincides with the non-interacting scenario (see Fig.~\ref{fig:spin_fa18}). Using the simulation results to inform the value of $\epsilon_{322}$, we confirm that this state cannot grow for large values of $f_a$ and sufficiently small values of $\alpha$. At $\alpha = 0.6$ (right panels) the condition is only marginally met: the growth of $\ket{433}$ is delayed by about an order of magnitude in time rather than prevented, and the later growth of the $m=4$ levels, which spins the black hole down further when GW emission is switched off, is suppressed. Similar conclusions are expected to hold for higher-order superradiant states.
%

\subsubsection{The zoo of populated modes}
Figure~\ref{fig:tab_tot} is the central result of this work, so it is worth walking through in detail.
The figure corresponds to $M = 10\,M_\odot$ evolved for a Salpeter time; the analogous figures for $10^4$ and $10^8\,M_\odot$ are shown in App.~\ref{app:evolutions}. A state is listed in a cell if its occupation exceeds $\epsilon_{\rm th} = 10^{-20}$ at any point during the evolution, and the colored squares beside it record the decay constants for which this happens. To assemble the basis for a given $(\alpha,f_a)$, one reads down a single column and keeps every state carrying the corresponding color. States colored red are those that would drive the spin-down in the non-interacting limit. The regimes described below are not separated by sharp boundaries, but they show how the complexity of the cloud grows along each axis.
The first trend follows directly from the superradiance condition~\eqref{eq:SR_cond}. As $\alpha$ grows the lower-$m$ levels stop being superradiant, so the effective ``ground state'' of the cloud climbs. The $\ket{211}$ entry survives only in the first columns, and each successive level takes over in turn: $\ket{322}$ persists to $\alpha\simeq0.6$, $\ket{433}$ to $\alpha\simeq1$, $\ket{544}$ to $\alpha\simeq1.3$, and $\ket{655}$ to the largest couplings we consider. The same drift appears within each row, where the dominant multipole shifts upwards: $\ket{411}\to\ket{422}\to\ket{433}$, and $\ket{541}\to\ket{533}\to\ket{544}$.
For $\alpha\lesssim0.15$, and independently of $f_a$, the cloud is well described by the familiar two-level system of $\ket{211}$ and $\ket{322}$. In the column we show, $\alpha=0.05$, $\ket{322}$ never has time to grow efficiently through superradiance and is populated almost entirely by self-interactions. This agrees with previous analyses. It is also the only part of the figure where the ``standard picture'' holds.
By $\alpha\simeq0.2$ additional states appear, and the decay constant takes over as the controlling parameter. At the largest $f_a$ the evolution stays comparatively simple, with only a handful of extra modes. As $f_a$ decreases, cascades of modes form. At $f_a\sim10^{14}\,\mathrm{GeV}$ (teal squares) one already finds sequences such as $\ket{322}\rightarrow \ket{544}\rightarrow\ket{988}$, together with several $\ket{n55}$ and $\ket{n66}$ overtones.
Between $\alpha\simeq0.2$ and $0.8$ the evolution grows steadily more intricate, though the late-time equilibrium is still typically shared among only four or five dominant modes. Here the $\ket{n66}$ and $\ket{n88}$ families reach highest, up to $n=45$, and at the lowest decay constants the $\ket{n\,16\,16}$ tower appears, also reaching high overtones. Above $\alpha\simeq0.8$ the $\ket{n\,10\,10}$ family becomes increasingly relevant, extending to $n=25$ at $\alpha=0.8$ and to $n\simeq45$ at $\alpha \geq 1.3$. For $\alpha \geq 1.3$, the $\ket{n \, 10 \, 10}$  tower also excites the $\ket{n\,20\,20}$ family to very high overtones. These families are excited by double-annihilation processes such as
\begin{equation}
\begin{aligned}
    \ket{433}\times\ket{433} &\to \ket{n66}\times {\rm BH}\,,\\
    \ket{544}\times\ket{544} &\to \ket{n88}\times {\rm BH}\,,\\
    \ket{655}\times\ket{655} &\to \ket{n\,10\,10}\times {\rm BH}\,,\nonumber
\end{aligned}
\end{equation}
whose rates stay surprisingly large even at very high principal quantum number, so that each channel populates a whole tower of overtones rather than a single state.
The richest dynamics occurs around $\alpha\gtrsim0.8$, where the $\ket{n33}$ and $\ket{n44}$ states are still strongly superradiant while self-interactions are already highly efficient, so that many excitation channels coexist. The cloud can then settle into equilibria involving eight or nine simultaneously populated modes, and the $\ket{n\,10\,10}$ family becomes most abundant at $\alpha = 1.3$. The figure records only which states are populated; their relative importance is best read off from the accompanying data \cite{AxionSR:zenodo}. A complementary, more quantitative view is provided by Fig.~\ref{fig:active-states} in App.~\ref{app:evolutions}, which counts the number of simultaneously active superradiant states, $N_{\rm active}$, across the $(\alpha,f_a)$ plane; the growth of this number with $\alpha$ and with decreasing $f_a$ quantifies the trends described above.
\section{Observational probes}
\label{sec:obsv}
\subsection{Extended clouds}
\begin{figure*}
    \centering
    \includegraphics[width=0.98\linewidth]{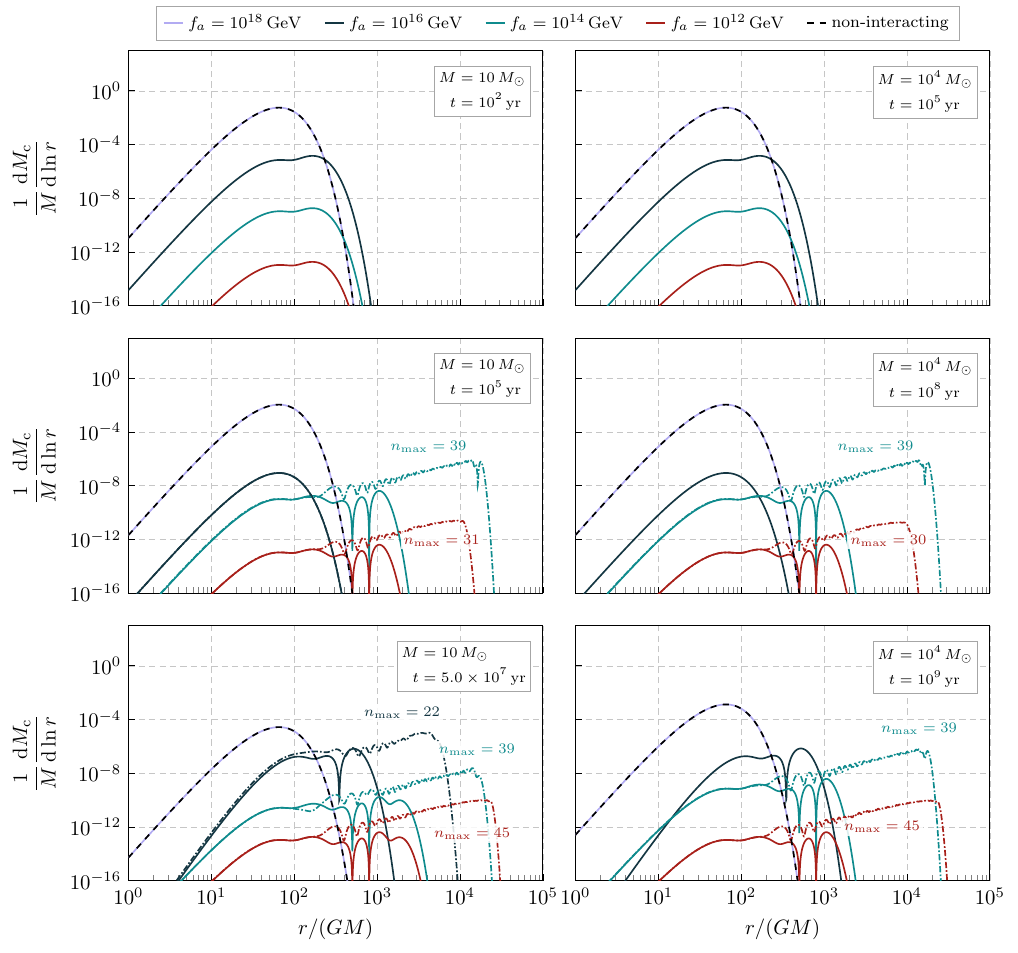}
    \caption{Radial mass profile $(1/M)\,\mathrm{d}M_{\rm c}/\mathrm{d}\ln r$ of the cloud for $M = 10\,M_\odot$ (left) and $10^4\,M_\odot$ (right), $\alpha = 0.4$ and $\tilde a_i = 0.95$, at three times and for different $f_a$, with $n_{\max} = 15$ (solid). Dashed shows the non-interacting limit. In the middle and bottom rows, dash-dotted curves show the $n_{\max} = 45$ runs; each label gives the highest $n$ of the levels that carry at least $10^{-4}$ of the cloud mass at that time. Profiles use hydrogenic wavefunctions, whose relativistic corrections, $\mathcal{O}\left((\alpha/n)^2\right)$, are negligible for the high-$n$ levels.}
    \label{fig:density_extended}
\end{figure*}

Populating higher-order states does not only change the bookkeeping of occupation numbers, but it also changes the shape of the cloud, and with it the observational phenomenology. The characteristic radius of a level scales as $r_n \simeq n^2 a_0$~\eqref{eq:bohr}, so a cloud whose mass is carried by $n\sim13-15$ levels sits a factor $(n/2)^2 \sim 40-60$ ($\sim20-25$) further from the black hole than the free-field $\ket{211}$ ($\ket{322}$) configuration. Figure~\ref{fig:density_extended} shows the resulting density profiles across the range of decay constants. Below we estimate the consequences of an extended cloud for GW emission and for the interaction with a binary companion, leaving a more detailed and quantitative treatment to future work; the spin-down is discussed in Sec.~\ref{sec:spin-down}.

\paragraph{Gravitational-wave emission.}
Consider first the continuous signal that superradiance searches target, in which two axions annihilate into a single graviton \cite{Palomba:2019vxe,Ng:2020jqd,LIGOScientific:2021rnv,Ellis:2026gwt}. The radiation is emitted at $\omega_{\rm GW} \simeq 2\mu$, so its wavelength $\lambda_{\rm GW} = \pi/\mu$ is set by the axion mass alone. The size of the source, on the other hand, grows with the level: the cloud spans
\begin{equation}
    \frac{r_n}{\lambda_{\rm GW}} \simeq \frac{n^2}{\pi\alpha}
\label{eq:GW_coherence}
\end{equation}
wavelengths. This ratio controls the emission. A source smaller than the wavelength radiates coherently; a source much larger than the wavelength does not, since its different parts emit out of phase and largely cancel. Even $\ket{211}$ is on the wrong side of this for any $\alpha \lesssim 1$ -- it covers two wavelengths at $\alpha = 0.6$ and six at $\alpha = 0.2$ -- and the resulting cancellation is why the emission carries the steep $\alpha^{14}$ scaling~\cite{Yoshino:2013ofa,Brito:2015oca,Brito:2017zvb}. The high-lying levels are even worse~\eqref{eq:GW_coherence}: the levels listed in Table~\ref{tab:minimal_sets} span anywhere from tens to $\sim 10^3$ wavelengths, and the cancellation is correspondingly more severe.
For a hydrogenic profile the emitted power scales as\footnote{The emitted graviton has $k = 2\mu \gg 1/r_n$, so the flat-space amplitude is the large-$k$ Fourier transform of the transverse-traceless part of $\partial_i\psi\,\partial_j\psi$ (App.~\ref{app:gw_matrix}). For a hydrogenic level this tail is set by the leading non-analytic term of the profile at the origin, which stems from the exponential $e^{-r/(n a_0)}$ and gives $\psi\,\partial_i\partial_j\psi \propto r^{2\ell-1}/n$; its Fourier transform at $k = 2\mu$ scales as $(\alpha/n)^{2\ell+2}\,n^{-2}$, which with the prefactor of Eq.~\eqref{eq:Gann} yields Eq.~\eqref{eq:gw_scaling}. For $n > \ell + 1$ the Laguerre polynomial introduces additional $n$-dependent prefactors that do not alter the scaling in $\alpha$. The $\ket{322}/\ket{211}$ power ratio obtained this way agrees with the literature value up to an $\mathcal{O}(1)$ factor.}
\begin{equation}
    \left.\frac{\mathrm{d}E_{\rm GW}}{\mathrm{d}t}\right|_{n\ell m}
    \sim \left(\frac{M_{\rm c}}{M}\right)^2 \alpha^2
    \left(\frac{\alpha}{n}\right)^{4\ell+8} .
    \label{eq:gw_scaling}
\end{equation}
The overall dependence $\alpha^{4\ell+10}$ reproduces the known scaling~\cite{Arvanitaki:2010sy,Brito:2017zvb,Siemonsen:2022yyf}, while the power of $\alpha/n$ is \eg $12$ for $\ket{211}$, $16$ for $\ket{322}$, and $56$ for $\ket{13\,12\,12}$. At fixed cloud mass, in other words, the high $n,\ell$ levels that populate the cloud radiate essentially nothing.

In the non-interacting scenario the cloud typically saturates in a single state, so the emission effectively tracks the total cloud mass. With self-interactions, by contrast, much of that mass sits in levels that do not radiate, and the annihilation signal is set by whatever occupation survives in the lowest populated level: $\ket{211}$ for $\alpha \lesssim 0.2$ (for $\tilde a_i = 0.95$ the $m = 1$ levels are stable for $\alpha \gtrsim 0.4$), $\ket{322}$ up to $\alpha \simeq 0.6$, and $\ket{433}$ up to $\alpha \simeq 1$ (Fig.~\ref{fig:tab_tot}). As long as the same level dominates the emission with and without self-interactions, the suppression relative to the non-interacting prediction is simply
\begin{equation}
    \frac{\dot E^{\rm SI}_{\rm GW}}{\dot E^{\rm free}_{\rm GW}}
    \simeq \left(\frac{\epsilon^{\rm SI}_{\rm low}}{\epsilon^{\rm free}_{\rm low}}\right)^2\,,
    \label{eq:gw_suppression}
\end{equation}
which can be read straight off the occupation histories in Figs.~\ref{fig:states_fa1e16} and~\ref{fig:states_fa1e14}. In the case of Fig.~\ref{fig:states_fa1e16} ($\alpha = 0.4$, $f_a = 10^{16}\,\mathrm{GeV}$, $M = 10\,M_\odot$), the $\ket{322}$ occupation is reduced by a factor $\sim 10^{4}-10^{5}$ relative to the free cloud over most of the evolution, suppressing the annihilation power by $\sim 10^{9}$ and the strain by $\epsilon^{\rm SI}_{\rm low}/\epsilon^{\rm free}_{\rm low}$. Emission from $\ket{544}$, whose occupation is comparable to that of $\ket{322}$, is further suppressed by the $n$-dependence in Eq.~\eqref{eq:gw_scaling} by a factor $\sim 10^{-12}$ and plays no role.
The cancellation is severe only because annihilation produces a very short wavelength. Transitions between levels instead emit at the difference of two nearby frequencies, $\Delta\omega \sim \mu\alpha^2/n^3$ for neighboring levels, and the associated wavelength exceeds the size of the cloud by a factor $\sim 2\pi n/\alpha$. These lines are only mildly (power-law) suppressed at high quantum numbers (App.~\ref{app:gw_matrix}), and an extended cloud offers many of them: several levels are populated at once, and each pair can radiate. The spectrum of these systems may therefore look less like a single line at $2\mu$ and more like a set of much lower-frequency lines whose pattern encodes which levels are occupied. For $M = 10\,M_\odot$ and $\alpha = 0.6$, for instance, the annihilation line sits at $3.9\,\mathrm{kHz}$, whereas the $\Delta n = 1$ transitions fall at $48\,\mathrm{Hz}$ ($n = 3 \to 2$), $8\,\mathrm{Hz}$ ($5 \to 4$) and $0.4\,\mathrm{Hz}$ ($13 \to 12$), reaching into the deci-Hz band, in line with the transition signals found in~\cite{Omiya:2024xlz}. Note that $n=45$ transitions will lead to GW line emission at frequencies which are more than an order of magnitude below these quoted ranges. Some preliminary work has been done to investigate the observability using two-state systems (with the $\ket{322} \to \ket{211}$ transition)~\cite{Baryakhtar:2020gao,Collaviti:2024mvh,Guo:2026qfb}; we leave a detailed analysis of the transition spectrum to future work.
\paragraph{Binary companions.}
An extended cloud is considerably more exposed to perturbations from a binary companion. The simplest way to see this is from direct overlap: a companion enters a level once $R \sim r_n \simeq n^2 GM/\alpha^2$. The $n \gtrsim 10$ levels populated across much of the parameter space (Fig.~\ref{fig:tab_tot} and Table~\ref{tab:minimal_sets}) are therefore reached at separations $\sim 50$ times larger than a $\ket{211}$ cloud, and the overtone towers found at the lowest decay constants, which hold a comparable share of the (then very light) cloud mass out to $r \sim 10^4\,GM$ (Fig.~\ref{fig:density_extended}), at separations $\sim 500$ times larger. Since $\Omega \propto R^{-3/2}$, these correspond to orbital frequencies a few hundred and $\sim 10^4$ times smaller, respectively. The high-$n$ structure found here thus moves cloud-binary interactions to a much earlier stage of the inspiral. Once inside, the companion can ionize the weakly bound outer levels: the continuum channel associated with the $g$-th orbital harmonic opens once $g\Omega \gtrsim |E_{b,n}| \simeq \mu\alpha^2/(2n^2)$, \ie~\cite{Baumann:2021fkf,Baumann:2022pkl,Tomaselli:2023ysb}
\begin{equation}
    \frac{R^{(g)}_{\rm ion}}{GM} \simeq
    \alpha^{-2}\left[4 g^2 (1+q)\, n^4\right]^{1/3}\,,
    \label{eq:r_ion}
\end{equation}
with $M_\star = qM$ the companion mass. This threshold grows with $n$, by a factor $(n/2)^{4/3}$. It grows more slowly than the levels themselves, however, $R^{(g)}_{\rm ion}/r_n \simeq [4g^2(1+q)]^{1/3} n^{-2/3}$, so the dominant low-$g$ channels open only once the companion orbits inside the level ($R \lesssim 0.5\,r_n$ for $n \simeq 13-15$ and $R \lesssim 0.2\,r_n$ for $n \approx 45$, taking $g = 2$ and $q \ll 1$).
The companion does not need to enter the cloud to perturb it, however. At larger separations its tidal field drives resonant transitions between bound levels~\cite{Baumann:2018vus,Berti:2019wnn,Baumann:2019ztm,Boskovic:2024fga,Tomaselli:2024bdd,Tomaselli:2024dbw}. For neighboring hydrogenic levels, $\Delta\omega \sim \mu\alpha^2/n^3$, and the resonance condition $|\Delta m|\,\Omega \simeq \Delta\omega$ gives
\begin{equation}
    R_{\rm res} \sim r_n \left[(1+q)(\Delta m)^2\right]^{1/3}\,,
    \label{eq:r_res}
\end{equation}
so the Bohr resonances lie at or beyond the level itself and move outwards with the cloud, $R_{\rm res} \propto n^2$. The same holds for fine ($\Delta\ell \neq 0$) and hyperfine ($\Delta m \neq 0$) resonances. A resonance can transfer the occupation of a level into one that is absorbed by the black hole, depleting the level well before the companion reaches it.
All of these effects scale with the cloud mass, which self-interactions reduce by orders of magnitude, and the relevant timescales depend strongly on the mass of the central black hole. In addition, levels depleted through these interactions change the occupation history and can allow other states to grow again through self-interactions; this interplay, explored so far in a two-mode setting~\cite{Takahashi:2024fyq}, is intricate and left for future work~\cite{Lucas_prep}.
\subsection{Evolution of the black hole spin}
\label{sec:spin-down}
\begin{figure*}
    \centering
    \includegraphics[width=0.98\linewidth]{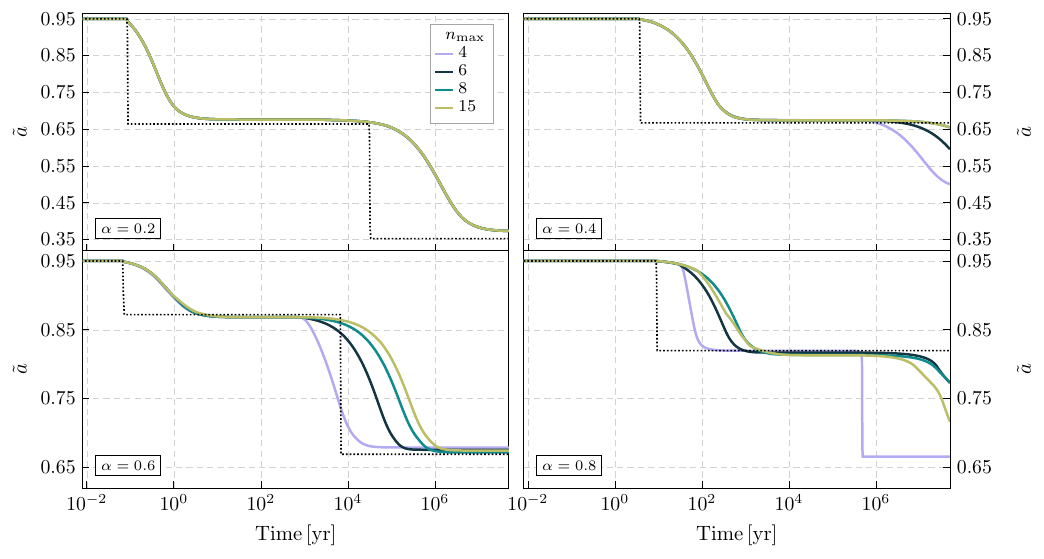}
    \caption{Spin evolution for $M = 10\,M_\odot$ and $f_a = 10^{16}\,$GeV for $n_{\max} = 4$, 6, 8 and 15, at four values of $\alpha$. Dotted lines show the non-interacting limit. The spin-down converges quickly with $n_{\max}$ at small $\alpha$, but requires up to $n_{\max} = 15$ at large $\alpha$.}
    \label{fig:spin-down-convergence}
\end{figure*}
\begin{figure*}
    \centering
    \includegraphics[width=0.98\linewidth]{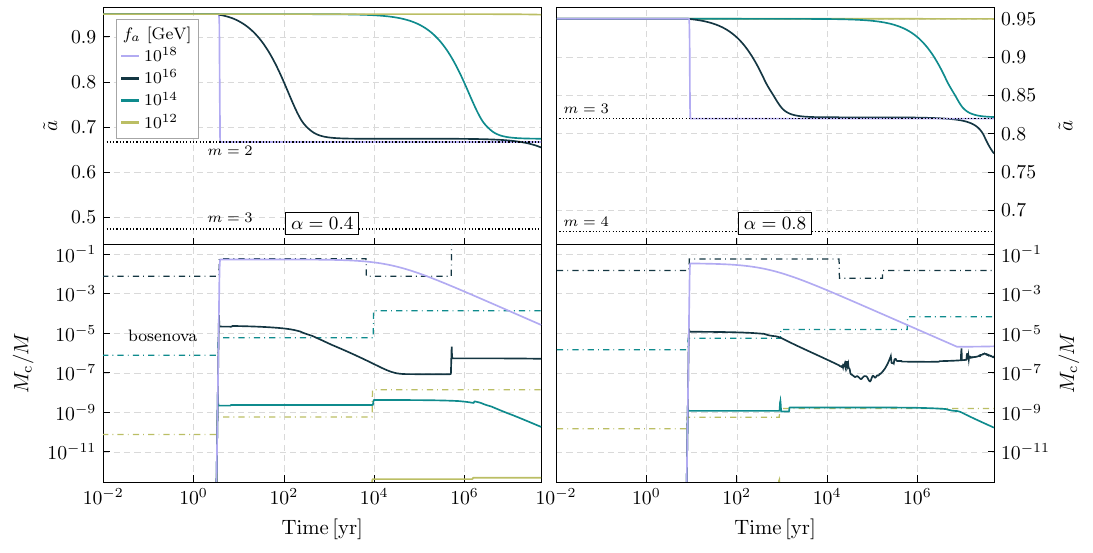}
    \caption{Spin (top) and total cloud mass (bottom) for $M = 10\,M_\odot$, $\tilde a_i = 0.95$ and $n_{\max} = 15$, for different $f_a$ at $\alpha = 0.4$ (left) and $0.8$ (right), averaged over the fast oscillations of the occupations. Dotted lines: saturation spins; dash-dotted lines: bosenova threshold of the dominant level.}
    \label{fig:spin-down-fa}
\end{figure*}

We now turn to a direct observable of superradiance: the spin-down of the black hole. The same convergence analysis can be phrased in terms of $\tilde a(t)$, and once the occupation numbers have converged the spin evolution necessarily has too. The converse does not hold. Across much of the parameter space the spin-down converges considerably faster, because scattering between widely separated levels is strongly suppressed: higher-order states can be populated efficiently through level mixing without ever feeding back on the low-lying modes that drive the spin-down. Raising $n_{\rm max}$ from $9$ to $10$, for instance, may populate $\ket{10\,88}$ in place of $\ket{988}$ and still leave $\ket{211}$ essentially untouched, since neither high-lying state interacts efficiently with it. Reference~\cite{Witte:2024drg} used exactly this argument to revisit spin-down constraints with an $n_{\rm max}=5$ truncation, arguing that it was sufficient to capture the $m=1,2$ spin-down up to $\alpha \approx 0.66$ for stellar-mass black holes.
Figure~\ref{fig:spin-down-convergence} shows the spin evolution for $M = 10\,M_\odot$ and $f_a = 10^{16}\,$GeV. At $\alpha = 0.2$ all truncations give the same final spin to better than $0.03\%$. At $\alpha = 0.4$, however, the low truncations overestimate the late-time spin-down ($n_{\max} = 5$ by $\approx 9\%$), and convergence requires $n_{\max} \simeq 8$. At $\alpha = 0.8$ the approach to the final spin depends non-monotonically on the truncation. With $n_{\max} = 4$ the black hole spins down past the $m = 3$ saturation spin within $10^5\,$yr, reaching $\tilde a \simeq 0.68$, the $m=4$ saturation spin. Truncations with $5 \leq n_{\max} \leq 8$ instead linger above the $m = 3$ saturation spin for more than $10^6\,$yr. For $n_{\max} = 15$, the black hole reaches the $m = 3$ saturation spin ($\tilde a \approx 0.82$) within $10^4\,$yr, before slowly declining further as the $m = 4$ levels grow. Across the full grid, such deviations appear mostly for $\alpha \gtrsim 0.8$, in agreement with Ref. \cite{Witte:2024drg}, with $\alpha = 0.4$ at $f_a = 10^{16}\,$GeV as the main exception. For supermassive black holes, the shorter rescaled evolution keeps the error of an $n_{\max} = 5$ truncation below $\approx 1.3\%$ everywhere. The saturation plateaus themselves are thus robust, even where the path to them is not: our results confirm existing bounds rather than weaken them, and extend their validity to larger $\alpha$.
Finally, Figure~\ref{fig:spin-down-fa} shows the spin and the total cloud mass for different values of $f_a$. Lowering $f_a$ strengthens every scattering channel, so the cloud settles into equilibria at lower occupation: the total cloud mass drops by many orders of magnitude as $f_a$ decreases, and sits far below the bosenova threshold of the dominant level (dash-dotted lines). The spin-down is correspondingly delayed but still reaches the same saturation spin (dotted lines) within a Salpeter time.
\section{Conclusions}\label{sec:conclusions}
In this work we have characterized the growth and evolution of an axion cloud around a rotating black hole across the $(\alpha, f_a)$ parameter space, for three benchmark black hole masses, with one question in mind: what is the minimal set of quasi-bound states required to trace the superradiant evolution on astrophysically relevant timescales? Any prediction of the observable consequences of axion superradiance rests on the answer.
The short answer is: many more than have been previously assumed. We find that a stable evolution requires 203 different states for $M = 10\,M_\odot$ over a Salpeter time, and 189 and 29 states for $M = 10^4$ and $10^8\,M_\odot$ over a Gyr, extending to principal quantum numbers $n \gtrsim 45$. These states are tabulated across the parameter space in Figure~\ref{fig:tab_tot} and Table~\ref{tab:minimal_sets}. We have restricted our analysis to an initial spin of $\tilde{a} = 0.95$; for a maximal initial spin even more levels could appear, and the analysis would have to be extended to larger values of $\alpha$, where yet higher-$n$ states are likely to appear. The superradiant and scattering rates, together with the methods used to obtain them, are publicly available \cite{AxionSR:github}; in the future we hope to extend this analysis to any black hole, of any mass and spin, formed at any point in cosmic history, and resolve the subset of non-convergent cases where states with $n > 45$ appear with large occupation.
Three points follow from this exercise. First, the spin-down is far more robust than the occupation numbers. It is controlled by the low-lying, most unstable levels, and scattering between widely separated states is weak, so the high-$n$ structure that dominates the cloud's mass leaves the spin evolution largely untouched. The truncation of Ref.~\cite{Witte:2024drg} is confirmed up to $\alpha \approx 0.6$; beyond that -- and most conspicuously near $\alpha\approx0.8$, where the approach to saturation depends non-monotonically on $n_{\rm max}$ -- the full state space is needed, but the final saturation spin is unchanged. We therefore confirm the robustness of existing spin-down bounds, while our results will allow these constraints to be extended to a wider range.
Second, because levels with high $n$ are populated, the cloud can be much more extended than usually assumed. These levels span so many gravitational wavelengths that their annihilation emission cancels almost completely [Eq.~\eqref{eq:gw_scaling}], so the monochromatic signal targeted by continuous-wave searches is set not by the total cloud mass but by whatever occupation survives in the lowest populated level (a quantity that is directly provided in our evolutionary diagrams). A binary companion, on the other hand, encounters such a cloud far earlier and at much lower frequency than is often assumed, suggesting astrophysical environments may play a role in curbing the growth of weakly bound states. In both cases, predictions built on a single saturated state will need revisiting.
Third, and most importantly, there are only a few corners of parameter space in which the problem becomes simple. The coupling enters only through the factor $(M_{\rm Pl}/f_a)^4$ multiplying every scattering rate, so $f_a$ alone controls the system (apart from the existence of an absolute time scale, which for fixed $\alpha$ is set by the black hole mass) -- and no value of it makes the system easy. Lower it, and the cloud settles almost at once into a low-occupation equilibrium shared among many states, so that the black hole never spins down appreciably within a Salpeter time. Raise it to the GUT scale, or even higher, and the free-field limit still does not fully return. A single saturated state or a two-level system forms only for $\alpha\lesssim0.15$, or transiently at early times. Self-interactions are thus not an optional refinement to axion superradiance: any axion with a conventional potential has them, and Bose enhancement ensures that they play a central role in the dynamics, however weak the coupling. Moreover, the complex (and occasionally highly oscillatory) dynamics makes evolving these systems extremely difficult in some regions of parameter space, which poses a challenge for future parameter inference.
Several directions follow naturally. The most immediate is to revisit spin-down constraints with the converged state space, using both stellar-mass and supermassive black hole spin measurements. Beyond that, the population histories computed here are the input required for any prediction of the GW phenomenology of these systems, whether that is monochromatic emission, level transitions, or the response of the cloud to a binary companion. Finally, constraints that assume the presence of fully grown, saturated clouds need to be reassessed, as the characteristic densities have likely been overestimated and the characteristic radii underestimated. We leave these directions for future work.

\vspace{0.5cm}
\noindent {\bf Acknowledgments.} We thank Lucas Gathercole for cross-checking the scattering rates and identifying the super-resonant channels. SJW and TS acknowledge support from a Royal Society University Research Fellowship (URF-R1-231065). This work is also supported by the Deutsche Forschungsgemeinschaft under Germany’s Excellence Strategy -- EXC 2121 ``Quantum Universe'' -- 390833306. TS acknowledges support from the Friends of the Institute for Advanced Study Fund and the Simons Foundation through the Black Holes and Strong Gravity program through Award No.~SFI-MPS-BH-00012593-10. EC acknowledges financial support provided under the European Union’s H2020 ERC Advanced Grant ``Black holes: gravitational engines of discovery'' grant agreement no. Gravitas–101052587, and from the Villum Investigator program supported by the VILLUM Foundation (grant no. VIL37766) and the DNRF Chair program (grant no. DNRF162) by the Danish National Research Foundation. The Tycho supercomputer hosted at the SCIENCE HPC center at the University of Copenhagen was used for supporting this work.

\appendix

\clearpage
\section{The superradiant spectrum}
\label{app:superradiance}
In this appendix, we provide the technical and numerical details on the two techniques adopted in this work to compute the superradiant spectrum. A large part of the spectrum is well-known and can be obtained using the standard Leaver continued-fraction method \cite{Leaver:1985ax, Dolan:2007mj}. This method, however, becomes unstable whenever the imaginary part of the eigenfrequency is small, as well as for high overtones. We overcome this obstacle by adopting a more precise Chebyshev-based algorithm, following the method introduced in \cite{Baumann:2019eav}. The two approaches are complementary: the Leaver method is computationally cheap and resolves the bulk of the spectrum efficiently, while the Chebyshev method, though more expensive, remains stable precisely where the former fails. In both cases, modes are computed with 77-digit precision using the \texttt{BigFloat} arithmetic of \texttt{Julia}. The corresponding code and data are made publicly available on GitHub~\cite{AxionSR:github}.
\subsection{Leaver method}
The Leaver method recasts the radial equation \eqref{eq:radial_angular} as an algebraic problem for the eigenfrequency, expanding the solution in a power series whose prefactor builds in the purely ingoing boundary condition at the horizon by construction \cite{Leaver:1985ax}. Explicitly, the radial function is written as
\begin{equation}
\label{eq:RLeaver}
R(r) = (r-r_+)^{-i\sigma} (r-r_-)^{i\sigma+\chi-1} e^{qr} \sum_{n=0}^{\infty} a_n \left( \frac{r-r_+}{r-r_-} \right)^n\,,
\end{equation}
where the parameters are defined as
\begin{align}\label{eq:param_Leaver}
\sigma &= \frac{2 r_+ GM (\omega - m \Omega_H)}{r_+ - r_-} \,, \\ q &= \pm \sqrt{\mu^2 - \omega^2} \,, \\ \chi &= \frac{GM (\mu^2 - 2 \omega^2)}{q} \,.
\end{align}
The sign of $q$ determines the asymptotic behavior of the solution: $\mathrm{Re}(q) < 0$ corresponds to quasi-bound states, which decay at spatial infinity, whereas $\mathrm{Re}(q) > 0$ corresponds to quasi-normal modes. In this work we focus on the quasi-bound spectrum.
Substituting the ansatz into the radial equation yields a three-term recurrence relation for the expansion coefficients,
\begin{align}
\alpha_0 a_1 + \beta_0 a_0 &= 0 \, , \\
\alpha_n a_{n+1} + \beta_n a_n + \gamma_n a_{n-1} &= 0 \quad (n>0) \, ,
\end{align}
with
\begin{align}
\alpha_n &= n^2 + (c_0 + 1)n + c_0 \, , \\
\beta_n  &= -2n^2 + (c_1 + 2)n + c_3 \, , \\
\gamma_n &= n^2 + (c_2 - 3)n + c_4 \, ,
\end{align}
where the coefficients $c_i$ depend on the black hole and field parameters (see, \eg \cite{Dolan:2007mj, Witte:2024drg} for their explicit form).
The eigenfrequencies are then determined by expressing the ratio of successive coefficients $a_n$ in terms of a continued fraction (using $a_1/a_0=-\beta_0/\alpha_0$):
\begin{equation}
\frac{\beta_0}{\alpha_0} - \frac{\gamma_1}{\beta_1 - \frac{\alpha_1 \gamma_2}{\beta_2 - \frac{\alpha_2 \gamma_3}{\beta_3 - \cdots}}} = 0\,.
\end{equation}
In practice, the continued fraction is truncated at a sufficiently large order $N$, and the roots of this equation are found numerically with a minimization algorithm. Convergence typically requires $N \sim 10^3-10^4$. Once the eigenfrequencies are obtained, the coefficients $a_n$ can be reconstructed recursively, and with that the full radial wavefunction~\eqref{eq:RLeaver}. 
\subsection{Chebyshev method}
\begin{figure*}
\centering
\includegraphics[width=0.98\linewidth]{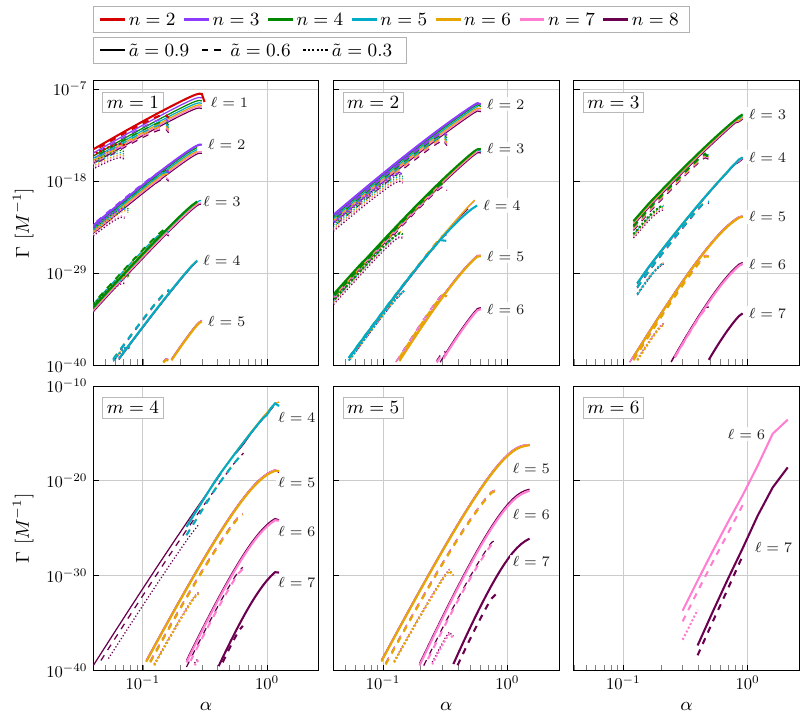}
\caption{Superradiant growth rates $\Gamma$ computed with the Chebyshev method for selected modes up to $n = 8$ (colors) and spins $\tilde a = 0.9$, $0.6$ and $0.3$ (line styles), for $m = 1$--$6$ (panels).}
\label{fig:SRrates}
\end{figure*}

The Chebyshev method proceeds by solving the Klein-Gordon equation directly as a boundary-value problem \cite{Baumann:2019eav}. Rather than expanding the scalar field in spheroidal harmonics, we decompose it as
\begin{equation}
\Phi(t,r,\theta,\phi) = F(r) \sum_{\ell} B_\ell(\zeta(r)) Y_{\ell m}(\theta,\phi) e^{-i\omega t}\, ,
\end{equation}
where the prefactor $F(r)$ explicitly extracts the singular asymptotic behavior of the solution. Explicitly,
\begin{equation}
F(r)= \left( \frac{r-r_+}{r-r_-} \right)^{iP_+} (r-r_-)^{-1+\nu-\frac{2\alpha^2}{\nu}} e^{-\frac{\mu\alpha}{\nu}(r-r_+)} \, ,
\end{equation}
so that the remaining functions $B_\ell$ are regular both at the horizon and at spatial infinity. Here
\begin{equation}
P_\pm \equiv \frac{ ma-2GMr_\pm\omega }{ r_+-r_-}\,,
\end{equation}
while the parameter $\nu$ is related to the frequency through
\begin{equation}
\omega = \mu \sqrt{ 1-\frac{\alpha^2}{\nu^2} }\, .
\end{equation}
Since the $B_\ell$ are now smooth from the horizon to infinity, we solve for them with a pseudospectral method based on Chebyshev interpolation. Instead of expanding the solution in terms of basis coefficients, which are numerically unstable at large truncation orders, we represent the functions directly through their values at the Chebyshev nodes
\begin{equation}
\zeta_n = \cos\left( \frac{\pi(2n+1)}{2(N+1)} \right)\,, \qquad n=0,\dots,N \, ,
\end{equation}
corresponding to the extrema of the Chebyshev polynomials. The scalar profile is then reconstructed through Lagrange cardinal polynomials associated with these nodes,
\begin{equation}
B_N(\zeta) = \sum_{k=0}^{N} B(\zeta_k) p_k(\zeta)\, ,
\end{equation}
where the cardinal functions satisfy $p_k(\zeta_n)=\delta_{kn}$. This representation is numerically more robust than directly evolving the Chebyshev coefficients, particularly when dealing with extremely small imaginary parts of the spectrum. The derivatives are computed through barycentric spectral differentiation matrices, which guarantee exponential convergence for smooth functions while maintaining high numerical stability.
To implement the interpolation, the semi-infinite radial domain must first be compactified into a finite interval $\zeta\in[-1,1]$ through a mapping $r\to\zeta(r)$. The choice of the map is crucial for the convergence properties of the interpolation. In particular, the truncation error depends sensitively on how the interpolation points sample the near-horizon and asymptotic regions. Simple compactifications tend to cluster points close to the horizon, leaving the far-region poorly resolved. This becomes problematic for high overtones, whose wavefunctions feature a non-trivial structure at large radii. For this reason, we adopt the mapping
\begin{equation}
\zeta(r)=\frac{r-\sqrt{4r_+(r-r_-)+r_-^2}}{r-r_-}\, ,
\label{eq:mapping}
\end{equation}
which distributes the interpolation points more evenly across the near, intermediate and asymptotic regions of the spacetime. In particular, while simpler mappings place the Bohr radius at $\zeta\simeq1-\mathcal{O}(\alpha^2)$, which then requires a large number of modes to resolve the far-region, the choice above maps it to $\zeta\simeq1-\mathcal{O}(\alpha)$, dramatically improving convergence \cite{Baumann:2019eav}.
In terms of the coordinate $\zeta$, the radial equation becomes a system of coupled differential equations for the functions $B_\ell(\zeta)$,
\begin{equation}
\begin{aligned}
&\left[ \frac{\mathrm{d}^2}{\mathrm{d}\zeta^2} + C^{(1)}_\ell(\zeta)\frac{\mathrm{d}}{\mathrm{d}\zeta} + C^{(2)}_\ell(\zeta) \right] B_\ell(\zeta) \\&+ \sum_{\ell'} C^{(3)}_{\ell,\ell'}(\zeta)\, B_{\ell'}(\zeta) = 0 \, ,
\end{aligned}
\end{equation}
where the coefficients $C^{(i)}$ depend on the background geometry and Clebsch-Gordan coefficients \cite{Lingetti:2022psy}. The angular decomposition induces couplings between neighboring multipoles, so that each $\ell$ mode interacts only with a finite subset of nearby angular harmonics.
In practice, the infinite tower of coupled equations is truncated at some value $\ell\leq \ell_{\max}$ sufficiently large to guarantee convergence. After discretization, the system reduces to a matrix equation
\begin{equation}
\mathbf{A}(\omega)\cdot \mathbf{B}=0\,,
\end{equation}
where $\mathbf{B}$ is an $(\ell_{\max}+1)(N+1)$-dimensional vector containing the values of the functions $B_\ell$ at the interpolation points. Non-trivial solutions exist only when $\det \mathbf{A}(\omega)=0$, yielding a nonlinear eigenvalue problem for the complex frequency $\omega$.
We solve this system using a nonlinear inverse iteration algorithm. Although this approach is computationally more expensive than the continued-fraction method, it is also much more stable when the imaginary part of the frequency is extremely small, and it allows us to accurately compute high overtones. Convergence is typically achieved for $\sim 90-120$ interpolation points and an angular truncation $\ell_{\max}\sim 4-6$.
While this method determines the eigenvalues accurately, reconstructing the corresponding eigenfunctions is considerably harder. The node distribution of the mapping~\eqref{eq:mapping}, while sufficient to determine the eigenvalue, still leaves the far region too poorly resolved to reconstruct the eigenfunction accurately. Recovering the wavefunctions directly from the interpolation would demand a prohibitively large number of nodes, and in practice the reconstruction is unreliable, particularly for high overtones.
To overcome this limitation, we exploit the fact that quasi-bound states in Kerr admit an analytic representation once the eigenfrequency is known. Starting from the separable ansatz of Eq.~\eqref{eq:ansatz}, the Klein-Gordon equation splits into angular and radial parts, the latter taking the form of a confluent Heun equation. Introducing the variable
\begin{equation}
    z \equiv -\frac{r-r_+}{r_+-r_-}\, ,
\end{equation}
and factoring out the asymptotic behavior,
\begin{equation}
    R(r)= e^{-ik(r-r_+)} z^{iP_+}(z-1)^{-iP_-} H(z)\,,
\end{equation}
where $k = iq$~\eqref{eq:param_Leaver} and the radial equation reduces to
\begin{equation}
\begin{aligned}
&\frac{\mathrm{d}^2H}{\mathrm{d}z^2} +\left(\alpha_H+\frac{1+\beta_H}{z} +\frac{1+\gamma_H}{z-1}\right)\frac{\mathrm{d}H}{\mathrm{d}z} \\&+\left(\frac{\mu_H}{z}+\frac{\nu_H}{z-1}\right)H=0\, ,
\end{aligned}
\end{equation}
whose solution is the confluent Heun function,
\begin{equation}
    H(z)\equiv\mathrm{HeunC}(\alpha_H,\beta_H,\gamma_H,\delta_H,\eta_H;z)\, .
\end{equation}
Here $\alpha_H,\beta_H,\gamma_H,\delta_H$ and $\eta_H$ are frequency-dependent parameters, with the coefficients $\mu_H,\nu_H$ of the radial equation mapping onto $\delta_H,\eta_H$ through the standard confluent-Heun relations; explicit expressions are given in \cite{Baumann:2021fkf}, App. E.
Therefore, once the eigenfrequency $\omega_{n\ell m}$ is determined numerically, the corresponding radial wavefunction can be reconstructed as
\begin{equation}
    R_{n\ell m}(r)=\mathcal{N}e^{-ik(r-r_+)}z^{iP_+}(z-1)^{-iP_-}H(z)\, ,
\end{equation}
where $\mathcal{N}$ is a normalization constant. We have verified that this procedure yields accurate wavefunctions across all regimes, cross checking that it reproduces the Leaver solutions where those are reliable, as well as the analytical hydrogenic wavefunctions in the small-$\alpha$ limit, even for extremely high overtones. We show the spectrum as obtained with the Chebyshev interpolation for some selected modes up to $n=8$ in Fig.~\ref{fig:SRrates}. 
\section{The non-relativistic limit}
\label{app:NonRel}
In this appendix we review the formalism describing self-interacting axion clouds in the non-relativistic limit. The material is not new; we include it for completeness and refer to \cite{Baryakhtar:2020gao} for a more detailed discussion.
We first describe the non-interacting scenario in this regime. In the non-relativistic limit $\alpha\ll1$, the radial Klein-Gordon equation in Kerr spacetime reduces to a Schr\"odinger-like form:
\begin{equation}
\begin{aligned}
&\biggl[ -\frac{1}{r^2}\frac{\mathrm{d}}{\mathrm{d}r}\left(r^2\frac{\mathrm{d}}{\mathrm{d}r}\right) -\frac{2\alpha\mu}{r} \\&+\frac{\ell(\ell+1)}{r^2} +(\mu^2-\omega^2) \biggr] R_{\ell m}=0 \, ,
\label{eq:hydroschroed}
\end{aligned}
\end{equation}
while the angular functions simply reduce to spherical harmonics with $\Lambda_{\ell m}= \ell(\ell+1)$. Note that this equation mirrors that of the hydrogen atom in quantum mechanics, with $\alpha$ playing the role of a ``gravitational fine-structure constant''. At leading-order in this limit, the field equations are insensitive to the presence of the horizon. Eq.~\eqref{eq:hydroschroed} can therefore be solved by imposing regular behavior at the origin and exponential decay at large radii. This discretizes the spectrum, leading to Bohr-like solutions:
\begin{equation}
R^{\rm far}_{n\ell} = C^{\rm far}_0 e^{-r/(na_0)} \left(\frac{2r}{na_0}\right)^\ell L^{2\ell+1}_{n-\ell-1} \left(\frac{2r}{na_0}\right) \, ,
\end{equation}
where we defined the Bohr radius $a_0\equiv (\alpha\mu)^{-1}$. At leading order in the hydrogen limit, the eigenfrequencies are real, and given by:
\begin{equation}
E_{n\ell m} \simeq \mu\Big(1-\frac{\alpha^2}{2n^2}\Big)\,.
\end{equation}
Let us now move to include self-interactions, which, as discussed in the main text, provide an inhomogeneous source term in the Klein-Gordon equation. Proceeding perturbatively in the coupling $\lambda$, one expands the scalar field as $\Phi=\sum_i \lambda^i \Phi^{(i)}$, so that the first-order correction $\Phi^{(1)}$ satisfies an inhomogeneous equation sourced by the zeroth-order bound states. 
In the fully relativistic problem, dissipation through the horizon renders the system non-Hermitian: the quasi-bound states no longer form a complete basis, the standard spectral decomposition breaks down, and solving for $\Phi^{(1)}$ requires a Green's function approach. The hydrogenic limit is far simpler. There the dynamics become insensitive to the horizon, dissipation is negligible, and the problem reduces to a Hermitian Schr\"odinger system. By the spectral theorem, the hydrogenic wavefunctions then form a complete orthonormal basis with a well-defined inner product, exactly as in ordinary quantum mechanics, and the first-order solution can be built directly as a mode expansion. Energy losses through the horizon are dominated by the bound component of the first-order perturbation. Using the completeness property, we can decompose the latter into eigenfunctions of the Schr\"odinger operator:
\begin{equation}
\label{eq:nonrelansatz}
\Phi^{(1)}= \sum_{n\ell m} c_{n\ell m}\psi_{n\ell m} + \sum_{\ell m}\int \mathrm{d}k c_{\ell m}(k)\psi_{k\ell m}\, ,
\end{equation}
where the first term corresponds to the discrete set of hydrogenic bound states, while the second describes continuum scattering states, where $k = \sqrt{\omega^2-\mu^2}$.
The continuum wavefunctions are given by Coulomb scattering states,
\begin{align}
R_{k\ell} =& \frac{2k\, e^{\pi/(2ka_0)} \left|\Gamma\!\left(\ell+1-\frac{i}{ka_0}\right)\right|} {(2\ell+1)!} e^{-ikr} (2kr)^\ell \\&\times \nonumber{}_1F_1\!\left( \ell+1+\frac{i}{ka_0},2\ell+2, 2ikr \right)\, ,
\end{align}
where ${}_1F_1$ denotes the confluent hypergeometric function.
Using the fact that the hydrogenic wavefunctions satisfy
\begin{equation}
\mathcal{O}\psi_{n\ell m} = -\frac{\alpha^2\mu^2}{n^2}\psi_{n\ell m}\,,
\qquad \mathcal{O}\psi_{k\ell m} = k^2\psi_{k\ell m}\, ,
\end{equation}
the equation of motion for $\Phi^{(1)}$ can be written as
\begin{multline}
\sum_{n\ell m} \left( \mu^2-\omega^2-\frac{\alpha^2\mu^2}{n^2}
\right) c_{n\ell m}\psi_{n\ell m} \\ + \sum_{\ell m}\int \mathrm{d}k\, \left( \mu^2-\omega^2+k^2 \right) c_{\ell m}(k)\psi_{k\ell m} = -\frac{1}{6}\left(\Phi^{(0)}\right)^3 \, .
\end{multline}
Projecting this equation onto the bound and continuum basis states, and using the orthogonality of the hydrogenic eigenfunctions, one obtains the corresponding expansion coefficients:
\begin{align}
c_{n\ell m} &= \frac{\xi}{ \mu^2-\omega^2-\frac{\alpha^2\mu^2}{n^2} } \int \mathrm{d}^3r \, \psi_1\psi_2\psi_3^*\psi_{n\ell m}^* \, ,\\ c_{\ell m}(k) &= \frac{\xi}{ 2\pi(\mu^2-\omega^2+k^2) } \int \mathrm{d}^3r \, \psi_1\psi_2\psi_3^*\psi_{k\ell m}^* \, ,
\end{align}
where $\xi=1/2$ for identical initial states and $\xi=1$ otherwise.
The corresponding energy fluxes can then be computed from the stress-energy tensor. Near the horizon, the dominant contribution arises from the $\ell=0$ bound component of $\Phi^{(1)}$, leading to an energy flux
\begin{equation}
\frac{\mathrm{d}E}{\mathrm{d}t}\Big|_{\rm BH} \simeq 4\alpha^2 \left(1+\sqrt{1-\tilde a^2}\right) \lambda^2 \left|R^{(1)}(r_+)\right|^2 \, ,
\end{equation}
where $R^{(1)}$ denotes the radial wavefunction of the induced bound oscillation.
Radiative losses to infinity arise from the continuum component of the first-order perturbation. Far from the black hole, the continuum wavefunctions behave as outgoing spherical waves, $\psi_{k\ell m}\sim e^{ikr}/r$. The coefficient $c_{\ell m}(k)$ contains a pole at
\begin{equation}
k=\sqrt{\omega^2-\mu^2}\, ,
\end{equation}
corresponding to the on-shell emitted non-relativistic mode. Evaluating the momentum integral in \eqref{eq:nonrelansatz} through the residue theorem isolates the outgoing contribution to the wavefunction, yielding
\begin{equation}
\begin{aligned}
\psi_{\rm nre}(r\to\infty) &\sim i\frac{\xi}{2k} \\&\times\frac{e^{ikr}}{r} \left[ \int \mathrm{d}r' \, \psi_1(r') \psi_2(r') \psi_3^*(r') \psi_{k\ell m}(r') \right] \, .
\end{aligned}
\end{equation}
The associated energy flux at infinity is then
\begin{equation}
\frac{\mathrm{d}E}{\mathrm{d}t}\Big|_{\infty} \simeq \omega\lambda^2 \frac{\xi^2}{2k} \left| \int \mathrm{d}r \,\psi_1(r)\psi_2(r)\psi_3^*(r)\psi_{k\ell m}(r) \right|^2 \, ,
\end{equation}
where $\psi_i$ denote the bound states entering the quartic interaction. With the fluxes at hand, one can then define scattering rates following the same procedure as in the main text.
A comparison between relativistic and non-relativistic scattering rates was presented in \cite{Witte:2024drg}. While the two approaches are in good agreement in the hydrogenic regime $\alpha\ll1$, relativistic corrections become increasingly important as $\alpha\to\mathcal{O}(1)$. In this regime, the rates -- particularly those associated with dissipation through the horizon -- can receive corrections as large as $\mathcal{O}(10-100)$.
\section{Gravitational-Wave Emission}
\label{app:gw_matrix}
In this appendix we sketch how the GW rate coefficients that
enter Eq.~\eqref{eq:general_rate}, $\gamma_{a \times b}^{{\rm GW}}$ and $\gamma_{a\rightarrow b}^{\rm GW}$, are obtained. These describe GW emission from annihilations and level transitions, respectively.  
\subsection{Annihilations}
For annihilations, the graviton momentum is given by $|\mathbf k|=\omega_a+\omega_b\simeq 2\mu$,
which corresponds to a wavelength that is small compared to the size of the cloud, and thus the usual multipole expansion does not apply. Instead, we use the flat-space GW power formula for a harmonic source $T_{ij}(\mathbf x)e^{-i\omega t}+\mathrm{c.c.}$,
\begin{equation}
\frac{\mathrm{d}P}{\mathrm{d}\Omega}
=\frac{G\omega^2}{\pi}
\Lambda_{ij,kl}(\hat{\mathbf k})
\tilde T_{ij}^*(\mathbf k)\tilde T_{kl}(\mathbf k)\,,
\label{eq:dpower}
\end{equation}
where
\begin{equation}
\tilde T_{ij}(\mathbf k)
=\int \mathrm{d}^3xT_{ij}(\mathbf x)e^{-i\mathbf k\cdot\mathbf x}\,,
\end{equation}
and $\Lambda_{ij,kl}$ is the transverse-traceless (TT) projector.
Only the gradient term $\partial_i\phi\,\partial_j\phi$ contributes after the TT projection. For annihilations between levels $a$ and $b$, the relevant spatial source is
\begin{equation}
\bar S_{ij}=
\begin{cases}
\partial_i\psi_a,\partial_j\psi_b+
\partial_j\psi_a,\partial_i\psi_b, & a\neq b,\\
\partial_i\psi_a,\partial_j\psi_a, & a=b\,.
\end{cases}
\end{equation}
The angular dependence of the emitted power is then captured by
\begin{equation}
F_{ab}(\kappa)=
\int\mathrm{d}\Omega_{\hat{\mathbf k}},
\Lambda_{ij,kl}(\hat{\mathbf k})
\tilde{\bar S}{ij}^*(\bar{\mathbf k})
\tilde{\bar S}{kl}(\bar{\mathbf k})\,,
\end{equation}
where the Fourier transform is evaluated in units of the Bohr radius, with $\kappa=ka_0=2/\alpha$.
Substituting into Eq.~\eqref{eq:dpower}, integrating over the emission directions, and relating the radiated power to the depletion of the level occupations gives
\begin{equation}
\gamma_{a\times b}^{\rm GW}(\alpha)
=\frac{\alpha^6}{2\pi}F_{ab}(2/\alpha)\,.
\label{eq:Gann}
\end{equation}
\subsection{Level Transitions}
For level transitions, the GW frequency is set by the energy splitting $\Delta\omega=\omega_a-\omega_b\ll\mu$. The wavelength is much larger than the cloud, so the multipole expansion applies. The radiated power is given by~\cite{Thorne:1980ru}
\begin{equation}
P=2G\sum_{L\geq 2}\Delta\omega^{2L+2}
\left(
c_L\sum_m|q_{Lm}|^2+
s_L\sum_m|J_{Lm}|^2
\right)\,,
\label{eq}
\end{equation}
where $q_{Lm}$ and $J_{Lm}$ are the mass and current multipole moments,
\begin{eqnarray}
q_{Lm}&=&\int\mathrm{d}^3x \, \rho \, r^L \, Y_{Lm}^*\,,\\ 
J_{Lm}&=&\frac{1}{L}\int\mathrm{d}^3x \, 
(\mathbf x\times\mathbf p)\cdot\nabla(r^LY_{Lm}^*)\,,
\label{eq:mass_currentGW}
\end{eqnarray}
with
\begin{eqnarray}
c_L&=&\frac{4\pi(L+1)(L+2)}
{(L-1)L[(2L+1)!!]^2}\,, \\ 
s_L&=&\frac{16\pi L(L+2)}
{(L-1)(L+1)[(2L+1)!!]^2}\,.
\end{eqnarray}
The contributions to the energy and momentum densities from the level-transition are given by
\begin{eqnarray}
\rho&=&\mu\sqrt{N_aN_b}\psi_a\psi_b^*\,, \\
\mathbf p&=&\frac{i}{2}\sqrt{N_aN_b}
\left(\psi_a\nabla\psi_b^* -\psi_b^*\nabla\psi_a\right) \,.
\end{eqnarray}
The angular integrals fix $m=m_a-m_b$ and restrict the allowed multipoles to
\begin{equation}
\max(2,|m_a-m_b|,|\ell_a-\ell_b|)
\leq L\leq\ell_a+\ell_b\,.
\end{equation}
Mass moments contribute only when $(-1)^{\ell_a+\ell_b+L}=+1$, while current moments require $(-1)^{\ell_a+\ell_b+L}=-1$.

Evaluating the moments with hydrogenic wavefunctions and expressing the power in terms of the dimensionless transition rate gives
\begin{eqnarray}
\gamma_{a\to b}^{\rm GW}
={}&\sum_{L,\mathrm{mass}}
K_L^{\rm mass} \, \alpha^{2-2L}\delta^{2L+1}\nonumber
\\ &+\sum_{L,\mathrm{current}}
K_L^{\rm curr} \, \alpha^{4-2L}\delta^{2L+1},
\label{eq:gw_transition}
\end{eqnarray}
where $\delta=\Delta\omega/\mu$ and
\begin{eqnarray}
K_L^{\rm mass}&=&2c_L\bar Q_L^2 \\ 
K_L^{\rm curr}&=&2s_L\bar J_L^2 \, .
\end{eqnarray}
The coefficients depend only on the radial and angular overlaps of the initial and final states. For the mass moments, one finds
\begin{equation}
\bar Q_L=
\int_0^\infty R_aR_b\bar r^{L+2}\mathrm{d}\bar r
\int\mathrm{d}\Omega
Y_{\ell_am_a}Y_{\ell_bm_b}^*Y_{Lm}^* \, .
\end{equation}
For the current moments, one can write $Y_{\ell m}=\Theta_{\ell m}(\theta)e^{im\varphi}$, and the resulting coefficients are then given by
\begin{eqnarray}
\bar J_L=\frac{\pi}{L}
\int_0^\infty R_aR_b\bar r^{L+1}\mathrm{d}\bar r \times \, \hspace{2cm} \\ \int_0^\pi\mathrm{d}\theta
\Big[
m(\Theta_a\Theta_b'-\Theta_b\Theta_a')\Theta_L -(m_a+m_b)\Theta_a\Theta_b\Theta_L'
\Big] \, . \nonumber
\end{eqnarray}
Here $R_i$ are the hydrogenic radial functions and $\bar r=r/a_0$. For $\ell_a+\ell_b$ odd, the leading mass contribution is the octupole, which enters at the same order in $\alpha$ as the current quadrupole. Both contributions must therefore be retained.
We compute the annihilation and transition rates in the non-relativistic limit, and include them in the evolution equations. To verify that relativistic corrections do not significantly alter these quantities, we compared with the relativistic rates obtained in~\cite{Omiya:2024xlz}.
\begin{figure}
    \centering    
    \includegraphics[width=0.98\linewidth]{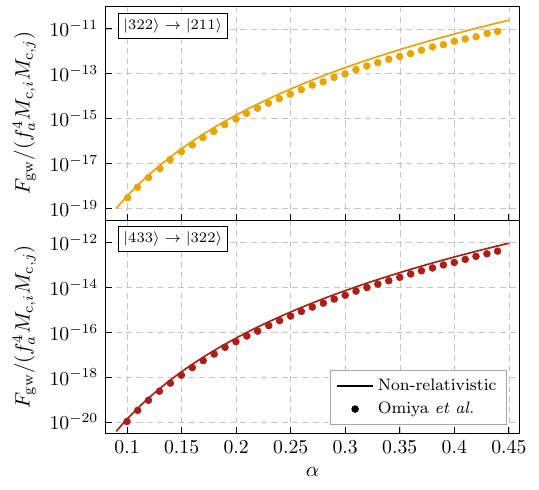}
    \caption{Gravitational-wave energy flux of the transitions $\ket{322}\to\ket{211}$ (top) and $\ket{433}\to\ket{322}$ (bottom) in the non-relativistic approximation adopted in this work (solid), compared with the relativistic results of Omiya \emph{et al.}~\cite{Omiya:2024xlz} (dots, multiplied by 2 to include both helicities). The non-relativistic treatment reproduces the overall magnitude of the relativistic fluxes even at large values of $\alpha$.}
\label{fig:comparison_transitions}
\end{figure}
As shown in Fig.~\ref{fig:comparison_transitions}, the discrepancy between the relativistic and non-relativistic rates remains moderate even at large $\alpha$, indicating that, unlike self-interaction rates, GW transition rates can be reliably estimated within the non-relativistic approximation.

\section{Additional evolutions}
\label{app:evolutions}
In this appendix we collect additional examples of the cloud and spin evolution, complementing Figs.~\ref{fig:states_fa1e16}, \ref{fig:states_fa1e14} and~\ref{fig:spin-down-convergence} in the main text. The full set of evolutions for all three black hole masses, together with the scripts used to produce them, is available in our GitHub repository and Zenodo folder~\cite{AxionSR:github,AxionSR:zenodo}.
\subsection{Cloud evolution}  
\begin{figure}
    \centering    
    \includegraphics[width=0.98\linewidth]{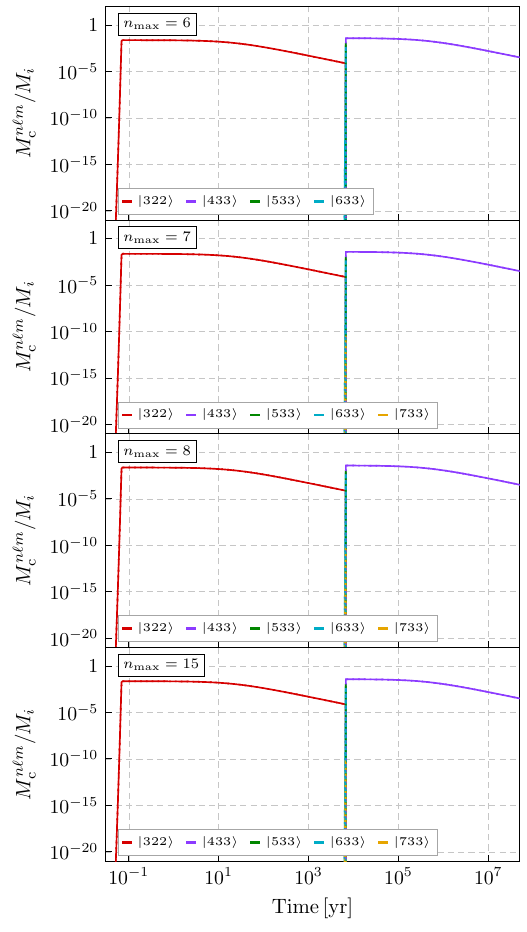}
    \caption{Same as Fig.~\ref{fig:states_fa1e16}, for $f_a = 10^{18}\,$GeV and $\alpha = 0.6$.}
    \label{fig:states_fa18}
\end{figure}

\begin{figure}
    \centering    
    \includegraphics[width=0.98\linewidth]{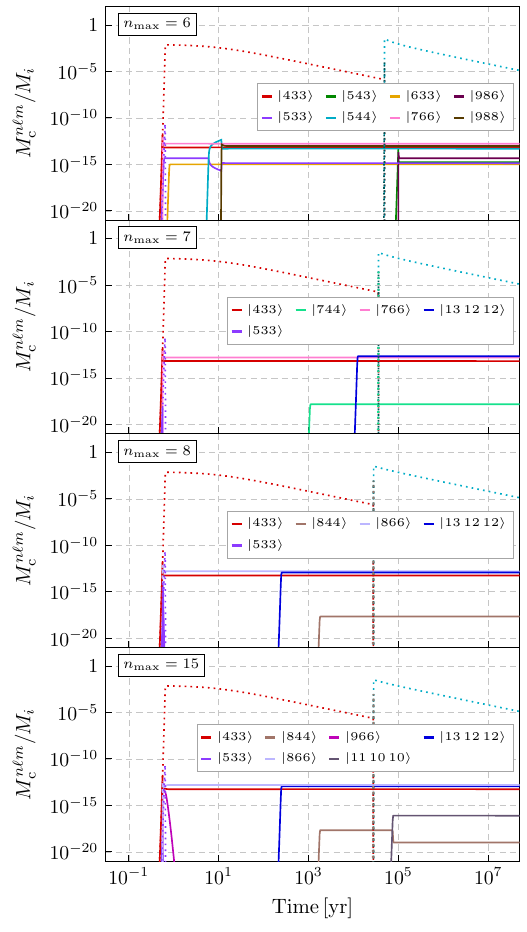}
    \caption{Same as Fig.~\ref{fig:states_fa1e16}, for $f_a = 10^{12}\,$GeV and $\alpha = 1$.}
    \label{fig:states_fa12}
\end{figure}

Figures~\ref{fig:states_fa18} and~\ref{fig:states_fa12} show the evolution of the populated levels at the two ends of the range of decay constants. At $f_a = 10^{18}\,$GeV and $\alpha = 0.6$ (Fig.~\ref{fig:states_fa18}), the evolution is already cloud converged at $n_{\max} = 7$ and the same as the non-interacting limit (this statement is not as trivial as it sounds, see the dashed lines in Fig.~\ref{fig:GW_impact}). At $f_a = 10^{12}\,$GeV and $\alpha = 1$ (Fig.~\ref{fig:states_fa12}), the strong coupling instead drives the cloud almost immediately into a quasi-steady configuration. There, many levels are populated simultaneously at $M_{\rm c}^{n\ell m}/M_i \sim 10^{-13}$, about eleven orders of magnitude below the non-interacting cloud.
\subsection{Spin evolution}  
\begin{figure*}
    \centering
    \includegraphics[width=0.98\linewidth]{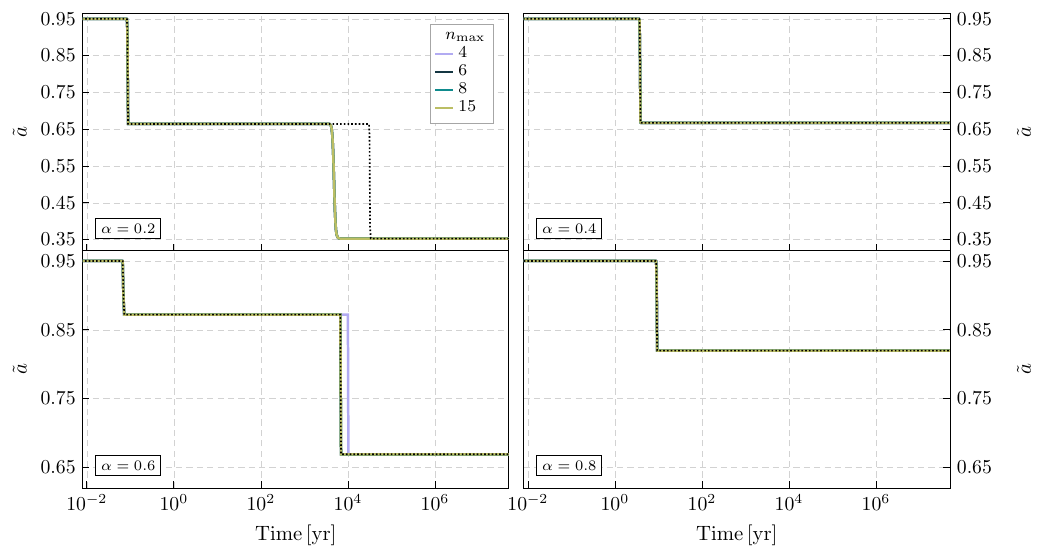}
    \caption{Same as Fig.~\ref{fig:spin-down-convergence}, for $f_a = 10^{18}\,$GeV.}
    \label{fig:spin_fa18}
\end{figure*}

\begin{figure*}
    \centering
    \includegraphics[width=0.98\linewidth]{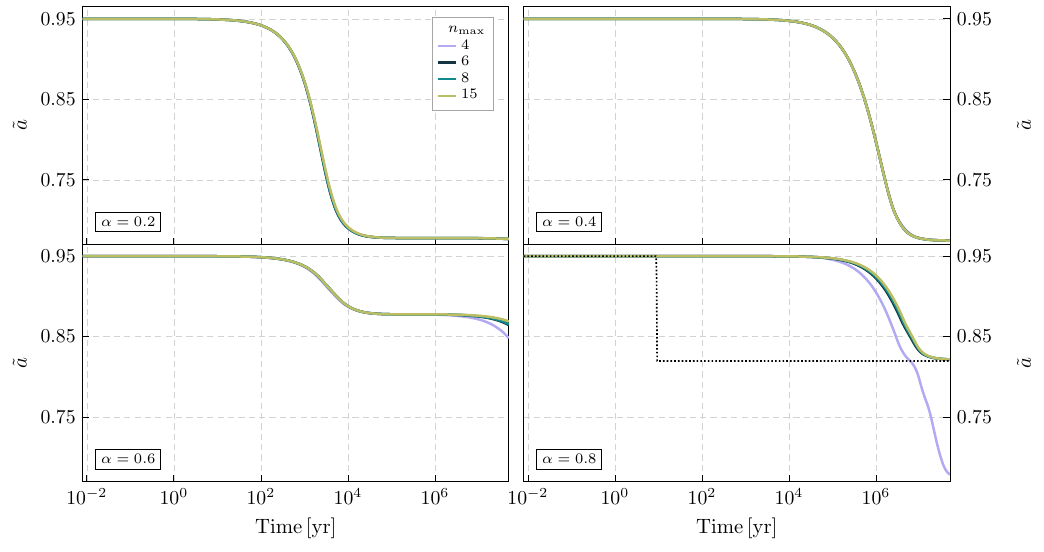}
    \caption{Same as Fig.~\ref{fig:spin-down-convergence}, for $f_a = 10^{14}\,$GeV.}
    \label{fig:spin_fa14}
\end{figure*}

\begin{figure*}
    \centering
    \includegraphics[width=0.98\linewidth]{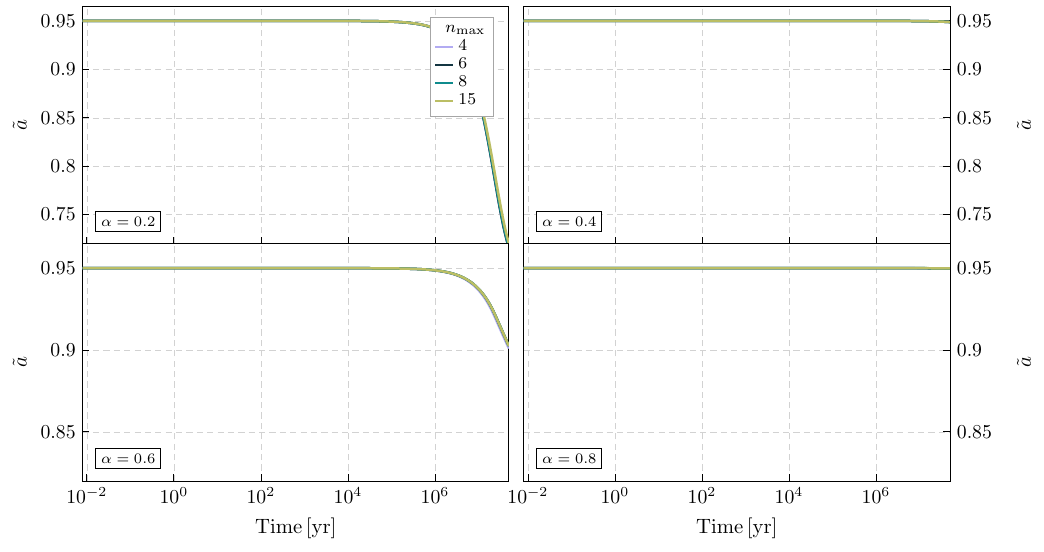}
    \caption{Same as Fig.~\ref{fig:spin-down-convergence}, for $f_a = 10^{12}\,$GeV.}
    \label{fig:spin_fa12}
\end{figure*}

Figures~\ref{fig:spin_fa18}--\ref{fig:spin_fa12} repeat the convergence study of Fig.~\ref{fig:spin-down-convergence} for $M = 10\,M_\odot$ and $f_a = 10^{18}$, $10^{14}$ and $10^{12}\,$GeV. At $f_a = 10^{18}\,$GeV all truncations agree, and the spin closely follows the non-interacting limit. The exception is $\alpha = 0.2$, where self-interactions trigger the $m = 2$ spin-down about five times earlier. At $f_a = 10^{14}\,$GeV the spin-down is delayed by several orders of magnitude in time, but still converged in $n_{\max}$. At $f_a = 10^{12}\,$GeV the black hole barely spins down: only at $\alpha = 0.2$ and $0.6$ does the spin drop appreciably within a Salpeter time.
\subsection{Higher black hole masses} 
\begin{figure*}
    \centering    
    \includegraphics[width=0.98\linewidth]{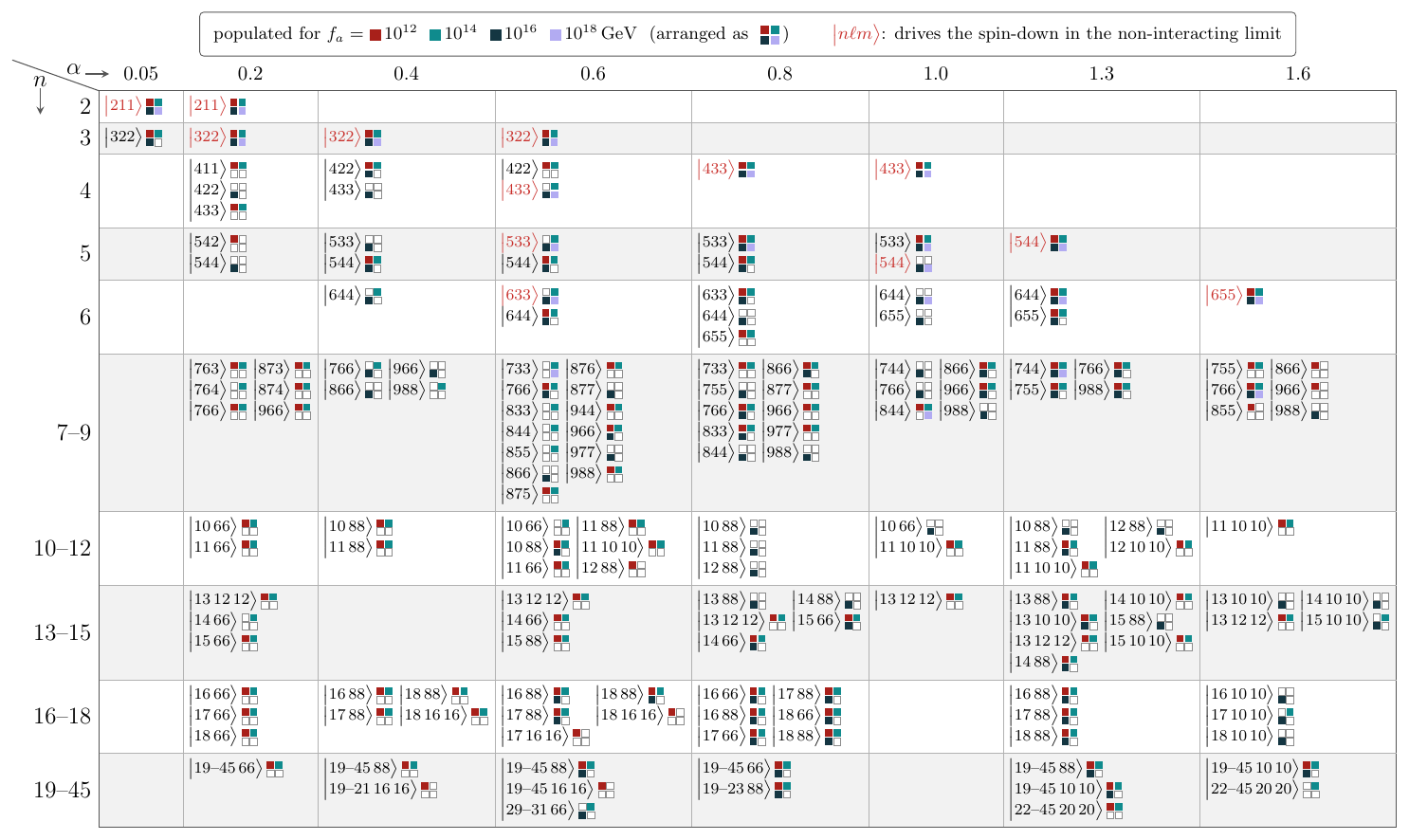}
    \caption{Same as Fig.~\ref{fig:tab_tot}, for $M = 10^4\,M_\odot$ evolved for 1\,Gyr.}
    \label{fig:table_M1e4}
\end{figure*}

\begin{figure*}
    \centering    
    \includegraphics[width=0.98\linewidth]{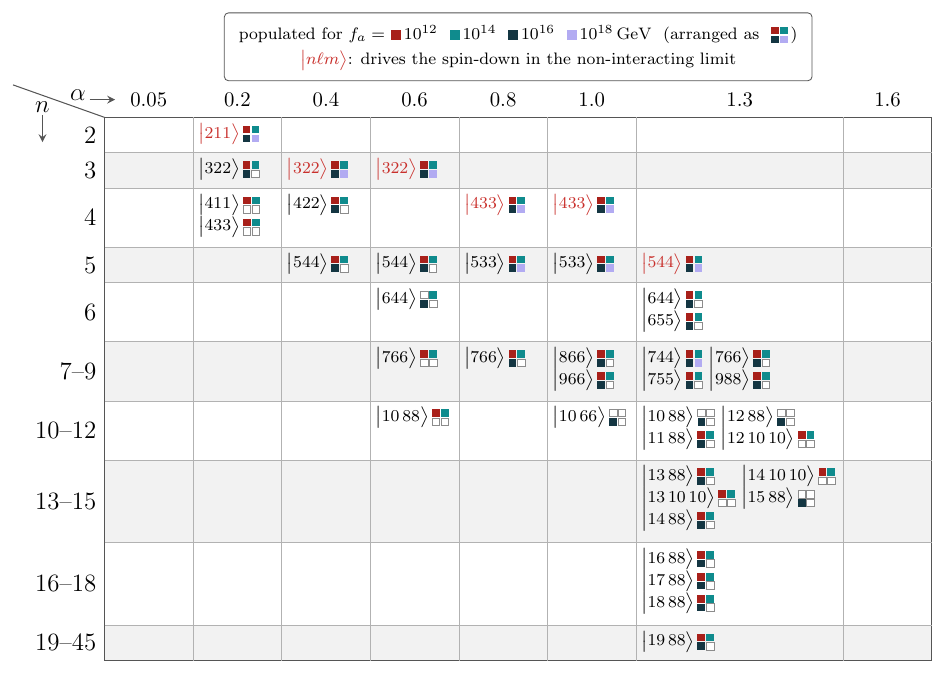}
    \caption{Same as Fig.~\ref{fig:tab_tot}, for $M = 10^8\,M_\odot$ evolved for 1\,Gyr. All runs are cloud-converged. No states grow within 1\,Gyr at $\alpha = 0.05$ and $1.6$.}
    \label{fig:table_M1e8}
\end{figure*}

Figures~\ref{fig:table_M1e4} and~\ref{fig:table_M1e8} show the populated states for $M = 10^4$ and $10^8\,M_\odot$, in the same format as Fig.~\ref{fig:tab_tot}. Because heavier black holes cover a shorter stretch of the rescaled evolution (Fig.~\ref{fig:overview_states}), these tables contain fewer states. Their states are a subset of those found for $M = 10\,M_\odot$. All runs for $M = 10^8\,M_\odot$ are cloud-converged. For $M = 10^4\,M_\odot$, the $|n\,16\,16\rangle$ tower appears again at $\alpha = 0.4$ and $0.6$.
\subsection{Number of active states}
\begin{figure*}
    \centering
    \includegraphics[width=0.98\linewidth]{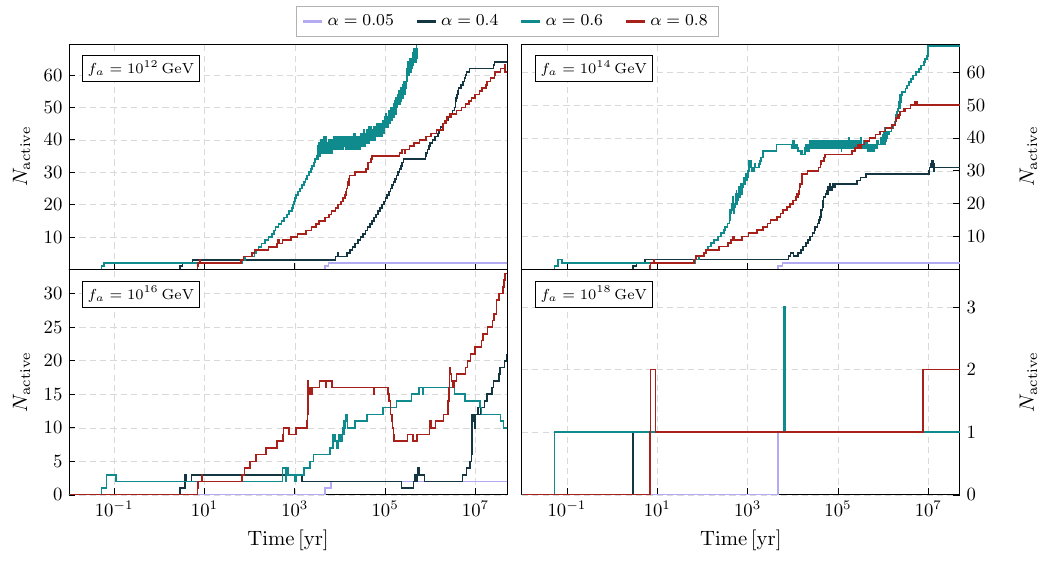}
    \caption{Number of simultaneously active superradiant states, $N_{\rm active}$, for $M = 10\,M_\odot$, $\tilde a_i = 0.95$ and four values of $f_a$ (panels).}
    \label{fig:active-states}
\end{figure*}

Finally, Fig.~\ref{fig:active-states} gives a simple measure of the complexity of the cloud: the number of superradiant states that are simultaneously populated, $N_{\rm active}(t)$, \ie with occupation above $\epsilon_{\rm th} = 10^{-20}$, for $M = 10\,M_\odot$ and four values of $f_a$.
\clearpage
\bibliography{References}

@article{Guo:2026qfb,
    author = "Guo, Yin-Da and Xu, Qi-Xuan and Brito, Richard and Cannizzaro, Enrico",
    title = "{Revisiting environmental effects on black hole quasibound-state spectra with relativistic perturbation theory}",
    eprint = "2608.14971",
    archivePrefix = "arXiv",
    primaryClass = "gr-qc",
    journal = "",
    month = "8",
    year = "2026"
}

@article{Lestingi:2026peq,
    author = "Lestingi, Jacopo and Sberna, Laura and Green, Stephen R.",
    title = {{Schr{\"o}dinger perturbation theory for black hole quasinormal modes}},
    eprint = "2607.19492",
    journal = "",
    archivePrefix = "arXiv",
    primaryClass = "gr-qc",
    month = "7",
    year = "2026"
}

@article{East:2022ppo,
    author = "East, William E.",
    title = "{Vortex String Formation in Black Hole Superradiance of a Dark Photon with the Higgs Mechanism}",
    eprint = "2205.03417",
    archivePrefix = "arXiv",
    primaryClass = "hep-ph",
    doi = "10.1103/PhysRevLett.129.141103",
    journal = "Phys. Rev. Lett.",
    volume = "129",
    number = "14",
    pages = "141103",
    year = "2022"
}

@article{Aswathi:2025nxa,
    author = "Aswathi, P. S. and East, William E. and Siemonsen, Nils and Sun, Ling and Jones, Dana",
    title = "{Ultralight boson constraints from gravitational wave observations of spinning binary black holes}",
    eprint = "2507.20979",
    archivePrefix = "arXiv",
    primaryClass = "gr-qc",
    doi = "10.1103/n5hr-zljn",
    journal = "Phys. Rev. D",
    volume = "112",
    number = "12",
    pages = "123048",
    year = "2025"
}

@article{May:2024npn,
    author = "May, Taillte and East, William E. and Siemonsen, Nils",
    title = "{Self-gravity effects of ultralight boson clouds formed by black hole superradiance}",
    eprint = "2410.21442",
    archivePrefix = "arXiv",
    primaryClass = "gr-qc",
    doi = "10.1103/PhysRevD.111.044062",
    journal = "Phys. Rev. D",
    volume = "111",
    number = "4",
    pages = "044062",
    year = "2025"
}

@article{Boskovic:2025ixx,
    author = "Bo{\v{s}}kovi{\'c}, Mateja and Porto, Rafael A. and Koschnitzke, Matthias",
    title = "{Trails of clouds in binary black holes}",
    eprint = "2512.17887",
    archivePrefix = "arXiv",
    primaryClass = "gr-qc",
    journal = "",
    reportNumber = "DESY-25-189",
    month = "12",
    year = "2025"
}

@article{Caputo:2025oap,
    author = "Caputo, Andrea and Franciolini, Gabriele and Witte, Samuel J.",
    title = "{Superradiance constraints from GW231123}",
    eprint = "2507.21788",
    archivePrefix = "arXiv",
    primaryClass = "hep-ph",
    reportNumber = "CERN-TH-2025-147, DESY-25-110",
    doi = "10.1103/f7w5-36ng",
    journal = "Phys. Rev. D",
    volume = "113",
    number = "8",
    pages = "083031",
    year = "2026"
}

@ARTICLE{1964ApJ...140..796S,
       author = {{Salpeter}, E.~E.},
        title = "{Accretion of Interstellar Matter by Massive Objects.}",
      journal = {\apj},
         year = 1964,
        month = aug,
       volume = {140},
        pages = {796-800},
          doi = {10.1086/147973},
       adsurl = {https://ui.adsabs.harvard.edu/abs/1964ApJ...140..796S}
}

@article{Gruzinov:2016hcq,
    author = "Gruzinov, Andrei",
    title = "{Black Hole Spindown by Light Bosons}",
    eprint = "1604.06422",
    journal = "",
    archivePrefix = "arXiv",
    primaryClass = "astro-ph.HE",
    month = "4",
    year = "2016"
}

@article{Yoshino:2015nsa,
    author = "Yoshino, Hirotaka and Kodama, Hideo",
    title = "{The bosenova and axiverse}",
    eprint = "1505.00714",
    archivePrefix = "arXiv",
    primaryClass = "gr-qc",
    reportNumber = "KEK-TH-1820, KEK-COSMO-169",
    doi = "10.1088/0264-9381/32/21/214001",
    journal = "Class. Quant. Grav.",
    volume = "32",
    number = "21",
    pages = "214001",
    year = "2015"
}

@article{Mathur:2020aqv,
    author = "Mathur, Anubhav and Rajendran, Surjeet and Tanin, Erwin H.",
    title = "{Clockwork mechanism to remove superradiance limits}",
    eprint = "2004.12326",
    archivePrefix = "arXiv",
    primaryClass = "hep-ph",
    doi = "10.1103/PhysRevD.102.055015",
    journal = "Phys. Rev. D",
    volume = "102",
    number = "5",
    pages = "055015",
    year = "2020"
}

@article{Caputo:2021efm,
    author = "Caputo, Andrea and Witte, Samuel J. and Blas, Diego and Pani, Paolo",
    title = "{Electromagnetic signatures of dark photon superradiance}",
    eprint = "2102.11280",
    archivePrefix = "arXiv",
    primaryClass = "hep-ph",
    doi = "10.1103/PhysRevD.104.043006",
    journal = "Phys. Rev. D",
    volume = "104",
    number = "4",
    pages = "043006",
    year = "2021"
}

@article{Cannizzaro:2022xyw,
    author = "Cannizzaro, Enrico and Sberna, Laura and Caputo, Andrea and Pani, Paolo",
    title = "{Dark photon superradiance quenched by dark matter}",
    eprint = "2206.12367",
    archivePrefix = "arXiv",
    primaryClass = "hep-ph",
    doi = "10.1103/PhysRevD.106.083019",
    journal = "Phys. Rev. D",
    volume = "106",
    number = "8",
    pages = "083019",
    year = "2022"
}

@article{Siemonsen:2022ivj,
    author = "Siemonsen, Nils and Mondino, Cristina and Egana-Ugrinovic, Daniel and Huang, Junwu and Baryakhtar, Masha and East, William E.",
    title = "{Dark photon superradiance: Electrodynamics and multimessenger signals}",
    eprint = "2212.09772",
    archivePrefix = "arXiv",
    primaryClass = "astro-ph.HE",
    doi = "10.1103/PhysRevD.107.075025",
    journal = "Phys. Rev. D",
    volume = "107",
    number = "7",
    pages = "075025",
    year = "2023"
}

@article{Noordhuis:2023wid,
    author = "Noordhuis, Dion and Prabhu, Anirudh and Weniger, Christoph and Witte, Samuel J.",
    title = "{Axion Clouds around Neutron Stars}",
    eprint = "2307.11811",
    archivePrefix = "arXiv",
    primaryClass = "hep-ph",
    doi = "10.1103/PhysRevX.14.041015",
    journal = "Phys. Rev. X",
    volume = "14",
    number = "4",
    pages = "041015",
    year = "2024"
}

@article{Dine:1982ah,
    author = "Dine, Michael and Fischler, Willy",
    editor = "Srednicki, M. A.",
    title = "{The Not So Harmless Axion}",
    reportNumber = "UPR-0201T",
    doi = "10.1016/0370-2693(83)90639-1",
    journal = "Phys. Lett. B",
    volume = "120",
    pages = "137--141",
    year = "1983"
}

@article{Abbott:1982af,
    author = "Abbott, L. F. and Sikivie, P.",
    editor = "Srednicki, M. A.",
    title = "{A Cosmological Bound on the Invisible Axion}",
    reportNumber = "PRINT-82-0695 (BRANDEIS)",
    doi = "10.1016/0370-2693(83)90638-X",
    journal = "Phys. Lett. B",
    volume = "120",
    pages = "133--136",
    year = "1983"
}

@article{Preskill:1982cy,
    author = "Preskill, John and Wise, Mark B. and Wilczek, Frank",
    editor = "Srednicki, M. A.",
    title = "{Cosmology of the Invisible Axion}",
    reportNumber = "HUTP-82-A048, NSF-ITP-82-103",
    doi = "10.1016/0370-2693(83)90637-8",
    journal = "Phys. Lett. B",
    volume = "120",
    pages = "127--132",
    year = "1983"
}

@article{Pani:2012bp,
    author = "Pani, Paolo and Cardoso, Vitor and Gualtieri, Leonardo and Berti, Emanuele and Ishibashi, Akihiro",
    title = "{Perturbations of slowly rotating black holes: massive vector fields in the Kerr metric}",
    eprint = "1209.0773",
    archivePrefix = "arXiv",
    primaryClass = "gr-qc",
    doi = "10.1103/PhysRevD.86.104017",
    journal = "Phys. Rev. D",
    volume = "86",
    pages = "104017",
    year = "2012"
}

@article{Cannizzaro:2023jle,
    author = "Cannizzaro, Enrico and Sberna, Laura and Green, Stephen R. and Hollands, Stefan",
    title = "{Relativistic Perturbation Theory for Black-Hole Boson Clouds}",
    eprint = "2309.10021",
    archivePrefix = "arXiv",
    primaryClass = "gr-qc",
    doi = "10.1103/PhysRevLett.132.051401",
    journal = "Phys. Rev. Lett.",
    volume = "132",
    number = "5",
    pages = "051401",
    year = "2024"
}

@article{Lingetti:2022psy,
    author = "Lingetti, Giuseppe and Cannizzaro, Enrico and Pani, Paolo",
    title = "{Superradiant instabilities by accretion disks in scalar-tensor theories}",
    eprint = "2204.09335",
    archivePrefix = "arXiv",
    primaryClass = "gr-qc",
    doi = "10.1103/PhysRevD.106.024007",
    journal = "Phys. Rev. D",
    volume = "106",
    number = "2",
    pages = "024007",
    year = "2022"
}

@article{Witte:2024drg,
    author = "Witte, Samuel J. and Mummery, Andrew",
    title = "{Stepping up superradiance constraints on axions}",
    eprint = "2412.03655",
    archivePrefix = "arXiv",
    primaryClass = "hep-ph",
    doi = "10.1103/PhysRevD.111.083044",
    journal = "Phys. Rev. D",
    volume = "111",
    number = "8",
    pages = "083044",
    year = "2025"
}

@article{Brito:2017zvb,
    author = "Brito, Richard and Ghosh, Shrobana and Barausse, Enrico and Berti, Emanuele and Cardoso, Vitor and Dvorkin, Irina and Klein, Antoine and Pani, Paolo",
    title = "{Gravitational wave searches for ultralight bosons with LIGO and LISA}",
    eprint = "1706.06311",
    archivePrefix = "arXiv",
    primaryClass = "gr-qc",
    doi = "10.1103/PhysRevD.96.064050",
    journal = "Phys. Rev. D",
    volume = "96",
    number = "6",
    pages = "064050",
    year = "2017"
}

@article{Baryakhtar:2020gao,
    author = "Baryakhtar, Masha and Galanis, Marios and Lasenby, Robert and Simon, Olivier",
    title = "{Black hole superradiance of self-interacting scalar fields}",
    eprint = "2011.11646",
    archivePrefix = "arXiv",
    primaryClass = "hep-ph",
    doi = "10.1103/PhysRevD.103.095019",
    journal = "Phys. Rev. D",
    volume = "103",
    number = "9",
    pages = "095019",
    year = "2021"
}

@article{Tomaselli:2023ysb,
    author = "Tomaselli, Giovanni Maria and Spieksma, Thomas F. M. and Bertone, Gianfranco",
    title = "{Dynamical friction in gravitational atoms}",
    eprint = "2305.15460",
    archivePrefix = "arXiv",
    primaryClass = "gr-qc",
    doi = "10.1088/1475-7516/2023/07/070",
    journal = "JCAP",
    volume = "07",
    pages = "070",
    year = "2023"
}

@article{Tomaselli:2024bdd,
    author = "Tomaselli, Giovanni Maria and Spieksma, Thomas F. M. and Bertone, Gianfranco",
    title = "{Resonant history of gravitational atoms in black hole binaries}",
    eprint = "2403.03147",
    archivePrefix = "arXiv",
    primaryClass = "gr-qc",
    doi = "10.1103/PhysRevD.110.064048",
    journal = "Phys. Rev. D",
    volume = "110",
    number = "6",
    pages = "064048",
    year = "2024"
}

@article{LIGOScientific:2021rnv,
    author = "Abbott, R. and others",
    collaboration = "LIGO Scientific, Virgo, KAGRA",
    title = "{All-sky search for gravitational wave emission from scalar boson clouds around spinning black holes in LIGO O3 data}",
    eprint = "2111.15507",
    archivePrefix = "arXiv",
    primaryClass = "astro-ph.HE",
    reportNumber = "P2100343",
    doi = "10.1103/PhysRevD.105.102001",
    journal = "Phys. Rev. D",
    volume = "105",
    number = "10",
    pages = "102001",
    year = "2022"
}

@article{Ng:2020jqd,
    author = "Ng, Ken K. Y. and Isi, Maximiliano and Haster, Carl-Johan and Vitale, Salvatore",
    title = "{Multiband gravitational-wave searches for ultralight bosons}",
    eprint = "2007.12793",
    archivePrefix = "arXiv",
    primaryClass = "gr-qc",
    reportNumber = "LIGO document number LIGO-P2000274",
    doi = "10.1103/PhysRevD.102.083020",
    journal = "Phys. Rev. D",
    volume = "102",
    number = "8",
    pages = "083020",
    year = "2020"
}

@article{Fan:2023jjj,
    author = "Fan, Kaiyuan and Tong, Xi and Wang, Yi and Zhu, Hui-Yu",
    title = "{Modulating binary dynamics via the termination of black hole superradiance}",
    eprint = "2311.17013",
    archivePrefix = "arXiv",
    primaryClass = "gr-qc",
    doi = "10.1103/PhysRevD.109.024059",
    journal = "Phys. Rev. D",
    volume = "109",
    number = "2",
    pages = "024059",
    year = "2024"
}

@article{Siemonsen:2022yyf,
    author = "Siemonsen, Nils and May, Taillte and East, William E.",
    title = "{Modeling the black hole superradiance gravitational waveform}",
    eprint = "2211.03845",
    archivePrefix = "arXiv",
    primaryClass = "gr-qc",
    doi = "10.1103/PhysRevD.107.104003",
    journal = "Phys. Rev. D",
    volume = "107",
    number = "10",
    pages = "104003",
    year = "2023"
}

@article{Tong:2022bbl,
    author = "Tong, Xi and Wang, Yi and Zhu, Hui-Yu",
    title = "{Termination of superradiance from a binary companion}",
    eprint = "2205.10527",
    archivePrefix = "arXiv",
    primaryClass = "gr-qc",
    doi = "10.1103/PhysRevD.106.043002",
    journal = "Phys. Rev. D",
    volume = "106",
    number = "4",
    pages = "043002",
    year = "2022"
}

@article{Boskovic:2024fga,
    author = "Bo\v{s}kovi\'c, Mateja and Koschnitzke, Matthias and Porto, Rafael A.",
    title = "{Signatures of Ultralight Bosons in the Orbital Eccentricity of Binary Black Holes}",
    eprint = "2403.02415",
    archivePrefix = "arXiv",
    primaryClass = "gr-qc",
    reportNumber = "DESY-24-030",
    doi = "10.1103/PhysRevLett.133.121401",
    journal = "Phys. Rev. Lett.",
    volume = "133",
    number = "12",
    pages = "121401",
    year = "2024"
}

@article{Duque:2023seg,
    author = "Duque, Francisco and Macedo, Caio F. B. and Vicente, Rodrigo and Cardoso, Vitor",
    title = "{Extreme-Mass-Ratio Inspirals in Ultralight Dark Matter}",
    eprint = "2312.06767",
    archivePrefix = "arXiv",
    primaryClass = "gr-qc",
    doi = "10.1103/PhysRevLett.133.121404",
    journal = "Phys. Rev. Lett.",
    volume = "133",
    number = "12",
    pages = "121404",
    year = "2024"
}

@article{Zouros:1979iw,
    author = "Zouros, T. J. M. and Eardley, D. M.",
    title = "{INSTABILITIES OF MASSIVE SCALAR PERTURBATIONS OF A ROTATING BLACK HOLE}",
    doi = "10.1016/0003-4916(79)90237-9",
    journal = "Annals Phys.",
    volume = "118",
    pages = "139--155",
    year = "1979"
}

@article{Starobinsky:1973aij,
    author = "Starobinsky, A. A.",
    title = "{Amplification of waves reflected from a rotating ``black hole''\,}",
    journal = "Sov. Phys. JETP",
    volume = "37",
    number = "1",
    pages = "28--32",
    year = "1973"
}

@article{Rosa:2017ury,
    author = "Rosa, Jo\~ao G. and Kephart, Thomas W.",
    title = "{Stimulated Axion Decay in Superradiant Clouds around Primordial Black Holes}",
    eprint = "1709.06581",
    archivePrefix = "arXiv",
    primaryClass = "gr-qc",
    doi = "10.1103/PhysRevLett.120.231102",
    journal = "Phys. Rev. Lett.",
    volume = "120",
    number = "23",
    pages = "231102",
    year = "2018"
}

@article{Blas:2020kaa,
    author = "Blas, Diego and Witte, Samuel J.",
    title = "{Quenching Mechanisms of Photon Superradiance}",
    eprint = "2009.10075",
    archivePrefix = "arXiv",
    primaryClass = "hep-ph",
    doi = "10.1103/PhysRevD.102.123018",
    journal = "Phys. Rev. D",
    volume = "102",
    number = "12",
    pages = "123018",
    year = "2020"
}

@article{Ikeda:2018nhb,
    author = "Ikeda, Taishi and Brito, Richard and Cardoso, Vitor",
    title = "{Blasts of Light from Axions}",
    eprint = "1811.04950",
    archivePrefix = "arXiv",
    primaryClass = "gr-qc",
    doi = "10.1103/PhysRevLett.122.081101",
    journal = "Phys. Rev. Lett.",
    volume = "122",
    number = "8",
    pages = "081101",
    year = "2019"
}

@article{Leaver:1985ax,
    author = "Leaver, E. W.",
    title = "{An Analytic representation for the quasi normal modes of Kerr black holes}",
    doi = "10.1098/rspa.1985.0119",
    journal = "Proc. Roy. Soc. Lond. A",
    volume = "402",
    pages = "285--298",
    year = "1985"
}

@article{Dolan:2007mj,
    author = "Dolan, Sam R.",
    title = "{Instability of the massive Klein-Gordon field on the Kerr spacetime}",
    eprint = "0705.2880",
    archivePrefix = "arXiv",
    primaryClass = "gr-qc",
    doi = "10.1103/PhysRevD.76.084001",
    journal = "Phys. Rev. D",
    volume = "76",
    pages = "084001",
    year = "2007"
}

@article{Spieksma:2023vwl,
    author = "Spieksma, Thomas F. M. and Cannizzaro, Enrico and Ikeda, Taishi and Cardoso, Vitor and Chen, Yifan",
    title = "{Superradiance: Axionic couplings and plasma effects}",
    eprint = "2306.16447",
    archivePrefix = "arXiv",
    primaryClass = "gr-qc",
    doi = "10.1103/PhysRevD.108.063013",
    journal = "Phys. Rev. D",
    volume = "108",
    number = "6",
    pages = "063013",
    year = "2023"
}

@article{Arvanitaki:2010sy,
    author = "Arvanitaki, Asimina and Dubovsky, Sergei",
    title = "{Exploring the String Axiverse with Precision Black Hole Physics}",
    eprint = "1004.3558",
    archivePrefix = "arXiv",
    primaryClass = "hep-th",
    doi = "10.1103/PhysRevD.83.044026",
    journal = "Phys. Rev. D",
    volume = "83",
    pages = "044026",
    year = "2011"
}

@article{East:2017ovw,
    author = "East, William E. and Pretorius, Frans",
    title = "{Superradiant Instability and Backreaction of Massive Vector Fields around Kerr Black Holes}",
    eprint = "1704.04791",
    archivePrefix = "arXiv",
    primaryClass = "gr-qc",
    doi = "10.1103/PhysRevLett.119.041101",
    journal = "Phys. Rev. Lett.",
    volume = "119",
    number = "4",
    pages = "041101",
    year = "2017"
}

@article{ZelDovich1972,
       author = {{Zel'Dovich}, Ya. B.},
        title = "{Amplification of Cylindrical Electromagnetic Waves Reflected from a Rotating Body}",
      journal = {Soviet Journal of Experimental and Theoretical Physics},
         year = 1972,
        month = jan,
       volume = {35},
        pages = {1085},
       adsurl = {https://ui.adsabs.harvard.edu/abs/1972JETP...35.1085Z}
}

@article{Lucas_prep,
	author = "Gathercole, Lucas and Spieksma, Thomas F. M. and Bo\v{s}kovi\'c, Mateja and Witte, Samuel J.",
    note = "\emph{In preparation.}",
    title = "",
    journal = "",
    year = ""
}

@article{Roy:2025qaa,
    author = "Roy, Soumen and Vicente, Rodrigo and Aurrekoetxea, Josu C. and Clough, Katy and Ferreira, Pedro G.",
    title = "{Scalar Fields around Black Hole Binaries in LIGO-Virgo-KAGRA}",
    eprint = "2510.17967",
    archivePrefix = "arXiv",
    primaryClass = "gr-qc",
    reportNumber = "MIT-CTP/5937",
    doi = "10.1103/fv9z-zkxx",
    journal = "Phys. Rev. Lett.",
    volume = "136",
    number = "19",
    pages = "191402",
    year = "2026"
}

@article{Thorne:1980ru,
    author = "Thorne, K. S.",
    title = "{Multipole Expansions of Gravitational Radiation}",
    doi = "10.1103/RevModPhys.52.299",
    journal = "Rev. Mod. Phys.",
    volume = "52",
    pages = "299--339",
    year = "1980"
}

@inproceedings{Baryakhtar:2022hbu,
    author = "Baryakhtar, Masha and others",
    title = "{Dark Matter In Extreme Astrophysical Environments}",
    booktitle = "{Snowmass 2021}",
    eprint = "2203.07984",
    archivePrefix = "arXiv",
    primaryClass = "hep-ph",
    month = "3",
    year = "2022"
}

@article{Cheng:2026dsq,
    author = "Cheng, Yu and Caputo, Andrea and Tomaselli, Giovanni Maria and Witte, Samuel J.",
    title = "{Stars Can Quench Superradiance Around Supermassive Black Holes}",
    eprint = "2610.40895",
    archivePrefix = "arXiv",
    primaryClass = "hep-ph",
    reportNumber = "DESY-26-146, CERN-TH-2026-274",
    journal  ="",
    month = "10",
    year = "2026"
}

@article{Omiya:2024xlz,
    author = "Omiya, Hidetoshi and Takahashi, Takuya and Tanaka, Takahiro and Yoshino, Hirotaka",
    title = "{Deci-Hz gravitational waves from the self-interacting axion cloud around a rotating stellar mass black hole}",
    eprint = "2404.16265",
    archivePrefix = "arXiv",
    primaryClass = "gr-qc",
    doi = "10.1103/PhysRevD.110.044002",
    journal = "Phys. Rev. D",
    volume = "110",
    number = "4",
    pages = "044002",
    year = "2024"
}

@article{Boskovic:2026eth,
    author = "Bo{\v{s}}kovi{\'c}, Mateja and Savi{\'c}, Nikola",
    title = "{Perturbing Gravitational Atoms: Negative Love, Resonant Tides and Shifted Resonances}",
    eprint = "2607.27174",
    archivePrefix = "arXiv",
    primaryClass = "gr-qc",
    reportNumber = "DESY-26-095",
    journal = "",
    month = "7",
    year = "2026"
}

@article{Collaviti:2024mvh,
    author = "Collaviti, Spencer and Sun, Ling and Galanis, Marios and Baryakhtar, Masha",
    title = "{Observational prospects of self-interacting scalar superradiance with next-generation gravitational-wave detectors}",
    eprint = "2407.04304",
    archivePrefix = "arXiv",
    primaryClass = "gr-qc",
    doi = "10.1088/1361-6382/ad96ff",
    journal = "Class. Quant. Grav.",
    volume = "42",
    number = "2",
    pages = "025006",
    year = "2025"
}

@article{ZelDovich1971,
       author = {{Zel'Dovich}, Ya. B.},
        title = "{Generation of Waves by a Rotating Body}",
      journal = {Soviet Journal of Experimental and Theoretical Physics Letters},
         year = 1971,
        month = aug,
       volume = {14},
        pages = {180},
       adsurl = {https://ui.adsabs.harvard.edu/abs/1971JETPL..14..180Z}
}

@article{Herdeiro:2021znw,
    author = "Herdeiro, Carlos A. R. and Radu, Eugen and Santos, Nuno M.",
    title = "{A bound on energy extraction (and hairiness) from superradiance}",
    eprint = "2111.03667",
    archivePrefix = "arXiv",
    primaryClass = "gr-qc",
    doi = "10.1016/j.physletb.2021.136835",
    journal = "Phys. Lett. B",
    volume = "824",
    pages = "136835",
    year = "2022"
}

@article{Berti:2005gp,
      author         = "Berti, Emanuele and Cardoso, Vitor and Casals, Marc",
      title          = "{Eigenvalues and eigenfunctions of spin-weighted
                        spheroidal harmonics in four and higher dimensions}",
      journal        = "Phys. Rev.",
      volume         = "D73",
      year           = "2006",
      pages          = "024013",
      doi            = "10.1103/PhysRevD.73.109902, 10.1103/PhysRevD.73.024013",
      note           = "[Erratum: Phys. Rev.D73,109902(2006)]",
      eprint         = "gr-qc/0511111",
      archivePrefix  = "arXiv",
      primaryClass   = "gr-qc",
      SLACcitation   = "%%CITATION = GR-QC/0511111;%%"
}

@article{Brito:2015oca,
      author         = "Brito, Richard and Cardoso, Vitor and Pani, Paolo",
      title          = "{Superradiance}",
      journal        = "Lect. Notes Phys.",
      volume         = "906",
      year           = "2015",
      pages          = "pp.1-237",
      doi            = "10.1007/978-3-319-19000-6",
      eprint         = "1501.06570",
      archivePrefix  = "arXiv",
      primaryClass   = "gr-qc",
      SLACcitation   = "%%CITATION = ARXIV:1501.06570;%%"
}

@article{Baumann:2018vus,
    author = "Baumann, Daniel and Chia, Horng Sheng and Porto, Rafael A.",
    title = "{Probing Ultralight Bosons with Binary Black Holes}",
    eprint = "1804.03208",
    archivePrefix = "arXiv",
    primaryClass = "gr-qc",
    reportNumber = "DESY-18-060, DESY 18-060",
    doi = "10.1103/PhysRevD.99.044001",
    journal = "Phys. Rev. D",
    volume = "99",
    number = "4",
    pages = "044001",
    year = "2019"
}

@article{Tsukada:2020lgt,
    author = "Tsukada, Leo and Brito, Richard and East, William E. and Siemonsen, Nils",
    title = "{Modeling and searching for a stochastic gravitational-wave background from ultralight vector bosons}",
    eprint = "2011.06995",
    archivePrefix = "arXiv",
    primaryClass = "astro-ph.HE",
    doi = "10.1103/PhysRevD.103.083005",
    journal = "Phys. Rev. D",
    volume = "103",
    number = "8",
    pages = "083005",
    year = "2021"
}

@article{Baumann:2022pkl,
    author = "Baumann, Daniel and Bertone, Gianfranco and Stout, John and Tomaselli, Giovanni Maria",
    title = "{Sharp Signals of Boson Clouds in Black Hole Binary Inspirals}",
    eprint = "2206.01212",
    archivePrefix = "arXiv",
    primaryClass = "gr-qc",
    doi = "10.1103/PhysRevLett.128.221102",
    journal = "Phys. Rev. Lett.",
    volume = "128",
    number = "22",
    pages = "221102",
    year = "2022"
}

@article{Tomaselli:2024dbw,
    author = "Tomaselli, Giovanni Maria and Spieksma, Thomas F. M. and Bertone, Gianfranco",
    title = "{Legacy of Boson Clouds on Black Hole Binaries}",
    eprint = "2407.12908",
    archivePrefix = "arXiv",
    primaryClass = "gr-qc",
    doi = "10.1103/PhysRevLett.133.121402",
    journal = "Phys. Rev. Lett.",
    volume = "133",
    number = "12",
    pages = "121402",
    year = "2024"
}

@article{Baumann:2021fkf,
    author = "Baumann, Daniel and Bertone, Gianfranco and Stout, John and Tomaselli, Giovanni Maria",
    title = "{Ionization of gravitational atoms}",
    eprint = "2112.14777",
    archivePrefix = "arXiv",
    primaryClass = "gr-qc",
    doi = "10.1103/PhysRevD.105.115036",
    journal = "Phys. Rev. D",
    volume = "105",
    number = "11",
    pages = "115036",
    year = "2022"
}

@article{Baumann:2019ztm,
    author = "Baumann, Daniel and Chia, Horng Sheng and Porto, Rafael A. and Stout, John",
    title = "{Gravitational Collider Physics}",
    eprint = "1912.04932",
    archivePrefix = "arXiv",
    primaryClass = "gr-qc",
    reportNumber = "DESY-19-221, DESY 19-221",
    doi = "10.1103/PhysRevD.101.083019",
    journal = "Phys. Rev. D",
    volume = "101",
    number = "8",
    pages = "083019",
    year = "2020"
}

@article{Zhang:2019eid,
    author = "Zhang, Jun and Yang, Huan",
    title = "{Dynamic Signatures of Black Hole Binaries with Superradiant Clouds}",
    eprint = "1907.13582",
    archivePrefix = "arXiv",
    primaryClass = "gr-qc",
    reportNumber = "Imperial/TP/2019/JZ/02",
    doi = "10.1103/PhysRevD.101.043020",
    journal = "Phys. Rev. D",
    volume = "101",
    number = "4",
    pages = "043020",
    year = "2020"
}

@article{Berti:2019wnn,
    author = "Berti, Emanuele and Brito, Richard and Macedo, Caio F. B. and Raposo, Guilherme and Rosa, Joao Luis",
    title = "{Ultralight boson cloud depletion in binary systems}",
    eprint = "1904.03131",
    archivePrefix = "arXiv",
    primaryClass = "gr-qc",
    doi = "10.1103/PhysRevD.99.104039",
    journal = "Phys. Rev. D",
    volume = "99",
    number = "10",
    pages = "104039",
    year = "2019"
}

@article{Baumann:2019eav,
    author = "Baumann, Daniel and Chia, Horng Sheng and Stout, John and ter Haar, Lotte",
    title = "{The Spectra of Gravitational Atoms}",
    eprint = "1908.10370",
    archivePrefix = "arXiv",
    primaryClass = "gr-qc",
    doi = "10.1088/1475-7516/2019/12/006",
    journal = "JCAP",
    volume = "12",
    pages = "006",
    year = "2019"
}

@article{Brito:2023pyl,
    author = "Brito, Richard and Shah, Shreya",
    title = "{Extreme mass-ratio inspirals into black holes surrounded by scalar clouds}",
    eprint = "2307.16093",
    archivePrefix = "arXiv",
    primaryClass = "gr-qc",
    doi = "10.1103/PhysRevD.108.084019",
    journal = "Phys. Rev. D",
    volume = "108",
    number = "8",
    pages = "084019",
    year = "2023"
}

@article{Ellis:2026gwt,
    author = {Ellis, Sebastian A. R. and Ning, Orion and Rodd, Nicholas L. and Sch{\"u}tte-Engel, Jan},
    title = "{Crowdsourcing Gravitational Waves from Superradiant Axions}",
    journal = "",
    eprint = "2603.15734",
    archivePrefix = "arXiv",
    primaryClass = "hep-ph",
    reportNumber = "RIKEN-iTHEMS-Report-26",
    month = "3",
    year = "2026"
}

@article{Arvanitaki:2016qwi,
    author = "Arvanitaki, Asimina and Baryakhtar, Masha and Dimopoulos, Savas and Dubovsky, Sergei and Lasenby, Robert",
    title = "{Black Hole Mergers and the QCD Axion at Advanced LIGO}",
    eprint = "1604.03958",
    archivePrefix = "arXiv",
    primaryClass = "hep-ph",
    doi = "10.1103/PhysRevD.95.043001",
    journal = "Phys. Rev. D",
    volume = "95",
    number = "4",
    pages = "043001",
    year = "2017"
}

@article{Tsukada:2018mbp,
    author = "Tsukada, Leo and Callister, Thomas and Matas, Andrew and Meyers, Patrick",
    title = "{First search for a stochastic gravitational-wave background from ultralight bosons}",
    eprint = "1812.09622",
    archivePrefix = "arXiv",
    primaryClass = "astro-ph.HE",
    doi = "10.1103/PhysRevD.99.103015",
    journal = "Phys. Rev. D",
    volume = "99",
    number = "10",
    pages = "103015",
    year = "2019"
}

@article{Palomba:2019vxe,
    author = "Palomba, Cristiano and others",
    title = "{Direct constraints on ultra-light boson mass from searches for continuous gravitational waves}",
    eprint = "1909.08854",
    archivePrefix = "arXiv",
    primaryClass = "astro-ph.HE",
    doi = "10.1103/PhysRevLett.123.171101",
    journal = "Phys. Rev. Lett.",
    volume = "123",
    pages = "171101",
    year = "2019"
}

@article{Hannuksela:2018izj,
    author = "Hannuksela, Otto A. and Wong, Kaze W. K. and Brito, Richard and Berti, Emanuele and Li, Tjonnie G. F.",
    title = "{Probing the existence of ultralight bosons with a single gravitational-wave measurement}",
    eprint = "1804.09659",
    archivePrefix = "arXiv",
    primaryClass = "astro-ph.HE",
    doi = "10.1038/s41550-019-0712-4",
    journal = "Nature Astron.",
    volume = "3",
    number = "5",
    pages = "447--451",
    year = "2019"
}

@article{Yuan:2022bem,
    author = "Yuan, Chen and Jiang, Yang and Huang, Qing-Guo",
    title = "{Constraints on an ultralight scalar boson from Advanced LIGO and Advanced Virgo\textquoteright{}s first three observing runs using the stochastic gravitational-wave background}",
    eprint = "2204.03482",
    archivePrefix = "arXiv",
    primaryClass = "astro-ph.CO",
    doi = "10.1103/PhysRevD.106.023020",
    journal = "Phys. Rev. D",
    volume = "106",
    number = "2",
    pages = "023020",
    year = "2022"
}

@article{Brito:2014wla,
    author = "Brito, Richard and Cardoso, Vitor and Pani, Paolo",
    title = "{Black holes as particle detectors: evolution of superradiant instabilities}",
    eprint = "1411.0686",
    archivePrefix = "arXiv",
    primaryClass = "gr-qc",
    doi = "10.1088/0264-9381/32/13/134001",
    journal = "Class. Quant. Grav.",
    volume = "32",
    number = "13",
    pages = "134001",
    year = "2015"
}

@article{Ferreira:2017pth,
    author = "Ferreira, Miguel C. and Macedo, Caio F. B. and Cardoso, Vitor",
    title = "{Orbital fingerprints of ultralight scalar fields around black holes}",
    eprint = "1710.00830",
    archivePrefix = "arXiv",
    primaryClass = "gr-qc",
    doi = "10.1103/PhysRevD.96.083017",
    journal = "Phys. Rev. D",
    volume = "96",
    number = "8",
    pages = "083017",
    year = "2017"
}

@article{Detweiler:1980uk,
    author = "Detweiler, Steven L.",
    title = "{KLEIN-GORDON EQUATION AND ROTATING BLACK HOLES}",
    doi = "10.1103/PhysRevD.22.2323",
    journal = "Phys. Rev. D",
    volume = "22",
    pages = "2323--2326",
    year = "1980"
}

@article{Yoshino:2013ofa,
    author = "Yoshino, Hirotaka and Kodama, Hideo",
    title = "{Gravitational radiation from an axion cloud around a black hole: Superradiant phase}",
    eprint = "1312.2326",
    archivePrefix = "arXiv",
    primaryClass = "gr-qc",
    reportNumber = "KEK-TH-1694",
    doi = "10.1093/ptep/ptu029",
    journal = "PTEP",
    volume = "2014",
    pages = "043E02",
    year = "2014"
}

@article{Weinberg:1977ma,
    author = "Weinberg, Steven",
    title = "{A New Light Boson?}",
    reportNumber = "HUTP-77/A074",
    doi = "10.1103/PhysRevLett.40.223",
    journal = "Phys. Rev. Lett.",
    volume = "40",
    pages = "223--226",
    year = "1978"
}

@article{Peccei:1977hh,
    author = "Peccei, R. D. and Quinn, Helen R.",
    title = "{CP Conservation in the Presence of Instantons}",
    reportNumber = "ITP-568-STANFORD",
    doi = "10.1103/PhysRevLett.38.1440",
    journal = "Phys. Rev. Lett.",
    volume = "38",
    pages = "1440--1443",
    year = "1977"
}

@article{Wilczek:1977pj,
    author = "Wilczek, Frank",
    title = "{Problem of Strong  $P$  and  $T$  Invariance in the Presence of Instantons}",
    reportNumber = "Print-77-0939 (COLUMBIA)",
    doi = "10.1103/PhysRevLett.40.279",
    journal = "Phys. Rev. Lett.",
    volume = "40",
    pages = "279--282",
    year = "1978"
}

@article{Svrcek:2006yi,
    author = "Svrcek, Peter and Witten, Edward",
    title = "{Axions In String Theory}",
    eprint = "hep-th/0605206",
    archivePrefix = "arXiv",
    reportNumber = "SLAC-PUB-11894",
    doi = "10.1088/1126-6708/2006/06/051",
    journal = "JHEP",
    volume = "06",
    pages = "051",
    year = "2006"
}

@article{Yoshino:2012kn,
    author = "Yoshino, Hirotaka and Kodama, Hideo",
    title = "{Bosenova collapse of axion cloud around a rotating black hole}",
    eprint = "1203.5070",
    archivePrefix = "arXiv",
    primaryClass = "gr-qc",
    reportNumber = "KEK-TH-1530",
    doi = "10.1143/PTP.128.153",
    journal = "Prog. Theor. Phys.",
    volume = "128",
    pages = "153--190",
    year = "2012"
}

@misc{AxionSR:zenodo,
  author       = {Cannizzaro, Enrico and Spieksma, Thomas F. M. and Witte, Samuel J.},
  title        = {{Data and figures for ``Toward the minimal set of states in axion superradiance''}},
  year         = {2026},
  publisher    = {Zenodo},
  version      = {1.0},
  doi          = {10.5281/zenodo.24118755},
  url          = {https://doi.org/10.5281/zenodo.24118755},
  note         = {Dataset}
}

@article{Takahashi:2024fyq,
    author = "Takahashi, Takuya and Omiya, Hidetoshi and Tanaka, Takahiro",
    title = "{Self-interacting axion clouds around rotating black holes in binary systems}",
    eprint = "2408.08349",
    archivePrefix = "arXiv",
    primaryClass = "gr-qc",
    reportNumber = "RUP-24-12",
    doi = "10.1103/PhysRevD.110.104038",
    journal = "Phys. Rev. D",
    volume = "110",
    number = "10",
    pages = "104038",
    year = "2024"
}

@article{Arvanitaki:2014wva,
      author         = "Arvanitaki, Asimina and Baryakhtar, Masha and Huang,
                        Xinlu",
      title          = "{Discovering the QCD Axion with Black Holes and
                        Gravitational Waves}",
      journal        = "Phys. Rev.",
      volume         = "D91",
      year           = "2015",
      number         = "8",
      pages          = "084011",
      doi            = "10.1103/PhysRevD.91.084011",
      eprint         = "1411.2263",
      archivePrefix  = "arXiv",
      primaryClass   = "hep-ph",
      SLACcitation   = "%%CITATION = ARXIV:1411.2263;%%"
}

@article{Arvanitaki:2009fg,
    author = "Arvanitaki, Asimina and Dimopoulos, Savas and Dubovsky, Sergei and Kaloper, Nemanja and March-Russell, John",
    title = "{String Axiverse}",
    eprint = "0905.4720",
    archivePrefix = "arXiv",
    primaryClass = "hep-th",
    doi = "10.1103/PhysRevD.81.123530",
    journal = "Phys. Rev. D",
    volume = "81",
    pages = "123530",
    year = "2010"
}

@misc{AxionSR:github,
  author = {Witte, Samuel J.},
  title = "{AxionSR}",
  year = {2026},
  howpublished = {\url{https://github.com/SamWitte/Axion_SR}}
}
\end{document}